\documentclass[reprint,superscriptaddress,showpacs,amssymb,amsmath,aps,prd,longbibliography]{revtex4-2}
\usepackage{graphicx}
\usepackage{xcolor}
\usepackage{amsmath}
\usepackage{amsthm}
\usepackage{bm}
\usepackage{bbm}
\usepackage{comment}
\excludecomment{Omitir}

\usepackage{mathdots}
\usepackage{lipsum}
\usepackage{verbatim}
\usepackage{natbib}
\usepackage{nccmath}

\usepackage[caption=false]{subfig}

\usepackage{amsfonts} 
\DeclareMathSymbol{\shortminus}{\mathbin}{AMSa}{"39}

\usepackage[colorlinks=true,citecolor=blue,linkcolor=magenta]{hyperref}

\begin{document}

\title{Spacetime quantum and classical mechanics with dynamical foliation}

\author{N. L.\  Diaz}
	
 \affiliation{Information Sciences, Los Alamos National Laboratory, Los Alamos, NM 87545, USA}
 \affiliation{Departamento de F\'isica-IFLP/CONICET,
		Universidad Nacional de La Plata, C.C. 67, La Plata (1900), Argentina}
	
	\author{J. M. Matera}
	\affiliation{Departamento de F\'isica-IFLP/CONICET,
		Universidad Nacional de La Plata, C.C. 67, La Plata (1900), Argentina}
	
	\author{R. Rossignoli}
	\affiliation{Departamento de F\'isica-IFLP/CONICET,
		Universidad Nacional de La Plata, C.C. 67, La Plata (1900), Argentina}
	\affiliation{Comisi\'on de Investigaciones Cient\'{\i}ficas (CIC), La Plata (1900), Argentina}
	\begin{abstract}
 The conventional phase space of classical physics treats space and time differently, and this difference carries over to field theories and quantum mechanics (QM). In this paper, the phase space is enhanced through two main extensions.
First, we promote the time choice of the Legendre transform to a dynamical variable. 
Second, we extend the Poisson brackets of matter fields 
to a spacetime symmetric form.
The ensuing ``spacetime phase space'' is employed to obtain an explicitly covariant version of Hamilton equations for relativistic field theories. A canonical-like quantization of the formalism is then presented in which 
the fields satisfy spacetime commutation relations and the foliation is quantum. 
In this approach, the classical action is also promoted to an operator  and retains explicit covariance through its 
nonseparability in the matter-foliation partition. The problem of establishing a correspondence between the new noncausal framework (where fields at different times are independent) and conventional QM 
is solved through a generalization of 
spacelike correlators to spacetime. In this generalization, the Hamiltonian is replaced by the action, and conventional particles 
by off-shell particles.
When the foliation is quantized, the 
previous map is recovered by conditioning on foliation eigenstates, in analogy 
with the Page and Wootters mechanism. 
We also 
provide 
an interpretation of the correspondence in which the causal structure of a given theory emerges from the quantum correlations between the system and an environment. This idea holds for general quantum systems and allows one to generalize the density matrix to an operator containing the information of correlators both in space and time. 

  \end{abstract}

	\maketitle

\section{Introduction}

Classical mechanics is built upon the Lagrangian and Hamiltonian formulations, both of which were developed before the advent of relativity and quantum mechanics (QM). Despite the revolutionary changes brought about by these later theories, the Lagrangian and Hamiltonian formalisms have remained largely unmodified. The Lagrangian approach 
has proven to be very well-suited in handling 
the spacetime symmetries revealed by Einstein's theories, while the Hamiltonian approach has widely inspired the QM framework and defines the canonical procedure to quantize a given theory. 
The use of the Hamiltonian formulation in relativity 
is less natural: 
the process of passing from a Lagrangian to a Hamiltonian involves selecting a specific time variable, which has the effect of singling out a particular observer. 
This explicit breaking of relativistic symmetries is inherited by the Hamiltonian phase space and carried over to the 
algebraic rules 
beneath every quantum theory.

At the same time, it is widely known that relativistic quantum field theories (QFTs) yield observer-independent predictions, even if a canonical Hamiltonian approach is followed. 
This important feature discussed
in the foundational years of QFTs \cite{dy.94} is also validated
by the related expressions in the Feynman's Path Integral (PI)
formulation \cite{Feynm.1948,feyn.05}, which emphasizes Lagrangians over Hamiltonians. The prize to pay by using PIs is that the conventional Hilbert space structure of canonical QM is replaced by
the use  
of “sums over histories” in classical configuration
space.

The previous seems to indicate that the asymmetries between space and time at the quantum level are not fundamental but rather an artifact of the canonical Hamiltonian formulation. 
One can then pose the problem of formulating QM in a manner that extends the familiar mathematical elements, such as states and operators, to be spacetime symmetric. Several discussions related to this issue, which apply both to relativistic and non-relativistic theories, have been advanced recently \cite{fit.15,ho.17,cot.18,giac.19,dia.19,diaz.21, diazp.21, hoh.21, fot.21, fav.22,pai.22}. These discussions highlight that the previous is an open and challenging problem of current interest: 
a genuine solution 
bears the potential to extend the insights associated with quantum correlations to the time domain. For example, the recent discussion about space emerging from entanglement \cite{van.10, Ca.17} cannot be extended straightforwardly to time (and then spacetime). 
Shifting to more applied areas, quantum computational protocols that employ quantum time ideas \cite{QT.15,b.16,b.18} to map temporal to spatial complexities have already been proposed \cite{barison2022variational, diaz2023parallel}. It is also clear that the issue is relevant in scenarios where general covariance comes into play, such as in quantum gravity, in which case the use of conventional QFTs techniques is no longer enough \cite{PaW.83, ish.85, ish.93, Ish.98,Ma.97,g.09,Ku.11, hoehn2023matter}.
The problem
calls for a critical revision  
of all the aspects involved, including the basic formulations of classical mechanics and, in particular, of the phase space of the Hamiltonian formulation.

In this work, we introduce a framework 
that seamlessly integrates relativistic covariance into an extended phase space which can be straightforwardly quantized. Our main focus is the case of special relativistic field theories, a scenario which  allows us to 
lay the foundations of a spacetime symmetric QM 
guided by Lorentz symmetry. 
Notably, several insights revealed by the relativistic case, including a map to conventional QM, can be applied 
to any quantum mechanical theory, nonrelativistic theories included.
As we remark throughout the manuscript, one can regard the final framework as an independent (spacetime symmetric) set of rules to formulate QM, and explore its consequences from the point of view of QM as a generalization of classical probability. This complementary point of view of our work, 
which seems to be particularly adequate to tackle the aforementioned foundational problems, is only 
preliminarily explored.

The  construction begins by
modifying the conventional phase space of Hamiltonian dynamics in two ways: In the first place, the time choice of the Legendre transformation which defines the Hamiltonian from a given Lagrangian is treated as dynamical.
Secondly, spacetime Poisson Brackets (PB) for matter fields that do not distinguish space and time are introduced. A straightforward way to recover classical dynamics by using the enhanced phase space and the classical action (written in terms of the enhanced phase space variables) is provided. 
The new versions of Hamilton equations 
are explicitly covariant, a feature which in conventional classical mechanics is only achieved
in configuration space. All these classical features are presented in section \ref{sec:stps} after a ``warm up'' example provided in section \ref{sec:spatial}.

A spacetime version of QM is then proposed in section \ref{sec:stqm} by replacing all PB with commutators (in the bosonic case; see remarks in section \ref{sec:concl}). 
A direct consequence is that the foliation is also quantized, allowing for a geometrical definition of spacetime transformations, that does not depend on the dynamics. The action is quantized as well yielding a ``spacetime quantum action'' operator, an object recently introduced in  \cite{diaz.21, diazp.21} (see also \cite{Sav.99}), here enhanced to take into account a dynamical foliation.
In this section, we also show how the diagonalization of free quantum actions leads to particles with general dispersion relation. The only difference between on-shell and off-shell particles is whether they  commute or not with the action. In both cases, their transformation properties are well defined, as induced by the transformation properties of the fields, momenta, foliation and action operators of relativistic theories.

In our framework, operators at different times commute, and time is treated as a geometrical ``index'' site, in complete analogy with space and indistinguishable from it at the algebra level. This raises the challenge of recovering conventional QM evolution (in a given foliation) from within what is essentially a non-causal framework. Notably, this problem can be solved as presented recently in \cite{diazp.21}. In section \ref{sec:stcorr}, we develop some of the ideas in \cite{diazp.21} further to establish a general correspondence between the spacetime formulation and conventional QFT through correlation functions at fixed foliation. The  classical limit is also analyzed and some possible connections with holography are pointed out. 

Furthermore, the previous emergence of time evolution admits a natural interpretation
in terms of a generalized pure state (non-orthogonal projector) involving an environment correlated with the given system. 
This mathematical object, which we may identify with a sensible generalization of the notion of state to spacetime,  codifies all  the information about the initial state, its evolution and the causal structure of the theory.
For free theories it can be built from a pair of conjugate entangled global vacua encompassing the system and an environment, and in general it is associated with a generalized purification involving the quantum action. 
We also comment on how the formulation gives new operational meaning to correlation functions, thus allowing the use of quantum computation protocols for their estimation. All of these features are described in section 
\ref{sec:stgstates} for a scalar field, while additional remarks for discrete spacetime and general systems are provided in the Appendices \ref{sec:Appurif} and \ref{sec:Apmap}.

Section \ref{sec:qfeffects} deals with the fact that invariant eigenstates of the action are entangled in the matter-foliation partition. As a consequence, the notion of particle becomes non-separable from the foliation. In particular, we show that ladder operators should be understood as foliation-controlled operators, thus having entangled eigenstates. Moreover, invariant eigenstates are necessarily entangled and have the structure underlying the Page and Wootters (PaW) mechanism \cite{PaW.83}. We exploit this analogy to introduce the notion of conditioning on foliation states, showing that for classical-like-states the correspondence of section \ref{sec:stcorr} is recovered.
 Finally, we show how the physical predictions of the theory transform properly and explicitly once the foliation is quantum.
The possibility of genuine quantum foliation effects is briefly considered as well.

A final discussion about the relevance of our results in different settings is provided in \ref{sec:concl}, together with
future perspectives
 regarding general covariance, its possible relevance in canonical quantum gravity, and in quantum foundational, as well as computational matters.

\section{Spacetime phase space formalism}\label{sec:stps}
\subsection{An introductory spatial analogy of the problem}\label{sec:spatial}
We begin our discussion by providing an example about how a Legendre transformation of the action and the ensuing phase space can hide an explicit spatial symmetry of a system.
Consider the following Lagrangian density $\mathcal{L}=\frac{1}{2}(\partial_t \phi)^2-\frac{1}{2}(\partial_x \phi)^2-\frac{1}{2}(\partial_y \phi)^2$. It is clear that the Lagrangian has a rotational symmetry in $(x,y)$ as part of its Lorentz symmetry which is manifest in the equations of motion $\partial_\mu \partial^\mu \phi=
\partial_t^2\phi-\partial_x^2\phi-\partial_y^2\phi=0$. Now let us introduce 
\begin{equation}
\begin{split}
    \mathcal{H}[\phi,\partial_t\phi,\partial_x \phi,\pi]:&=\partial_y \phi \frac{\partial \mathcal{L}}{\partial (\partial_y \phi )}-\mathcal{L}\\&=-\frac{1}{2}\pi^2-\frac{1}{2}(\partial_t \phi)^2+\frac{1}{2}(\partial_x \phi)^2
\end{split}\label{1}
\end{equation}
which is a ``Hamiltonian'' density defined by the Legendre transform which replaces $-\partial_y \phi\to \pi$. While this is certainly a poorly motivated change of variables, $\mathcal{H}$ should conserve the complete information of the system, as the transformation is invertible.  
In fact, a direct use of the equation of motion yields the equations
\begin{equation}\label{eq:hamiltony}
    \begin{split}
    \partial_y \pi&=-\frac{\partial \mathcal{H}}{\partial \phi}
    =-\partial_t^2\phi+\partial_x^2\phi\\
        \partial_y \phi&=\frac{\partial \mathcal{H}}{\partial \pi}
        =-\pi\,,
    \end{split}
\end{equation}
which have the form of Hamilton equations in the new variables. Clearly, after deriving the first equation with respect to $\partial_y$, the second equation yields $\partial_\mu \partial^\mu \phi=0$ back. 
One can also obtain \eqref{eq:hamiltony} from a variation of the action in phase space variables \footnote{Notice that the variation in phase space variables of the action ${S\!=\!\int dt dx dy(\pi\phi_{y} -{\cal H})}$, $\!\!{\text{with} \;\mathcal{H}}$ in \eqref{1},  is well defined   (here ${\phi_{x^\mu}\!\equiv\!\partial_\mu\phi}$). In fact, one obtains
${\delta S=\int dtdxdy\,[ (\phi_y+\pi)\delta \pi- (\pi_y+\phi_{tt}-\phi_{xx})\delta\phi]}+\left.\int  dxdt\, \pi \delta \phi\right|_{y_i}^{y_f}+\left.\int dxdy\, \phi_t \delta \phi \right|_{t_i}^{t_f}-\left.\int dydt\, \phi_x \delta \phi \right|_{x_i}^{x_f}$. All the boundary terms vanish under standard assumptions, namely $\delta\phi(t_i)=\delta\phi(t_f)=0$ and $\frac{\partial \mathcal{L}}{\partial \phi_i}\to 0$  for large $|x_i|$ in noncompact manifolds, or $\delta\phi=0$ in all the boundaries of a compact spacetime \cite{banados2016short}. 
Thus no ``differentiability'' problem \cite{regge1974role} arises. Notice, however, that $\cal H$ in \eqref{1} is not positive definite. }. 
Notice that the rhs of \eqref{eq:hamiltony} can be written in terms of Poisson brackets (PB), i.e., $-\frac{\partial \mathcal{H}}{\partial \phi}=\{\pi,H\}$, 
$\frac{\partial \mathcal{H}}{\partial \pi}=\{\phi,H\}$, where $H=\int dtdx\,\mathcal{H}$ is the ``Hamiltonian'' 
and the canonical PBs  are here 
\begin{equation}
    \{\phi(t,x),\pi(t',x')\}=\delta(t-t')\delta(x-x')
\end{equation}
at fixed $y$ (with the other PBs vanishing). 
The quantum versions (commutators) of such ``rotated'' PBs were recently used in \cite{har.23} to define ``timelike  entanglement''.

Interestingly, we see that one can recover the proper equations of motion from PBs which treat  $t$ on equal footing with $x$. 
In addition, these canonical relations are explicitly preserved by a Lorentz transformation of the form $\phi(t,x)\to \phi(t',x')$, $\pi(t,x)\to \pi (t',x')$, which are now treated as conventional symplectic transformations, in analogy with rotations.

Instead, rotations in the $(x,y)$ plane can no longer be treated in their natural geometrical character: from the phase space point of view, a rotation involves the ``evolution'' parameter $y$; its description becomes formally a dynamical problem. 
Note also that only the $x$ derivative appears in $\mathcal{H}$, and the symmetry is hidden.
This is of course an artifact introduced by the ``spatial'' Legendre transformation and the associated phase space structure. In fact, in the conventional Hamiltonian formulation based on $\pi=\partial_t \phi$ the PBs
$\{\phi(x,y),\pi(x',y')\}=\delta(x-x')\delta(y-y')$ are explicitly preserved by a rotation. Conversely, in this conventional approach we can no longer treat boost transformations as symplectic transformations.

 A clear problem with the previous Legendre transformation is its
selection of a particular direction in space.
Clearly, a second ``observer'' can choose any other direction $y'$ and construct its own phase space and canonical PBs at a fixed $y'$.
Yet, there is no simple rule relating the two constructions that does not involve dynamical information unless $y=y'$ (even if the initial conditions are imposed at fixed $t$, the phase spaces do not include $y$ ($y'$)).
To connect these two different phase spaces we need to somehow keep track of the momentum's choice. 
In addition, in order to unify them one must consider an extension of the PBs which includes all spatial dimensions. These are the main modifications to the conventional Hamiltonian approach which we develop in the next section for spacetime. 

Notice that another possibility is to include a second momentum in the $x$ direction and deal with a multisymplectic structure. We do not pursue this different approach which has been explored by other authors \cite{de.30,wey.35,got.91,kan.98,ish.02} and whose  quantization is not straightforward \cite{kan.98}. Moreover, it has recently been shown that one way to quantize these theories is to construct a canonical momentum from a poly-momentum first \cite{chester2024quantization}. One might use this route to relate our work to multisimplectic ideas; however, our proposal is independent of these constructions as it does not require such a preliminary step.

\subsection{Covariant Legendre transform}

The situation we have described in space, parallels the conventional space-time asymmetry which originates from separating the role of space and time just as
(\ref{1}) separates $x$ and $y$ in our previous ``experiment''.
In this section we generalize the conventional definition of momentum through a covariant Legendre transformation in order to eliminate the need for a preferred choice of time.

The key idea is that the conventional momentum conjugate to a given field $\phi$ can be written as 
\begin{equation}\label{eq:momen}
    \pi=\frac{\partial \mathcal{L}}{\partial (n^\mu \partial_\mu \phi)}\,
\end{equation}
for $n^\mu=\eta^{\mu 0}$, with the convention for the metric $\eta^{\mu \nu}=\text{diag}(1,-1,\dots,-1)$. But this choice of $n^\mu$ is arbitrary, the only requirement for a flat space-time and foliation being a {\it timelike} vector $n^\mu n_\mu =1$, such that it describes inertial observers (spacelike vectors will not be considered in the rest of this work).

For example, in the $1+1$ dimensional case we can separate time and space by choosing a basis $n^\mu$, $n^\mu_1$ with
 $n^\mu_1 n_\mu=0$ and $n^\mu_1 n_{1 \mu}=-1$. One general parameterization is provided by inertial observers in relative speed $v=\tanh\eta$ to a ``rest'' reference frame ($n^\mu\equiv \eta^{\mu 0}$) such that their choice corresponds to
 \begin{align}
     n^\mu=(\cosh\eta, \sinh\eta),\; n_1^\mu=(\sinh\eta, \cosh\eta)\,.\nonumber
 \end{align}
Now we can introduce a covariant $\mathcal{H}=\mathcal{H}[\phi,\pi,n^\mu_1\partial_\mu \phi]$ as the $n$-dependent Legendre transformation of $\mathcal{L}$ defined as follows:
\begin{equation}\label{eq:legen}
\mathcal{H}[\phi,\pi,n^\mu_1\partial_\mu \phi]:= \pi n^\mu \partial_\mu \phi-\mathcal{L}\,.
\end{equation}
The Hamiltonian density $\mathcal{H}$ is a function of the momentum $\pi$ defined as in (\ref{eq:momen}) but by an arbitrary direction $n^\mu$,  and the derivatives which are orthogonal to that direction (in this case there is only one). Note that this is not a multisymplectic formalism: just one momentum has been introduced, we simply retain the information of the time choice.

To write (\ref{eq:legen}) explicitly, one needs $\partial_\rho \phi$ in terms of the perpendicular derivatives. For the $1+1$ case, these are easily obtained as
\begin{equation}\label{eq:deriv}
    \partial_\rho \phi= n_\rho n^\mu\partial_\mu \phi- n_{1 \rho}n^\mu_{1}\partial_\mu \phi= n_\rho \pi- n_{1 \rho}n^\mu_{1}\partial_\mu \phi\,.
\end{equation}
It is now straightforward to rewrite any $\mathcal{L}$ as a function of the new variables.

As a concrete example consider a scalar field with Lagrangian density $\mathcal{L}=\frac{1}{2}(\partial_\mu \phi)^2 -\frac{1}{2}m^2\phi^2$.  By using Eq.\ (\ref{eq:deriv}) for writing $(\partial_\rho \phi)^2=(n_\rho\pi)^2+(n_{1\rho} n^\mu_1 \partial_\mu \phi)^2
$, the covariant Hamiltonian density  for a timelike $n^\mu$ can be written as 
\begin{equation}
    \mathcal{H}=\frac{1}{2}\pi^2+\frac{1}{2}(n^\mu_1 \partial_\mu \phi)^2+\frac{1}{2} m^2 \phi^2\,,\label{eq:kgH}
\end{equation}
where we are omitting the argument of $\mathcal{H}$ for ease of notation.
For $n^\mu=(1,0)$ one recovers the usual Hamiltonian density $\mathcal{H}=\frac{1}{2}\pi^2+\frac{1}{2}(\partial_1 \phi)^2+\frac{1}{2} m^2 \phi^2$. Instead, for general $n^\mu$, the contraction of the indices indicates Lorentz symmetry. Notice also that for a timelike $n^\mu$ the Hamiltonian density is positive.

The Hamilton equations corresponding to $\mathcal{H}$ have the same form as before with the time derivatives $\partial_t$ generalized to $n^\mu \partial_\mu$. This can be easily seen by applying the principle of least action in phase space \footnote{assuming standard boundary conditions, namely asymptotically vanishing fields for large $|\textbf{x}|$, in agreement with $n^\mu$ timelike}
to \begin{equation}\label{eq:action}
    \mathcal{S}=\int d^{d+1}x\, \left(\pi n^\mu \partial_\mu \phi-\mathcal{H}\right)\,,
\end{equation}
a result which holds for general fields, theories, and $D=d+1$ dimensions.
For the Hamiltonian (\ref{eq:kgH}) one obtains 
\begin{subequations}
\label{eq:hamiltonkg}
\begin{align}
 n^\mu \partial_\mu \pi&= \shortminus  \frac{\partial \mathcal{H}}{\partial \phi}=(n^\mu_1 n^\nu_1 \partial_\mu  \partial_\nu-m^2)\phi\\
n^\mu \partial_\mu \phi&= \;\,  \frac{\partial \mathcal{H}}{\partial \pi}=\pi
\end{align}
\end{subequations}
automatically implying (by acting with $n^\mu \partial_\mu$ on the second equation)
\begin{equation}
\label{KG}
    [\underbrace{(n^\mu n^\nu-n^\mu_1 n^\nu_1)}_{\eta^{\mu \nu}} \partial_\mu \partial_\nu+m^2]\phi=0
\end{equation}
which is just the Klein-Gordon equation.
Clearly, the conventional Hamiltonian density also yields this covariant second order equation for $\phi$, however, it does not provide separated first order covariant equations for $\phi$ and $\pi$ as the ones obtained in (\ref{eq:hamiltonkg}).

Moreover, the covariant aspect of $\mathcal{H}$ is new and not only formal: under Lorentz transformations one has 
\begin{subequations}\label{eq:transf}
    \begin{align}
        \phi(x)&\to \phi(\Lambda x)\label{eq:transfphi}\\
        n^\mu &\to \Lambda^{\mu}_{\;\;\nu}n^\nu\label{eq:transfn}\\
        \pi(x)&\to \pi(\Lambda x)\label{eq:transfpi}
    \end{align}
\end{subequations}
where equations (\ref{eq:transfn}-\ref{eq:transfpi}) are a novelty of the formalism, while (\ref{eq:transfphi}) 
holds for a scalar field.
The transformation law of $\pi$ follows from
$\pi=n^\mu \partial_\mu \phi$ assuming (\ref{eq:transfphi}) and (\ref{eq:transfn}). 
The important novelty is that under these transformations the Hamiltonian density transforms as
\begin{equation}
    \mathcal{H}(x)\to \mathcal{H}(\Lambda x)\,,
\end{equation} i.e. it is \emph{a Lorentz scalar}.
This is compatible with the new relation between $\mathcal{H}$ and the energy-momentum tensor which is easily found to be $\mathcal{H}=n_\mu n_\nu T^{\mu \nu}$.

All previous properties hold in arbitrary $d+1$ dimensions with the covariant $\mathcal{H}$ always defined as in Eq.\ (\ref{eq:legen}). 
For example, the $d+1$ generalization of (\ref{eq:kgH}) is
\begin{equation}
\label{eq:kgd1}
    \mathcal{H}=\frac{1}{2}\pi^2+\frac{1}{2}(n^\mu n^\nu -\eta^{\mu\nu}) \partial_\mu \phi \partial_\nu \phi+\frac{m^2}{2} \phi^2
\end{equation}
with the tensor $n^\mu n^\nu -\eta^{\mu\nu}$ projecting onto the $d$ spatial directions $n^\mu_{\;i}$ orthogonal to $n^\mu$, such that the central term in \eqref{eq:kgd1} is $\sum_{i=1}^d( n^\mu_{\;i}\partial_\mu\phi)^2$. This can be easily seen by noting that the complete ``reference frame'' axes can be written as $n^\mu_{\;\alpha}\equiv\partial x'^\mu/\partial x^\alpha$ with $n^\mu_{\;0}\equiv n^\mu$ and using
\begin{equation}\label{eq:tensor}
    \eta^{\mu \nu}=\frac{\partial x'^\mu}{\partial x^\alpha}\frac{\partial x'^\nu}{\partial x^\beta}\eta^{\alpha \beta}=n^\mu n^\nu -\sum_{i=1}^d n^\mu_{\; i}n^\nu_{\;i}
\end{equation}
for $x'^\mu$ related to $x^\mu$ through a Lorentz transformation. This allows to write $\mathcal{H}$ as a function of $n^\mu$ only (rather than of all $n^\mu_{\; \alpha}$).

The new transformation properties also imply
the invariance of the ``integrated'' Legendre transform 
\begin{equation}
  \mathcal{P}_0:=  \int d^{d+1}x\, \pi\, n^\mu \partial_\mu \phi\,.
\end{equation}
As a consequence, the action in phase space variables (\ref{eq:action}) has always a Lorentz invariant expression as well. For the scalar field example, one obtains
\begin{align}
       \mathcal{S}=&\int d^{d+1}x\, \Big[\pi n^\mu \partial_\mu \phi-\tfrac{1}{2}\pi^2-\tfrac{1}{2}(n^\mu n^\nu -\eta^{\mu\nu}) \partial_\mu \phi \partial_\nu \phi  \nonumber\\&-\tfrac{1}{2}m^2 \phi^2\Big]\,.\label{eq:kgaction}
\end{align}
In contrast, the conventional  action in phase space,  ${\mathcal{S}=\int d^{d+1}x\, \big(\pi \dot{\phi}-\frac{1}{2}\pi^2-\frac{1}{2}(\nabla \phi)^2-\frac{1}{2}m^2\phi^2\big)}$ hides the Lorentz symmetry as it corresponds to choosing a time direction  $n^\mu=\eta^{\mu 0}$ in (\ref{eq:kgaction}).

In general, it is also feasible to leave the length $n^\mu n_\mu$ 
 arbitrary (but nonzero), without affecting the final Klein-Gordon equation (see Appendix \ref{sec:Apa}).  
Let us also mention that the treatment of nonscalar fields can be developed along the same lines presented in this section, by simply adapting the transformation rules \eqref{eq:transf}. This is shown in the case of a Dirac field in Appendix \ref{sec:apdirac}. Therein additional main body results are exemplified for this field, while the principal example in the main body is the Klein-Gordon field.

\subsection{Spacetime Symplectic Structure}
The conventional phase space associated with our previous construction corresponds to canonical algebras satisfied at fixed hypersurfaces by matter fields. For each choice of $n^\mu$, 
a symplectic structure should be defined.
On the other hand, our objective is to keep $n^\mu$ general and to promote it to a ``dynamical'' variable, in the sense explained below Eq.\ \eqref{eq:nalg}, which involves a foliation-algebra.

In order to keep the matter-foliation algebras separated,
we introduce another element in the formalism: we extend the phase space by treating each field in space time and its conjugate momentum as independent canonical variables satisfying
\begin{equation}\label{eq:extpb}
    \{\phi(x),\pi(y)\}=\delta^{(d+1)}(x-y)\,.
\end{equation}
 The PBs are defined as usual but encompass all variables
\begin{equation}
    \{f,g\}=\int d^{d+1}x\, \left(\frac{\delta f}{\delta \phi(x)}\frac{\delta g}{\delta \pi (x)}-\frac{\delta g}{\delta \phi(x)}\frac{\delta f}{\delta\pi (x)}\right)\,
\end{equation}
in perfect spacetime symmetry
\footnote{From the mathematical perspective we can identify the ensuing phase space $\Omega$ with the limit $N\to \infty$ of the direct product $\Omega \equiv  \omega_t^{\times N}$ for $\omega_t$ the traditional phase space defined at fixed $t$. This is precisely the mathematical structure conventionally applied to fields in space such that $\omega_t\equiv\omega_{tx}^{\times M}$ for $M$ spatial slices and $\omega_{tx}$ the phase space of a single oscillator.  In summary, we may write $\Omega \equiv  \omega_{tx}^{\times N\cdot M}$.} and independent on how one foliates spacetime. 

This extended symplectic structure enables a straightforward treatment of spacetime symmetries: Note first that 
Eq.\ (\ref{eq:extpb}) implies 
  \begin{align}
  \label{eq:cp0}
  \{\phi, \mathcal{P}_0\}=n^\mu \partial_\mu \phi  \,,\;\;\;\;\;\;\;\;\;\;\;\;\;\;\; \{\pi, \mathcal{P}_0\}&=n^\mu \partial_\mu \pi\,
\end{align}
meaning that $\mathcal{P}_0$, the Legendre transformation integrated in time, generates time translations in the $n^\mu$ direction.
In this framework the time translations are geometrical and independent of evolution. This is reflected by the fact that $\mathcal{P}_0$ generates the transformations and not the Hamiltonian (this point is further discussed when evolution is considered in Section \ref{sec:physub}).

For $n^\mu=(1,0,\dots)$ we can also write $\{\phi,\mathcal{P}_\mu\}=\partial_\mu \phi$ for $\mathcal{P}_\mu=\int d^{d+1}x\, \pi \partial_\mu \phi$ which for $\mu=1,\dots ,d+1$ is just the conventional momentum carried by the field integrated in time.
In addition, \begin{equation}\mathcal{L}_{\mu\nu}:=\int d^{d+1}x\, \pi (x_\mu \partial_\nu-x_\nu \partial_\mu)\phi\end{equation} generates the Lorentz transformations
\begin{subequations}\label{eq:transpb}
\begin{align}
      \{\mathcal{L}_{\mu\nu},\phi\}&=-(x_\mu \partial_\nu-x_\nu\partial_\mu)\phi\\
         \{\mathcal{L}_{\mu\nu},\pi\}&=-(x_\mu \partial_\nu-x_\nu\partial_\mu)\pi\,.
         \end{align}
\end{subequations}
Through exponentiation of the previous transformations finite general Poincaré transformations are thus obtained.
In particular, the transformation properties of $\phi$ and $\pi$ in Eq.\ (\ref{eq:transf}) are recovered. The addition of spin is straightforward but introduced in the Appendix \ref{sec:apdirac} 
for simplicity.

In order to obtain the transformation law of $n^\mu$ in a similar fashion an additional symplectic structure 
may be defined: we introduce a conjugated momentum $\kappa_\nu$ such that
\begin{equation}\label{eq:nalg}
    \{n^\mu,\kappa_\nu\}=\eta^\mu_{\;\nu}\,.
\end{equation} 
One can impose $n^\mu n_\mu-1 \approx 0$ as a weak constraint. 
It is now feasible to introduce $l_{\mu\nu}=n_\nu \kappa_\mu-n_\mu \kappa_\nu$ such that 
\begin{equation}
    \{l_{\alpha\beta},n_\mu\}=n_\alpha \delta_{\mu\beta}-n_\beta \delta_{\mu\alpha}\,.
\end{equation}
Then 
\begin{equation}\label{eq:jmunu}{\cal J}_{\mu\nu}:=\mathcal{L}_{\mu\nu}+l_{\mu\nu}\end{equation} generates the complete transformation (\ref{eq:transf}). 
Within this formalism the statement of a Lorentz invariant theory becomes explicit (see e.g. the action in Eq.\ (\ref{eq:kgaction})):
\begin{equation}\label{eq:jsconm}
    \{\mathcal{S},\mathcal{J}_{\mu\nu}\}=0\,.
\end{equation}
Notice that $\{\phi,n^\mu\}=\{\phi,\kappa^\mu\}=\{\pi,n^\mu\}=\{\pi,\kappa^\mu\}=0$ such that $\{\mathcal{L}_{\mu\nu},l_{\alpha\beta}\}=0$, in other words the algebras are independent. On the other hand, $\mathcal{S}$ has a nonvanishing PB with all variables except with $n^\mu$, in particular $\{\mathcal{S},\mathcal{L}_{\mu\nu}\}=-\{\mathcal{S},l_{\mu\nu}\}\neq 0$. Additionally, it should be noted that the generators $\mathcal{J}_{\mu\nu}$ are independent of the Hamiltonian, meaning that we have successfully separated the coordinate transformations from the dynamics.

\vspace{0.1cm}

The introduction of a symplectic structure associated with $n^\mu$ provides the final piece for an elegant treatment of spacetime symmetries within a phase space framework. Yet, at first sight, it seems unjustified physically since no associated dynamic has been introduced. On second thought, similar situations arise in many physical scenarios: consider for example a particle in an external magnetic field $\textbf{B}$ with coupling $H_{\text{int}}\propto -\textbf{B}\cdot \textbf{M}$ for $\textbf{M}$ the magnetic moment vector associated with the particle. It is clear that $H_{\text{int}}$ exhibits rotational symmetry, even if we do not associate a symplectic structure with $\textbf{B}$ that implements rotations. We can, however, treat $\textbf{B}$ as a formal dynamical field in an additional phase space and define a total (product) rotation operator $R^{\text{tot}}$ which also rotates ${\bf B}$ such that  $\{H_{\text{int}},R^{\text{tot}}\}=0$, even if no momentum dependent terms appear
 in $H_{\text int}$ ($\{H_{\text int},{\bf B}\}={\bf 0}$).  
The genuine Hamiltonian description of the field has  an associated symplectic structure which may coincide with the formal one, but can be ignored  when treated as an external source.

\vspace{0.1cm}

We can speculate that a similar situation may arise in future investigations with $\mathcal{S}\to \mathcal{S}+\mathcal{S}_{n^\mu}$ for $\mathcal{S}_{n^\mu}$ including $\kappa^\mu$ just as $H_{\text{int}}\to H_{\text{int}}+H_{A^\mu}$ makes $A^\mu$ and $\textbf{B}$ dynamical. While we have introduced a foliation phase space for mathematical convenience, 
a theory of a dynamical metric and associated foliations may provide a genuine dynamical description of $n^\mu$ (see also the quantum discussion in sections \ref{sec:expart}-\ref{sec:qfeffects}). Conversely, the considerations in this work point to its existence.

\subsection{Equations of motion from extended brackets}\label{sec:physub}
In the new framework a ``timeless'' picture emerges: all spacetime variables, including  $t\equiv x^\mu n_\mu$, are site indices of independent fields in spacetime. There is no variable parameterizing evolution and no causal structure is assumed a priori. Yet, any dynamical information is to  be encoded within the extended phase space itself given that it already contains ``time''.

Remarkably, the new symplectic structure provides an elegant way to introduce evolution: 
the definition of the extended brackets yields 
\begin{align}\label{eq:pbhamiltonian}
   \frac{\partial \mathcal{H}}{\partial \pi}&=\Big\{\phi, \medop{\int} d^{d+1}x\,\mathcal{H}\Big\}\,,\;\;   -  \frac{\partial \mathcal{H}}{\partial \phi}=\Big\{\pi, \medop{\int} d^{d+1}x\,\mathcal{H}\Big\}\,.\end{align}
As a consequence, the action $\mathcal{S}$ defined in (\ref{eq:action})
 naturally emerges as the difference between Eqs.\ \eqref{eq:pbhamiltonian} and 
\eqref{eq:cp0} in such a way that 
\begin{subequations}
\label{eq:hamiltonaction}
\begin{align}
 n^\mu \partial_\mu \pi+  \frac{\partial \mathcal{H}}{\partial \phi}&=\{\pi,\mathcal{S}\}\\
n^\mu \partial_\mu \phi-   \frac{\partial \mathcal{H}}{\partial \pi}&=   \{\phi,\mathcal{S}\}\,.
\end{align}
\end{subequations}
When set equal to zero they are precisely the Hamilton equations. 
We may define a ``physical subspace'' (or subvariety) as 
\begin{align}\label{eq:physconstra}
   \{\pi(x),\mathcal{S}\}= \{\phi(x),\mathcal{S}\}
   \approx 0\,,
\end{align}
imposed for all spacetime points $x$.
In this formulation, these should be regarded as weak equalities with evolution emerging  from the constraints themselves. They impose an equality between displacements in time, as generated by $\mathcal{P}_0$ (Eqs.\ (\ref{eq:cp0})), and the transformation generated by the Hamiltonian.

For instance, for the Klein Gordon field the action is given by (\ref{eq:kgaction}) which, with the addition of a potential term $\mathcal{H}\to \mathcal{H}+\mathcal{V}(\phi)$,   yields
    \begin{align}\label{eq:physsubcl}
  \{\pi(x),\mathcal{S}\}&=n^\mu \partial_\mu \pi -(n^\mu n^\nu-\eta^{\mu\nu}) \partial_\mu \partial_\nu \phi+m^2\phi+\mathcal{V}'(\phi)
  \nonumber\\
   \{\phi(x),\mathcal{S}\}&=n^\mu \partial_\mu \phi -\pi\,.
\end{align}
When these are equaled to zero they become the Hamilton equations (see (\ref{eq:hamiltonkg})) implying 
\begin{equation}
    (\partial_\mu \partial^\mu +m^2)\phi+\mathcal{V}'(\phi)=0\,.
\end{equation}
Interestingly, the relation $\pi=n^\nu\partial_\nu \phi$ 
is compatible with 
\begin{equation}\label{eq:plpb}
    \{\kappa_\mu,\mathcal{L}(x)\}=(\pi (x)-n^\nu \partial_\nu \phi(x))\partial_\mu \phi(x)\approx 0
\end{equation}
with $\mathcal{L}$ the Lagrangian density in \eqref{eq:kgaction} (such that ${\mathcal{S}=\int d^{d+1}x\, \mathcal{L}(x)}$). 
Then, since $\{n^\mu,\mathcal{S}\}=0$ is fulfilled trivially, while  $
\{\kappa^\mu,\mathcal{S}\}\approx 0$ follows from  \eqref{eq:plpb}, any function in the foliation phase space ``commutes'' with the action (in the physical subspace). If some dynamical part $\mathcal{S}_{n^\mu}$ were added to $\mathcal{S}$ ($\mathcal{S}\to \mathcal{S}+\mathcal{S}_{n^\mu}$), the foliation action $\mathcal{S}_{n^\mu}$ would determine the equations of motion of the foliation independently from the original action of matter fields $\mathcal{S}$.

Some comments about units are in order: the fields have now rescaled units with a factor $T^{-1/2}$ because of the additional time delta. This means that a time parameter $\tau$ may be introduced for multiplying  $\phi$, $\pi$ by $\sqrt{\tau}$. For quadratic actions this means an overall factor $\tau$ such that $\tau \mathcal{S}$ is adimensional in agreement with an interpretation of $\mathcal{S}$ as generator in ``$\tau$ evolution''. Equation (\ref{eq:physconstra}) may then be identified as the set of conditions defining $\tau$-constants of motion and the extended PB \eqref{eq:extpb} with a canonical algebra at ``equal $\tau$'' in a $d+2$ theory. Notice however that this analogy does not extend to the foliation algebra. We consider that it is more appropriate to treat the formalism as describing $D=d+1$ theories through a new set of rules rather than  $d+2$ theories in a canonical approach (see however the remarks in section \ref{sec:stcorr}).  
In general, if $\phi, \pi$ satisfy the equations of motion arising from a rescaled $\mathcal{S}$, then $\sqrt{\tau} \phi, \sqrt{\tau}\pi$ have the correct units and satisfy the conventional equations of motion. In this section, we simply set $\tau \equiv 1$ but this parameter has important consequences in the quantum case. 

\vspace{0.1cm}

Let us mention that in Appendix \ref{sec:apdirac} the case of the Dirac action is also developed. Therein we show how to recover Dirac's equation from the previous constraints.  Interestingly,  
the equation in its Hamiltonian form exhibits Lorentz covariance explicitly for $n^\mu$ general. This agrees with 
Eq.\ \eqref{eq:jsconm}, holding for Dirac's action and $\mathcal{J}^{\mu\nu}$ including the spin angular momentum.

\vspace{0.1cm}
  
Let us also notice that the present formalism 
can be applied to any classical system and not only fields: one ``promotes'' variables $q_i, p_j$ satisfying $\{q_i,p_j\}=\delta_{ij}$ to $q_i(t), p_j(t)$ such that 
\begin{equation}
    \{q_i(t),p_j(t')\}=\delta_{ij}\delta(t-t')\,.
\end{equation}
To recover evolution one then introduces an action ${\mathcal{S}=\int dt\, (p_i\dot{q}_i-H)}$ and imposes
\begin{subequations}
\begin{align}
    \{q_i,\mathcal{S}\}&=\dot{q}_i-\frac{\partial H}{\partial p_i}\approx 0\\
    -\{p_i,\mathcal{S}\}&=\dot{p}_i+\frac{\partial H}{\partial q_i}\approx 0\,.
\end{align}    
\end{subequations}
One recognizes again the Hamilton equations imposed as constraints.

It is straightforward to see that the physical subspaces are invariant under transformation symmetries of the action. In fact, the generator $G$ of any such symmetry satisfies $\{G,\mathcal{S}\}=0$. Then, for a function $F[\phi,\pi]$ (or $F[q,p]$) within the physical subspace the Jacobi identity implies
\begin{equation}
 \{F,\mathcal{S}\}=0\Rightarrow   \{\{G,F\},\mathcal{S}\}=0\,,
\end{equation}
i.e. the transformed $F$ is also in the physical subspace. An example is provided by Lorentz symmetry for the scalar field, as described in (\ref{eq:jsconm}).

\vspace{0.2cm}

\section{Spacetime quantum mechanics}\label{sec:stqm}
\subsection{Extended quantization}
\begin{figure*}[t]
    \centering
\includegraphics[width=\linewidth]{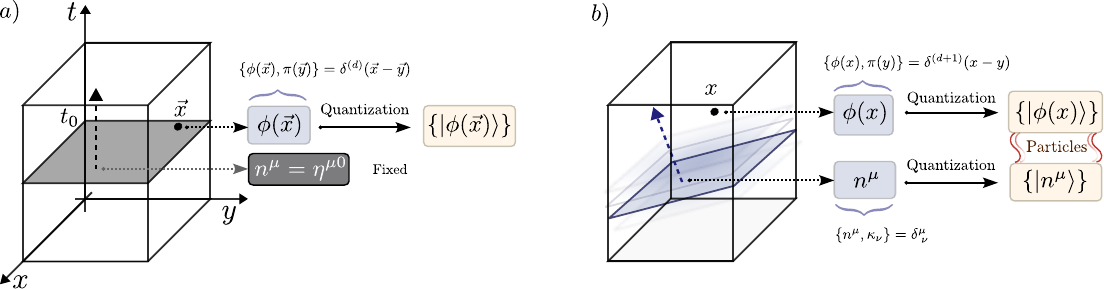}
    \caption{\textbf{Standard phase space \& quantization vs the spacetime approach}. a) In Hamiltonian classical mechanics a symplectic structure is defined for a fixed choice of time. The quantization is thus performed in a given $d$ dimensional hypersurface by promoting $\phi(\textbf{x})$ and $\pi(\textbf{x})$ to quantum operators. One possible basis of the ensuing Hilbert space is given by field configurations in the hypersurface, detoned by $|\phi(\textbf{x})\rangle$. b) In the spacetime approach, both Poisson brackets \&  commutators are spacetime symmetric and the foliation is ``dynamical''. A basis of the Hilbert space is given by the tensor product between spacetime configurations of the field $|\phi(x)\rangle$ and the foliation eigenstates $|n\rangle\equiv |n^0,n^1,\dots n^d\rangle$. General operators, such as the spacetime quantum actions and ensuing ladder operators (associated with extended off-shell particles)  are nonseparable in the matter-foliation partition. Their explicit covariant features become feasible in the complete Hilbert space only.}
    \label{fig:fig1}
\end{figure*}
The first step in the conventional canonical quantization of a Hamiltonian theory is to promote the canonical PBs to canonical commutators. We impose the same to the extended algebra (\ref{eq:extpb}) implying (we set $\hbar\equiv 1$)
\begin{equation}\label{eq:extalg}
    [\phi(x),\pi(y)]=i\delta^{(d+1)}(x-y)\,,
\end{equation}
with the other commutators vanishing (we have also assumed a bosonic algebra). Then, any function of the phase space variables is also promoted to an operator (up to the usual ordering ambiguities). Remarkably, in the extended scheme this means that not only the Hamiltonian, but also the action $\mathcal{S}$ expressed as in (\ref{eq:action}), are  now promoted.

It is worth remarking that $\phi(x)=\phi(t,\textbf{x})$ is not the field operator evolved in the Heisenberg picture, but for each time an independent field and associated momentum is present. In particular,
\begin{equation}
    [\phi(t,\textbf{x}),\phi(t',\textbf{x}')]=0
\end{equation}
even for causally-connected regions. This is a stronger statement than microcausality, in fact, there is no causal connection between fields (and momenta) at different spacetime points. Accordingly, one possible basis for this Hilbert space is provided by states $|\phi(x)\rangle$ representing field configurations in \emph{spacetime}, such that 
\begin{equation}\label{eq:fieldbas}
    \hat{\phi}(x)|\phi(x)\rangle=\phi(x)|\phi(x)\rangle\,,
\end{equation}
with $\langle \phi(x)|\phi'(x)\rangle=\delta${\large$^{\infty}$}$[\phi(x)-\phi'(x)]$ equivalent to the continuum limit of $\prod_{x=(t,{\bf x})}\delta[\phi_{x}-\phi'_x]$. We can also regard these states as ``quantum trajectory'' states of conventional field eigenstates at a given time $|\phi(\textbf{x})\rangle$ 
 in the sense that $|\phi(x)\rangle \equiv \otimes_t |\phi_t(\textbf{x})\rangle$ (with $\langle \phi({\bf x})|\phi'({\bf x})\rangle=\delta^{\infty}[\phi({\bf x})-\phi'({\bf x})]$, see also \cite{diaz.21,diazp.21} for a more detailed discussion). In other words, the Hilbert space which arises from (\ref{eq:extalg}) is isomorphic to a tensor product of copies in time of the traditional Hilbert space (this statement becomes rigorous only after a proper discretization, see Appendix \ref{sec:Apmap}). This is valid for a bosonic algebra, the fermionic case can be developed along similar lines with commutators replaced by anticommutators. 

As in the classical Hamiltonian case, the advantage of the extended algebra is the explicit and geometric treatment of spacetime symmetries. 
In fact, with these definitions one can promote $\mathcal{P}_0$ and $\mathcal{L}_{\mu\nu}$ to operators such that a quantum version of Eqs.\ (\ref{eq:cp0}, \ref{eq:transpb}) is readily obtained by replacing $\{\;,\;\}\to -i[\;,\;]$. Then e.g.  $\mathcal{P}_0$ generates geometric time translations such as $e^{i\tau \mathcal{P}_0}\phi(x)e^{-i\tau \mathcal{P}_0}=\phi(x^0+\tau,\textbf{x})$ for $\tau\in \mathbbm{R}$ and $n^\mu=\eta^{\mu0}$.

In addition, we promote (\ref{eq:nalg}) to
\begin{equation}    [n^\mu,\kappa_\nu]=i\delta^{\mu}_{\; \nu}
\end{equation}
as an algebra independent of the matter fields, such that 
\begin{equation} \!\!\!\![\phi(x),n^\mu]=[\phi(x),\kappa^\mu]=[\pi(x),n^\mu]=[\pi(x),\kappa^\mu]=0\,.\end{equation}
We can now 
introduce the total angular momentum operator $\mathcal{J}_{\mu\nu}={\cal L}_{\mu\nu}+l_{\mu\nu}$, as in Eq.\ (\ref{eq:jmunu}), with both ${\cal L}$ and $l$ now promoted to operators.  
 Then, within the complete Hilbert space we can write the transformation of the operators $\phi$, $n^\mu$  and $\pi$ in the unified form  
\begin{subequations}\label{eq:qtransf}
    \begin{align}
        \mathcal{U}^\dag(\Lambda)\phi(x) \mathcal{U}(\Lambda)&= \phi(\Lambda x)\\
        \mathcal{U}^\dag(\Lambda) n^\mu  \mathcal{U}(\Lambda)&= \Lambda^{\mu}_{\;\;\nu}n^\nu\\
         \mathcal{U}^\dag(\Lambda)\pi(x) \mathcal{U}(\Lambda)&= \pi(\Lambda x)\,,
    \end{align}
\end{subequations}
where 
\begin{equation}
     \mathcal{U}(\Lambda):=\exp(i \omega_{\mu\nu}\mathcal{J}^{\mu\nu}/2)
\end{equation}
is the unitary Lorentz operator corresponding to the  transformation $\Lambda=e^{\omega}$ ($x'^\mu=\Lambda^{\mu}_{\;\nu}x^\nu$).  Equations (\ref{eq:qtransf}) are of course the quantum version of (\ref{eq:transf}). It is worth noting that the definition of $\mathcal{U}(\Lambda)$ does not involve the Hamiltonian, meaning that it is theory independent.

The final version of the Hilbert space that includes the ``quantum foliation'' is depicted in Figure \ref{fig:fig1} and has one basis of the form
\begin{equation}
    \{|\phi(x)\rangle \otimes |n\rangle\}
\end{equation}
 for $\hat{n}^\mu|n\rangle=n^\mu |n\rangle$ ($|n\rangle\equiv |n^0\ldots n^d\rangle$) and \begin{equation}\label{eq:basistrans}
 \mathcal{U}(\Lambda)|\phi(x)\rangle\otimes|n\rangle=|\phi(\Lambda^{-1}x)\rangle\otimes|\Lambda^{-1}n\rangle\,.
 \end{equation} 
 Of course, in the foliation sector more general states $|\psi\rangle=\int dn\, \psi(n) |n\rangle$ ($dn\equiv dn^0\dots dn^d$) are possible including e.g. momentum eigenstates, coherent states and Fock states. One can also implement the condition $n^\mu n_\mu \approx 1$ as the quantum constraint $(n^\mu n_\mu-1)|\psi\rangle=0$ which only allows the superposition of $n^2=1$ states (implicitly assumed throughout this section).

Notably, a general state will clearly exhibit entanglement between the matter-foliation partition. This feature emerges naturally from the formalism even when no
physical mechanism has been imposed (we have not considered interactions between the matter-foliation sectors). In particular, the quantum action $\mathcal{S}$ is not a product operator but rather a controlled-like operator, i.e.
\begin{equation}\label{eq:scontrol}
{\cal S}\equiv {\cal S}(\hat{n}^\mu)=\int dn \,{\cal S}(n^\mu)\otimes |n\rangle\langle n|\,.
\end{equation}  
This fact has consequences which are discussed in sections \ref{sec:expart} and \ref{sec:qfeffects}. For the moment, we remark that it is precisely because of this structure that we can write  
\begin{equation}\label{eq:actioninv}
[\mathcal{S},\mathcal{J}_{\mu\nu}]=0\,,
\end{equation}
indicating the covariance of the action explicitly while
\begin{equation}
[\mathcal{S},\mathcal{L}_{\mu\nu}]=-[\mathcal{S},l_{\mu\nu}]\neq 0\,.
\end{equation}

Note also that the Hilbert of $n$ is isomorphic to a $d+1$ particle. Interestingly, this observation suggests possible connections with the recent Quantum Reference Frame transformations \cite{giac.19} where the notion of quantum particle's rest frame is defined (however the constraint
emphasizes important mathematical and interpretational differences, at least at this stage of development).

It is important to mention that in the context of the consistent-history approach to QM \cite{ish.93}, the need for a quantized foliation has also been reported \cite{ish.02}, a result which unfortunately has not attracted much attention or further development. While the treatment in \cite{ish.02} has been different (both classically and in its quantized version) the reasons for its introduction are the same: a proper treatment of Lorentz transformations in a QFT with extended algebra (also a characteristic of Isham's approach to continuum histories \cite{Ish.98}).

\subsection{Extended Particles}\label{sec:expart}
Having introduced the proper kinematical framework, we begin to discuss how to introduce dynamics within the formalism. A basic observation is that since fields at different spacetime points are independent, no causality is assumed a priori, and evolution cannot correspond to a parameterized unitary transformation as usual: while ``$t$'' is a parameter, it has a completely different meaning that in conventional QFT. It is here treated as a ``site'' index just as ``$\textbf{x}$''.
Yet, we would like to recover the same predictions of traditional QMs concerning evolution, at least under reasonable assumptions such as conventional Hamiltonians and ``classical'' foliations (we postpone most of the discussion about effects related to a quantum foliation to section \ref{sec:qfeffects}).

As suggested by the classical case, evolution should arise from the
the action $\mathcal{S}$, now a quantum operator. Consider as a concrete example the Klein-Gordon action (\ref{eq:kgaction}) with $\phi,\pi, n^\mu$ operators. Let us discuss first its diagonalization. Being a quadratic operator for each fixed $n^\mu$ (see Eq.\ (\ref{eq:scontrol})) its diagonal form is easily achieved: we expand the fields as
\begin{subequations}\label{eq:fieldsexp}
    \begin{align}
            \phi(x)&=\int \frac{d^{D}p}{(2\pi)^{D}}\frac{1}{\sqrt{2 E_{p}(n)}}\,\left(a(p) e^{-i px} + \text{H.c.}\right)\\
            \pi(x)&=\int \frac{d^{D}p}{(2\pi)^{D}}(-i)\sqrt{\frac{E_p(n)}{2}}\,\left( a(p) e^{-i px}- \text{H.c.}\right)
    \end{align}
\end{subequations}
for $a^\dag(p)$, $a(p)$ extended creation(annihilation) operators satisfying
\begin{equation}\label{eq:algpart}
    [a(p),a^\dag(p')]=(2\pi)^{(D)}\delta^{(D)}(p-p')\,,
\end{equation}
with other commutators vanishing. In these expressions 
\begin{equation}D=d+1\,,\end{equation} and we have defined
\begin{equation}
    E_p(n):=\sqrt{p^\mu p^\nu (\eta_{\mu\nu}-n_\mu n_\nu)+m^2}
\end{equation}
for $n^\mu$ an operator: a function $F$ of the operators $n^\mu$ should be interpreted as $F[n]\equiv \int dn\,F[n] |n\rangle \langle n|$ 
(for simplicity we have here worked within 
the subspace $(n^\mu n_\mu-1)|\psi\rangle=0$; see section \ref{sec:qfeffects} and Appendix \ref{sec:Apa}).

In terms of these extended ladder operators, the action \eqref{eq:kgaction} has the normal form
\begin{equation}\label{eq:diagaction}
    \mathcal{S}=\int \frac{d^{D}p}{(2\pi)^D}\, \big(p^\mu n_\mu-E_p(n)\big)a^\dag(p)a(p)\,,
\end{equation}
where we have dropped a ``constant'' related to the vacuum energy (interestingly, the term arising as usual from normal ordering the operators remains an operator in the foliation sector; see section \ref{sec:qfeffects}). Notice the two different contributions to the ``normal frequencies'' $p^\mu n_\mu-E_p(n)$, with $E_p(n)$ associated with the Hamiltonian density $\mathcal{H}$ (see also section \ref{sec:qfeffects})  while 
\begin{equation}\label{eq:p0diagonal}
     \mathcal{P}_0=\int \frac{d^{D}p}{(2\pi)^D}\, p^\mu n_\mu\,a^\dag(p)a(p)\,.
\end{equation}
One can show that this normal form of $\mathcal{P}_0$ is not unique \cite{diaz.21}. For $n^\mu \equiv \eta^{\mu0}$ one has $p^\mu n_\mu-E_p(n)\to p^0-\sqrt{p^2+m^2}$, i.e. $E_p(n)\equiv E_{\textbf{p}}=\sqrt{\textbf{p}^2+m^2}$ the conventional relativistic energy.

The previous diagonalization of $\mathcal{S}$ mimics the expressions of conventional QFTs concerning the diagonalization of a free Hamiltonian in $d=D-1$ dimensions. However, important differences should be noted: the expansion of the fields in particle operators is completely \emph{off-shell}, with $p^0$ unrelated to $\textbf{p}$. Yet, the quantity $E_p(n)$ which appears in the normal form of the Hamiltonian part of $\mathcal{S}$ is positive for all $p$, allowing the expansion (\ref{eq:fieldsexp}).
The positivity follows from Eq.\ (\ref{eq:tensor}) which implies $p^\mu p^\mu(\eta_{\mu\nu}-n_\mu n_\nu)=\sum_i(p_\mu n^\mu_{\;i})^2$.

Moreover, on-shell, i.e. for $p^0=E_{\textbf{p}}=\sqrt{\textbf{p}^2+m^2}$ a direct computation yields $E_p(n)=E_{\Lambda\textbf{p}}$ for $\Lambda$ defined as the Lorentz transformation that brings a normalized $n^\mu$ to the ``canonical'' time direction $n^\mu=\eta^{\mu 0}$. In other words, $E_p(n)$ on-shell corresponds to the energy measured by the observer with axis $n^\mu$.

A basic consistency requirement for the extended, in general off-shell, particles is that different inertial observers agree on their notion (e.g. their number) and properties (after transforming their momenta). For this to be fulfilled, in consistency with the transformation rules of the fields, and their expansion in extended modes, it is \emph{crucial} that $n^\mu$ is an operator such that
\begin{equation}
    \mathcal{U}^\dag(\Lambda) E_p(n) \mathcal{U}=E_p(\Lambda n)\,.
\end{equation}
 In fact, by noting that $E_{p}(\Lambda^{-1} n)=E_{\Lambda p}(n)$ and that $d^Dp$ is an invariant measure,  one easily finds
\begin{equation}\label{eq:atransf}
\mathcal{U}^\dag(\Lambda)a(p)\mathcal{U}(\Lambda)=a(\Lambda p)\,.
\end{equation}
In summary, the extended particles transform properly (even off-shell) because $E_p(n)$ is also affected by the quantum transformation. This requires a quantum $n^\mu$. 

One can gain further insight by noting that this requires $[a(p),l_{\alpha\beta}]\neq 0$,
which is only possible if the creation/annihilation operators act non trivially in the foliation Hilbert space. This can be seen explicitly by inverting the relations (\ref{eq:fieldsexp}). The result is
\begin{equation}\label{eq:aexp}
    a(p)=\int d^Dx\,e^{ipx}\Big(\sqrt{\frac{E_p(n)}{2}}\phi(x)+\frac{i}{\sqrt{2E_p(n)}}\pi(x)\Big)\,,
\end{equation}
where we recall that $n^\mu$ is an operator and as a consequence $[a(p),\kappa^\alpha]\neq 0$.
The very notion of quantum particle, as an excitation of the (extended) fields, becomes inseparable from the quantum foliation. We revisit and expand upon this point in section \ref{sec:qfeffects}.

We now return to the notion of physical subspace suggested by the classical discussion of section \ref{sec:physub}. By using the classical version of the expansion (\ref{eq:fieldsexp}), one can show that the constraints  (\ref{eq:physconstra}) imposed for all times are equivalent to $\{\mathcal{S},a(p)\}\approx 0$ imposed for all $p$ (and its conjugate). 
We can impose half of these infinite constraints at the quantum level by requiring that physical states are annihilated by the conditions, namely
\begin{equation}\label{eq:physsub}
    [\mathcal{S},a(p)]|\Psi\rangle_{\rm phys}=0\,.
\end{equation}
This requires that the only particles present in the physical subspace are those on-shell, as it follows from $[\mathcal{S},a(p)]=-[p^\mu n_\mu -E_p(n)]a(p)$ which vanishes only for $p^0=E_{\textbf{p}}$. An on-shell particle is physical in any reference frame as it follows from 
$
    [\mathcal{J}_{\mu\nu},\mathcal{S}]=0
$
and Jacobi's identity. 

In this simple free case, it is straightforward to recover dynamical information from  physical states. For example, for quadratic theories under a translation in time 
on-shell ladder operators ``move through time'' as if they were evolving. This fact can be employed to obtain conventional transition amplitudes from the extended formalism, as shown in \cite{diaz.21}. 
Another interesting feature to notice is that single particle (sp) states have the form of the Page and Wootters (PaW) states \cite{PaW.83,QT.15}, as shown in \cite{dia.19,diaz.21,gio.23}.   In this sense, one can state that the excitations of the fields, in their extended approach, are particles formulated as in quantum time/string-inspired formalisms \cite{Schb.01}.

In the free case the physical subspace has a clear interpretation as the linear space of particles on-shell (see also the results in \cite{di.19, dia.19} regarding the normalization of states). Nonetheless, as interactions come into play, the notion of particle becomes less clear, and the meaning of physical subspaces as well.  
In the following, we develop a much more powerful approach to map extended quantities to standard quantum evolution which holds for interacting theories. The concept of physical subspace appears again naturally when considering scattering processes in which case the external particles are (roughly speaking) regarded as asymptotically free in the usual sense.

\subsection{Spacetime correlators and map to conventional QM at a fixed foliation}\label{sec:stcorr}
Besides particles, another key element of QFTs (and QMs in general) are correlators. Conventional correlators are associated  with spacelike separations between operators. For hermitian operators such correlators can be interpreted as  the mean value of an observable. Instead, correlators involving timelike separated observables do not correspond to hermitian operators, but usually appear associated with transition amplitudes, e.g. in perturbation theory.
In this section, we show in complete generality how the extended formalism allows to recover both  
in a unifying way. This introduces a general correspondence between the spacetime version of QM and the conventional approach.

Let us recall first that quadratic operators are fully determined by their basic contractions (Wick's theorem). In the diagonal case, one has essentially the correlator 
\begin{equation}\label{eq:contr}
    \langle a^\dag_k a_l\rangle:=\frac{{\rm Tr}\big[\exp(-\sum_i \lambda_i a^\dag_i a_i)a^\dag_k a_l\big]}{{\rm Tr}\big[\exp(-\sum_i \lambda_i a^\dag_i a_i)\big]}=\frac{1}{\exp(\lambda_k)-1}\delta_{kl}\,.
\end{equation}
Here the indices $k,l$ are ``space-like separated'', in the sense that the operators $a^\dag_k$, $a_l$ are not evolved in the given reference frame and correspond to orthogonal modes.  We have also assumed a bosonic algebra $[a_k,a^\dag_l]=\delta_{kl}$, the fermionic case is analogous.
Similarly, one may consider instead the position-momentum correlators $\langle q_i q_j\rangle$ and $\langle p_i p_j\rangle$ for $[q_i,p_j]=i\delta_{ij}$, which correspond to the mean value of hermitian operators.
 
The extended algebra (\ref{eq:extalg}) allows us to apply equation (\ref{eq:contr}) to both space and time indices. Equivalently, we can apply it to off-shell correlators, as now permitted by (\ref{eq:algpart}). Remarkably, when we use it in conjunction with a quadratic action operator $\mathcal{S}$, and ``insert'' operators at different points in time,
the propagators of conventional QM naturally emerge. Conversely, one could ``rediscover'' the operator $\mathcal{S}$ as the only quadratic operator whose spacetime contractions are the conventional free propagators.

This result, recently proven in \cite{diazp.21} without $n^\mu$, provides a general map between conventional QM in $d$ dimensions and the extended formulation with algebras in $D=d+1$ dimensions.  It also leads directly to a redefinition of the Path Integral (PI) formulation as a trace involving the quantum action $\mathcal{S}$. 
We provide here a new derivation 
of the map particularly suited for field theories thus unveiling new features. In the following subsection \ref{sec:stgstates}, we further develop it by providing an interpretation in terms of generalized states and ``pseudo'' correlations.
In this section, we consider a classical $n^\mu$. The extension to a quantum $n^\mu$ is developed in section \ref{sec:qfeffects}, building on the classical foliation case.

We want to exponentiate the action operator, which is not adimensional, so we introduce a time or inverse energy scale $\tau$ (regarded as a positive real parameter for convenience) and define $\mathcal{S}_\tau=\tau \mathcal{S}$ in the free case. 
Additional comments on this new ``coordinate'' are made at the end of this section, while a complementary time-slice approach is presented in Appendix \ref{sec:Apmap}. 
Similarly, in order to consider fields with correct units we will add a factor $\sqrt{\tau}$ for each operator. 
We indicate the extended correlators of $e^{i\mathcal{S}_\tau}$ as
\begin{equation}\label{eq:Scorr}
    \langle \mathcal{O} \rangle :=  \frac{{\rm Tr}\,e^{i\mathcal{S}_\tau} \mathcal{O}} {{\rm Tr}\,e^{i\mathcal{S}_\tau}}\,.
\end{equation}
\vspace{0.08pt}

By recognizing that for a quadratic $\mathcal{S}$ equation (\ref{eq:Scorr}) can be regarded as a particular instance of (\ref{eq:contr}), any spacetime correlator can be easily obtained. We consider the example of the Klein Gordon action which in its diagonal form 
 (\ref{eq:diagaction}), characterized by off-shell particles, yields

\begin{widetext}
    \begin{align}\label{eq:momcorr}
  \langle a^\dag(p)a(k)\rangle\!=\!\frac{{\rm Tr}\big[\exp\big\{i\tau \int \frac{d^Dp'}{(2\pi)^D}\, (p'^0-E_\textbf{p'}+i\epsilon)a^\dag(p')a(p')\}a^\dag(p)a(k)\big]}{{\rm Tr}\big[\exp \big\{i\tau \int \frac{d^Dp'}{(2\pi)^D}\, (p'^0-E_\textbf{p'}+i\epsilon)a^\dag(p')a(p')\big \}\big]}
  \!=\!\frac{1}{\exp \{-i\tau (p^0-E_\textbf{p}+i\epsilon)\}-1}(2\pi)^D\delta^{(D)}(p-k)\,,
\end{align}
\end{widetext}
 where we have replaced $E_\textbf{p}\to E_\textbf{p}-i\epsilon$ and assumed for simplicity $n^\mu=\eta^{\mu 0}$ (the general case corresponds to $p^0-E_\textbf{p}\to p^\mu n_\mu -E_p(n)$). 
It is interesting to consider the small $\tau$ case of this expression. One has
 $\langle a^\dag(p)a(k)\rangle= \frac{1}{\tau}\frac{i}{(p^0-E_\textbf{p}+i\epsilon)}(2\pi)^D\delta^{(D)}(p-k)+\mathcal{O}(\tau)$ whose Fourier transform in $p^0$ yields a Heaveside theta function in the conjugate variable, i.e. in the time variable.

 It is important to notice that (\ref{eq:momcorr}) cannot correspond to a genuine correlator (having the ``spacelike'' form of Eq.\ (\ref{eq:contr})) in traditional QM, essentially because the extended formalism has  extra indices. In other words, since time is an ``index site'' indicating independent field operators, $p^0$ becomes also a label and denotes independent ladder operators: most correlators in the extended Hilbert space do not correspond to a single ``contraction'' of traditional QM. Only for those with operators inserted at a single time slice a one-to-one identification is possible. As we will now show, evolution emerges from this apparent ``redundancy''.

 Equation (\ref{eq:momcorr}) is the basic off-shell momentum space correlator from which spacetime-localized correlators can be obtained. The latter are defined by the expansion (\ref{eq:fieldsexp}). 
 In particular, 
 it is straightforward to compute
    \begin{align}\label{eq:feynproptau}
   \langle\phi(x)\phi(y)\rangle= \frac{1}{\tau}\int \frac{d^Dp}{(2\pi)^D}\frac{i}{p^2-m^2+i\epsilon}e^{-ip (x-y)}+\mathcal{O}(\tau)
\end{align}
where we used $\frac{i}{p^0-E_\textbf{p}+i\epsilon}-\frac{i}{p^0+E_\textbf{p}-i\epsilon}\equiv 2E_\textbf{p}\frac{i}{{p^2-m^2+i\epsilon}}$ and we are considering a small $\tau$. One immediately recognized the expression of the Feynman propagator which allows us to write
\begin{equation}\label{eq:feynprop}
  \lim_{\tau\to 0}  \langle\sqrt{\tau}\phi(x)\sqrt{\tau}\phi(y)\rangle= \langle 0| \hat{T} \phi_H(x)\phi_H(y)|0\rangle\,.
\end{equation}
On the right-hand side, $\phi_H(\textbf{x},t):=e^{iHt}\phi(\textbf{x})e^{-iHt}$ is the conventional (non-extended) field operator in the Heisenberg picture and $|0\rangle$ is the usual ground state of the free Klein Gordon Hamiltonian $H$, while $\hat{T}$ denotes time ordering.  On the left-hand side, the operators are not evolved with some evolution operator, instead, their ``position in time'' has determined the amount of evolution:
the lhs of (\ref{eq:feynprop}) can always be understood as a correlator such as the one in Eq.\ (\ref{eq:contr}), even for $|x-y|$ time-like in which case evolution emerges. Notice also that instead of considering the small (positive) $\tau$ limit, which reflects the intuition of a discrete spacetime (see Appendix \ref{sec:Apmap}), one might consider integrating around loops in the complex plane and exploiting the pole structure of correlators. 

The previous results also define the proper treatment of interacting field theories: consider $\mathcal{S}_\tau\to \mathcal{S}_\tau+\mathcal{S}_{\text{int}}[\sqrt{\tau}\phi]$ for $\mathcal{S}_{\text{int}}[\sqrt{\tau}\phi]$ having the classical functional form on the fields, e.g. for a classical action $S_{\text{int}}=-\int d^Dx\,\frac{\lambda}{4!} \phi^4$ one has $\mathcal{S}_{\text{int}}=-\int d^Dx\,\frac{\lambda}{4!} \tau^2 \phi^4$. Then in the small $\tau$ limit,  the ``interacting'' correlator of fields, defined by considering the whole action in (\ref{eq:Scorr}), has the following expansion: 
\begin{equation}\label{eq:corrpert}
\begin{split}\lim_{\tau \to 0}\langle\sqrt{\tau}\phi(x)\sqrt{\tau}\phi(y)\rangle_{\text{int}}
      &=\frac{\langle 0| \hat{T} e^{iS_{\text{int}}[\phi_I]}\phi_I(x)\phi_I(y)|0\rangle}{\langle 0|\hat{T}e^{iS_{\text{int}}[\phi_I]}|0\rangle}\,.
\end{split}
\end{equation}
The equality is a direct consequence of (\ref{eq:feynprop}) and Wick's theorem (for Gaussian ``states'') applied to the free part of the action, with the interacting part  expanded perturbatively for small $\tau$. For this reason, the evolution which emerges is the one that would correspond to the interacting picture, i.e. $\phi_I(t,\textbf{x})=e^{iH_0t} \phi(\textbf{x})e^{-iH_0t}$ for $H_0$ the free Klein-Gordon Hamiltonian. 
One also recognizes in the rhs of (\ref{eq:corrpert}) the perturbative expansion of the interacting correlator $\langle GS| \hat{T} \phi_H(x)\phi_H(y)|GS\rangle$, with $|GS\rangle$ the ground state of the interacting Hamiltonian. Assuming as usual the validity of perturbation theory,  we conclude that
\begin{equation}\label{eq:map}
 \!\!   \lim_{\tau \to 0}\langle\sqrt{\tau}\phi(x)\sqrt{\tau}\phi(y)\rangle_{\text{int}}=\langle GS| \hat{T} \phi_H(x)\phi_H(y)|GS\rangle\,.
\end{equation}

From these expressions scattering amplitudes can be computed as usual e.g. by using the LSZ reduction formula \cite{srednicki2007}. One can show that the $d+1$ dimensional Fourier transform involved translates to inserting on-shell ladder operators in the correlators. In other words, scattering amplitudes are proportional to correlators of the form $\langle \prod_i a(k_i) \prod_j a^\dag(p_j) \rangle_{\text{int}}$ for $p_j$ ($k_i$) the ``in'' (``out'') momenta (see also \cite{diazp.21} and section \ref{sec:stgstates}). 

Note also that for a fixed but general $n^\mu$ one just needs to make the replacement $p^0-E_\textbf{p}\to p^\mu n_\mu -E_p(n)$ in Eq.\ (\ref{eq:momcorr}). Equations  (\ref{eq:feynproptau}-\ref{eq:map}) remain unchanged.

Before proceeding further, we would like to remark that the similarities between the previous expressions and the PIs ones are not a coincidence. While the previous results have been obtained from Hilbert space techniques, associated with the algebra of operators, and are thus independent of Feynman's approach, one can evaluate the previous traces explicitly on a given basis. If one chooses the spacetime basis of field configurations $|\phi(x)\rangle$ (see Eq.\ (\ref{eq:fieldbas})) Feynman PIs  emerge, as shown in \cite{diazp.21}.
In this sense, the formalism is embedding the PI formulation in a Hilbert space.

It is also interesting to discuss how the extended classical formalism of the previous section can be recovered in the limit $\hbar\to 0$. In first place, let us notice that 
\begin{equation}
    \begin{split}
        [\phi(x),\mathcal{S}_\tau]&=i\dot{\phi}(x)-[\phi(x),\int d^Dz\, \mathcal{H}]\\
        -[\pi(x),\mathcal{S}_\tau]&=i\dot{\pi}(x)+[\pi(x),\int d^Dz\, \mathcal{H}]
    \end{split}
\end{equation}
have the form of \emph{Heisenberg equations} if set to zero (and absorbing the $\tau$ factors in the fields). Notably, since the cyclicity of the trace implies $\langle [\dots,\mathcal{S}_\tau]\rangle \propto {\rm Tr}\{e^{i\mathcal{S}_\tau}[\dots,\mathcal{S}_\tau]\}=0$  for any operator, we have $\langle [\phi(x),\mathcal{S}_\tau]\rangle=\langle [\pi(x),\mathcal{S}_\tau]\rangle=0$, which according to our map (holding for small $\tau$) agrees with Heisenberg equations in conventional QM. 
On the other hand,
by following a similar argument as in the standard PI formulation, for $\hbar\to 0$ the only contributions to the trace come essentially from extreme classical configurations of the action (see also \cite{diazp.21}). 
At the same time, since the form of the extended commutators and extended PBs is the same we can write
\begin{equation}
    0=\langle [\dots, \mathcal{S}_\tau]\rangle\overset{\hbar\to 0}{\sim} \{\dots, \mathcal{S}\}|_{\text{On-shell}}
\end{equation}
where the commutator is applied to any extended quantum operator, and the PB to the associated extended-phase space function (the ordering becomes irrelevant in the small $\hbar$ limit). The latter is first computed according to the extended algebra \eqref{eq:extpb} and then evaluated at a solution of the equations of motion. 
We conclude that the quantum result $\langle [\dots, \mathcal{S}_\tau]\rangle=0$, together with Eqs.\ \eqref{eq:hamiltonaction},  imply
Hamilton equations for $\hbar\to 0$. Interestingly, they emerge as a limit of spacetime QM through the extended PBs of spacetime classical mechanics.

The previous results establish a basic connection between the extended and conventional QFTs at zero temperature (i.e. associated with the ground state of the Hamiltonian in question). It is also interesting to briefly mention how thermal propagators arise for a finite time window of length $T$. Essentially, the diagonalization of the free Klein Gordon action now yields
\begin{equation}\label{eq:diagactionwindow}
    \mathcal{S}=\frac{1}{T}\sum_n \int \frac{d^dp}{(2\pi)^d}(w_n-E_\textbf{p})a_n^\dag(\textbf{p})a_n(\textbf{p})\,,
\end{equation}
for $w=2\pi n/T$, the Matsubara frequencies, here arising from the diagonalization of $\mathcal{P}_0$. We are also assuming a compactified time (periodic conditions) such that (\ref{eq:algpart}) is replaced by $[a_n(\textbf{p}),a_{n'}^\dag(\textbf{k})]=T\delta_{nn'}(2\pi)^d\delta^{d}(\textbf{p}-\textbf{k})$ while the expansions (\ref{eq:fieldsexp}) hold by replacing the integral in $p^0$ with a sum over $n$ (with also $(2\pi)^{-1}\to T^{-1}$).

If one now considers $E_\textbf{p}\to -iE_\textbf{p}$, it is straightforward to see that (\ref{eq:feynprop}) is replaced by the Matsubara expansion of the (thermal) correlator \cite{kh.09}. The corresponding temperature is $\beta \equiv T$.

If one also discretizes time in $N=T/\epsilon$ steps, results such as (\ref{eq:feynprop}) become exact for operators ``inserted'' at time commensurable with $\epsilon$. One also dispenses with $\tau$ which is replaced by the time step $\epsilon$ (see Appendix \ref{sec:Apmap} for the details and the definition of the quantum action for discrete spacetime). Moreover, since all spacelike correlators are obtained from the quantum action by simply considering e.g. operators in the initial slice
\begin{equation}\label{eq:ptrace}
    e^{-\beta H}={\rm Tr}_{t\neq 0}\, e^{i\mathcal{S}}\,,
\end{equation}
i.e. we can recover the conventional thermal state from the quantum action by considering a partial trace over all times except the initial slice (we are assuming a Wick rotation of the Hamiltonian part of the action; this does not affect $\mathcal{P}_0$). Here $\beta \equiv T$. This also implies $Z:={\rm Tr}\, e^{-\beta H}={\rm Tr}\, e^{i\mathcal{S}}$. Interestingly, partial traces over arbitrary spacetime regions can be considered in the extended formalism. In principle, only those associated with spacelike hypersurfaces correspond to
conventional quantum states (and real entropies, see section \ref{sec:stgstates}) but the partial trace is well-defined in general \cite{diazp.21}.

We finally mention that if one is interested in states (or transitions) besides thermal ones or ground states, these can be specified by adding a projector on the ``initial slice'', as developed in \cite{diazp.21} and shown in Appendix \ref{sec:Apmap}.

Let us also mention that 
in \cite{diazp.21} a version with finite $\tau$ has also been constructed, which may be employed to define rigorously the limit $\tau\to 0$ (a result which is not required here). Instead, for large $\tau$ one can rewrite the map as an asymptotic mean value of more complicated ($\tau$ evolved) operators \cite{diazp.21}. 
The previous mean value may then be associated to a $D+1=d+2$ theory with spacetime volume $\propto d^{d+2}x=d\tau d^{d+1}x$, essentially by considering $\tau$ as an evolution parameter in the conventional sense.

\subsection{Spacetime generalized states}\label{sec:stgstates}
Arguably, the most fundamental element of the mathematical framework of QM is the notion of state. Conventional pure states encode all the information about a quantum system at a given time. 
The state is thus associated with physical predictions at a specific moment, as determined by QM axioms. 

While we have established a general map between quantities of the extended formalism to quantities involving conventional states, this map relies on the operator $e^{i\mathcal{S}_\tau}$ which is clearly not a state nor a density matrix \footnote{In \cite{diazp.21} it is shown that for large $\tau$ one can rewrite the map as an asymptotic mean value of more complicated ($\tau$ evolved) operators. Here we are more concerned with a notion of state for arbitrary (even small) $\tau$.}.  Yet, since in principle, all predictions of the system at different times can be extracted from the spacetime  correlators, one could argue that some notion of ``spacetime state'' may be assigned to the previous map. Conversely, if a notion of spacetime state can be properly defined it should be related to it.

Equation (\ref{eq:momcorr}), which is essentially the Bose-Einstein distribution with the role of the thermal state replaced by $e^{i\mathcal{S}_\tau}$, suggests an interesting course of action: 
one may consider some sort of purification of $e^{i\mathcal{S}_\tau}$ such as the ones considered in thermofield dynamics to treat thermal effects in QFTs with zero temperature techniques \cite{kh.09}. Therein thermal traces are replaced by mean values on properly defined ``enlarged'' pure states. This idea is further reinforced by the fact that by considering the free $\mathcal{S}$ as the $\tau$ evolution generator in a $d+2$ theory, $e^{i\tau \mathcal{S}}$ takes the role of a $d+1$ ``thermal'' state with imaginary temperature $-i\tau$. Moreover,  Eq.\ (\ref{eq:ptrace}) explicitly shows that the information of conventional $d$ thermal states for arbitrary Hamiltonians can also be contained in $e^{i\mathcal{S}_\tau}$.
While a thermofield-like approach is strictly speaking not necessary (one can use the previous map in the form of subsection \ref{sec:stcorr}), it leads to interesting insights on the nature of the non-hermitian operator $e^{i\mathcal{S}_\tau}$.

A ``purification'' of $e^{i\mathcal{S}_\tau}$
is easily obtained by considering two different states living in a duplicated  Hilbert space. Considering for simplicity the $T\to\infty$ limit and the free Klein Gordon theory, we indicate the ``environment'' operators as e.g. $\tilde{a}(p)$ (with $[\tilde{a}(p),\tilde{a}^\dag(k)]=(2\pi)^D\delta^{(D)}(p-k)$)  and environment states as $|\tilde\Psi\rangle$.  
Then, by considering a partial trace over the environment $E$, we can express $e^{i\cal S_\tau}$ as a reduced generalized state: 
\begin{equation}
    \frac{e^{i\mathcal{S}_\tau}}{{\rm Tr  
 \,e^{i\mathcal{S}_\tau}}}={\rm Tr}_E\,R_\tau\,,\;\;R_\tau:=\frac{|\Omega_\tau\rangle \rangle \langle\langle\overline{\Omega}_\tau|}{\langle\langle \overline{\Omega}_\tau| \Omega_\tau\rangle\rangle}\,,
 \label{Rdef}
\end{equation}
where we have introduced the two distinct global pure states 
\begin{align}
    |\Omega_\tau\rangle\rangle&:=\exp\Big [\int \frac{d^Dp}{(2\pi)^D}\, e^{i\tau(p^0-E_p+i\epsilon)/2}a^\dag(p)\tilde{a}^\dag(p)\Big]|\Omega\rangle\rangle \nonumber\\
|\overline{\Omega}_\tau\rangle\rangle&:=\exp\Big[\int \frac{d^Dp}{(2\pi)^D}\, e^{i\tau(E_p-p^0+i\epsilon)/2}a^\dag(p)\tilde{a}^\dag(p)\Big]|\Omega\rangle\rangle\,,
        \label{eq: states2}
\end{align}
with $|\Omega\rangle\rangle=|\Omega\rangle |\tilde{\Omega}\rangle$ the  global vacuum  and $a(p)|\Omega\rangle=0$, $\tilde a(p)|\tilde\Omega\rangle=0$ $\forall p$ (even off-shell). The states \eqref{eq: states2} are in fact  system-environment entangled Bogoliubov vacua of global annihilation operators (see Appendix \ref{sec:Appurif} for details and proof of \eqref{Rdef}). 

In \eqref{Rdef} we have defined the nonorthogonal (nonhermitian) projector $R$ ($R^2=R$) having trace  $1$ (and hence a single nonzero eigenvalue 1), such that it can be considered as a generalization of the notion of pure state. We also notice that \begin{equation}
  {\rm Tr \,e^{i\mathcal{S}_\tau}}  =\langle\langle \overline{\Omega}_\tau| {\Omega}_\tau\rangle\rangle\,,
\end{equation}
which is nonzero.

Interestingly, this kind of generalization of the traditional purification has been recently introduced \cite{nak.21} in the context of the dS/CFT correspondence to define a notion of time-like entanglement \cite{tak.23, har.23, nar.22, chu2023time}  (in conventional, non-extended QM where there is no action operator). It has also been employed to define a dual quantity (a pseudo entropy) to minimal area surfaces in time-dependent spacetimes \cite{nak.21}, according to the AdS/CFT correspondence \cite{mal.99}.
 The fact that these generalized states emerge naturally both in those contexts and in the present spacetime version of QM,  may be an indicator that they are in fact  
 required in any (sufficiently general) extension of the  notion of state to the time domain \footnote{There are other useful extensions of the notion of state associated with ``quantum time'' (see e.g \cite{QT.15,b.16,di.19,dia.19}). However, none of these allow a partial trace over time regions \cite{diazp.21}.}. One observation in support of this hypothesis is that, contrary to conventional states, they lead to complex entropies \cite{nak.21,nar.22,tak.23}, which may be related to the pseudo-Riemannian nature of classical spacetime (we recall the conjectures of space emerging from entanglement \cite{ryu.06,van.10,Ca.17}).

\begin{figure}[t!]
    \centering
    \includegraphics[width=0.48\textwidth]{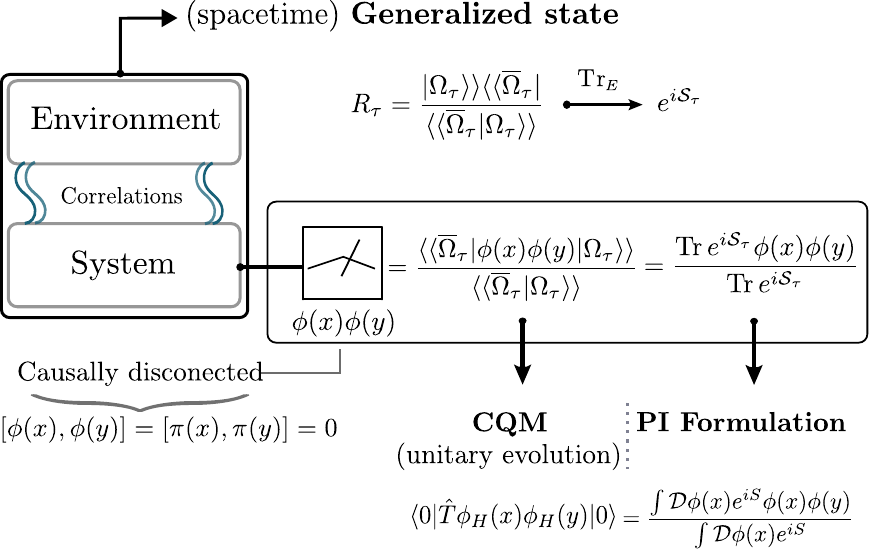}
    \caption{ 
    \textbf{Scheme of the correspondence between the QM formulations (fixed $n^\mu)$}. In the spacetime formulation, we can codify all the information about a given system and its evolution in generalized states, encompassing an environment correlated with the system. By ``measuring'' on the system only (see the remarks on weak values) conventional propagators and Feynman rules are recovered. The example of the Feynman propagator is depicted, which corresponds to the hermitian observable $\phi(x)\phi(y)$. Contrary to canonical QM (CQM) where $[\phi_H(x),\phi_H(y)]\neq 0$ inside the light-cone, in the spacetime formulation every field $\phi(x)$ is independent from the others and $[\phi(x),\phi(y)]=0$ for any spacetime points (a much stronger statement than microcausality). 
    The information about evolution and causality is contained in the generalized system-environment state $R_\tau$ and one can think that it emerges from the (``generalized/pseudo'') correlations between the two. Since the environment is ignored, one can also work with the partial ``state'' of the system $e^{i\mathcal{S}_\tau}$ directly, as described in section \ref{sec:stcorr}. It was shown in \cite{diazp.21} that particular evaluations of the ensuing traces lead to the PI formulation (see also Appendix \ref{sec:Apmap}). }
    \label{fig:fig2}
\end{figure}

With these results at hand, we can write spacetime correlators (\ref{eq:Scorr}) as
\begin{equation}\label{eq:weakv}
    \langle \mathcal{O}\rangle= \frac{\langle\langle \overline{\Omega}_\tau| \mathcal{O}\otimes \mathbbm{1}_E|\Omega_\tau\rangle \rangle}{\langle \langle \overline{\Omega}_\tau|\Omega_\tau\rangle \rangle}={\rm Tr} \big[ R_\tau\, \mathcal{O}\otimes \mathbbm{1}_E \big] \,.
\end{equation}
For instance, Feynman's propagator can be written as $\langle \langle \overline{\Omega}_\tau|\phi(x)\phi(y)|\Omega_\tau\rangle\rangle\propto\langle 0| \hat{T}\phi_H(x)\phi_H(y)|0\rangle$ (see \eqref{eq:feynprop}).
Interestingly, we see that evolution emerges from correlations between the system and (the so far) abstract environment: since there is no indication in the operators $\phi(x), \phi(y)$ on whether $x,y$ are space or time variables it is clear that all the information about the causal structure of the theory is encoded in the entangled states $|\Omega_\tau\rangle\rangle$, $|\overline{\Omega}_\tau\rangle\rangle$, as represented in Figure \ref{fig:fig2}. The correlations responsible for the time evolution emergence are precisely the ones which the recently introduced pseudo entropies aim to quantify (e.g. $S(R_\tau)=-{\rm Tr}R_\tau \log R_\tau=0$ but the subsystem ``state'' $e^{i\mathcal{S}_\tau}$ is not a projector). 
The complete information about  conventional spacelike correlations of the state $|0\rangle$ is also encoded in $R_\tau$,  since the correlators $\langle 0|\phi(\textbf{x})\phi(\textbf{y})|0\rangle$, $\langle 0|\pi(\textbf{x})\pi(\textbf{y})|0\rangle$ are particular cases of \eqref{eq:weakv} corresponding to $\mathcal{O}=\phi(\textbf{x},t_0)\phi(\textbf{y},t_0)$, $\mathcal{O}=\pi(\textbf{x},t_0)\pi(\textbf{y},t_0)$, i.e. to the insertion of operators on spatial slices. These appear e.g. in the definition of spacelike entanglement \cite{cas.09}.

\begin{figure}[h!]
    \centering
    \includegraphics[width=.7\columnwidth]{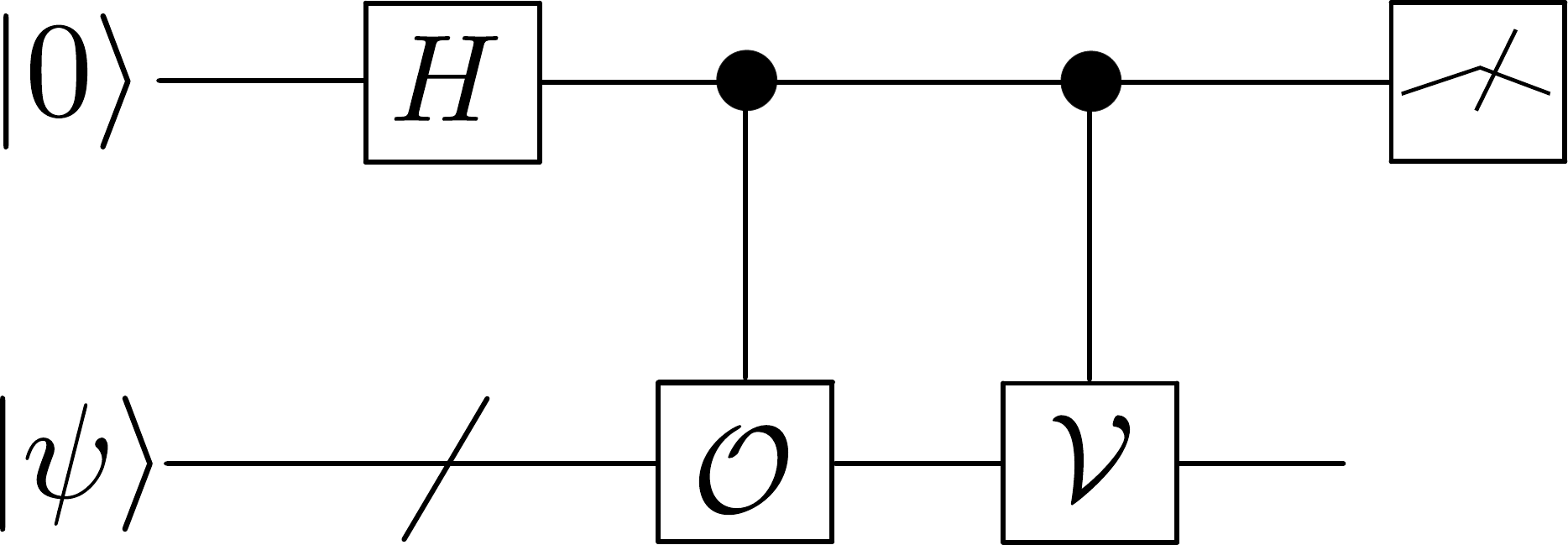}
    \caption{\textbf{Quantum circuit for computing a quantity of the form} $\langle \varphi|\mathcal{O}|\psi\rangle$. The scheme is a  Hadamard test where measurements are performed in the ancillary qubit (on top) to estimate the real and imaginary parts of $\langle \psi| \mathcal{V}\mathcal{O}|\psi\rangle$, and where we choose $\mathcal{V}$ so that  
    $|\varphi\rangle=\mathcal{V}^\dag|\psi\rangle$. By using states $|\psi\rangle$, $|\varphi\rangle$ that define a generalized state, one can compute spacetime correlation functions (see also Appendix \ref{sec:Apmap}). }
    \label{fig:circ}
\end{figure}

We also notice that a quantity of the type \eqref{eq:weakv} for hermitian operators also appears in conventional QM where it is denoted as \emph{weak value} \cite{yak.88}. A spacetime correlator can then be understood as the weak value of  $\mathcal{O}\otimes \mathbbm{1}_E$  for $\mathcal{O}$ hermitian. We recall that while e.g. $\hat{T}\phi_H(x)\phi_H(y)$ is not hermitian for a timelike separation, $\mathcal{O}=\phi(x)\phi(y)$ is always an observable. As a consequence, one can use existent  techniques (see e.g. \cite{dres.14} for ways of measuring weak values) to access (\ref{eq:weakv}) through measurements. This is an interesting result on its own since it provides timelike correlators a direct operational meaning, a remark that holds for general quantum systems, as discussed in Appendix \ref{sec:Apmap}. In addition, it is well-known how to compute such quantities in quantum computers (see Figure \ref{fig:circ}; for a recent development about quantum circuits measuring weak values see \cite{wag.23})

Another case of interest is the evaluation of scattering amplitudes. Equation (\ref{eq:map}) suggests to consider $\mathcal{O}=a(k_1)a(k_2)\dots a(k_m)e^{i\mathcal{S}_{int}}a^\dag(p_1)a^\dag(p_2)\dots a^\dag(p_n)$. In fact, for on-shell momenta $\langle \mathcal{O}\rangle$ is proportional to the $S$ matrix elements (the proportionality must be chosen to agree with the LSZ formula, such that it ``amputates'' external lines).
To show this it is sufficient to notice that 
$\langle \langle \overline{\Omega}_\tau|\phi(x) a^\dag(p)|\Omega_\tau \rangle\rangle\propto e^{-ipx}$
and use the results of the previous subsection to recover the \emph{Feynman rules} in position space. A direct application of the usual techniques can then be used to obtain finite physical predictions. 
As a final remark, we notice that by using Eq.\ (\ref{eq:weakv}) one can write the elements of the scattering matrix as a transition amplitude between states created by on-shell (extended) modes on the global vacua \eqref{eq: states2}. These states can also be explicitly related to the physical states defined in Eq.\ (\ref{eq:physsub}) by writing $|\Omega_\tau\rangle\rangle$, $|\overline{\Omega}_\tau\rangle\rangle$ as Bogoliubov transformations on the product vacuum $|\Omega\rangle \rangle$ (see Appendix \ref{sec:Apmap}).

\section{Matter-Foliation entanglement} \label{sec:qfeffects} 

\subsection{Particles as foliation controlled operators}
Having discussed the classical case, its quantization, and how to establish a general map to conventional QM for a fixed  $n^\mu$ (with an interpretation of evolution emerging from generalized spacetime states), we dedicate a final section to lay the foundations 
to handle a  full quantum foliation.

Before proceeding with a mathematical exposition, it is worth dedicating some discussion on why considering a full quantum $n^\mu$ could be physically relevant beyond the consistency of the formalism. We recall that the introduction of an algebra associated with $n^\mu, p^\nu$ was a mathematical necessity: a proper transformation rule of the extended off-shell particles can only be achieved if the foliation is modified by the transformation, in turn requiring a quantum $n^\mu$. As we show below, this is fundamental when considering expectation values as well: by taking into account the transformation properties of the foliation we can prove the explicit covariance of all mean values, conditioned to classical values of the foliation.
On the other hand, there is nothing preventing the use of more general foliation states. 
While it is reasonable to suspect that this may be pointing to something deeper physically, gaining further insight requires additional development like the application of the extended quantization scheme to dynamical spacetimes (beyond the scope of the present work).

On the other hand, it is easy to come up with scenarios in which some notion of quantum uncertainty is assigned to observers. Since the formalism provides a rigorous framework containing this feature, it is interesting to explore it even if for this reason only. Many such scenarios can be constructed by appealing to the argument that in practice observers need to perform measurements to establish their own notion of space and time. Since those measurements are fundamentally described by QM then one may conclude that a quantum uncertainty is inherited. This argument can be found in the literature in different contexts \cite{aha.84,rov.91,cas.18,giac.19,fav.20,hoh.21,pai.22}, usually related to some generalization of QM and in relation to the ``problem of time''.
One particularly interesting observation is that according to the cosmological principle \cite{Wein.72} a cosmic foliation (or cosmic time) can be defined such that the universe looks homogeneous and isotropic at each moment. At an early stage of the universe, where quantum effects may become important these hypotheses need not hold, and the foliation could become ``fuzzy''.

We now return to the example of the free Klein Gordon theory and the notion of extended particles introduced in \ref{sec:expart} but focus on a completely quantum $n^\mu$. Similar ideas hold for other field theories. Let us first notice that the operator $a(p)$ we have introduced can be properly written as
\begin{equation}\label{eq:atimesn}
    a(p)=\int dn\, a(p,n)\otimes |n\rangle\langle n|
\end{equation}
for $a(p,n)$ the annihilation operator obtained by replacing the operators $n^\mu$ in Eq.\ (\ref{eq:aexp}) with the fixed value $n$. Each $a(p,n)$ is a genuine annihilation operator satisfying $[a(p,n),a^\dag(p',n)]=(2\pi)^D\delta^{(D)}(p-p')$, while strictly \begin{equation}[a(p),a^\dag(p')]=[a(p,n),a^\dag(p',n)]\otimes\hat{\mathbbm{1}}_n\end{equation} 
which was implicit in \eqref{eq:algpart}.  Here $\hat{\mathbbm{1}}_n$ is to be read as the projector on the subspace generated by those $|n\rangle$ with $n^\mu$  timelike. 
 Note, however, that we can let $||n||^2>0$ arbitrary in these  expressions. 
   All expressions in section \ref{sec:expart} hold for 
  $E_p(n)=||n||\sqrt{\left(\frac{n^\mu n^\nu}{||n||^2}-\eta^{\mu\nu}\right)p_\mu p_\nu +m^2}$, as shown in Appendix \ref{sec:Apa}. 

Equation \eqref{eq:atimesn} reveals that $a(p)$ has the form of a controlled operator in which the values of the foliation states determine which $a(p,n)$ acts (one can compare this with a control-not operation between qubits $U_{\text{control-not}}=\sum_{n=0,1}(\sigma_x)^n\otimes |n\rangle\langle n|$ for $\sigma_x$ the $x$-Pauli matrix acting on the controlled qubit). This way of writing $a(p)$ makes its transformation properties more clear: the operator $a(p,n)$ acts on the matter sector and is transformed with $\mathcal{U}_\phi(\Lambda):=\exp(i\omega_{\mu\nu}{\cal L}^{\mu\nu}/2)$ the boost operator that transforms the fields but not the foliation. In fact, the complete transformation of $a(p)$ can be understood as
\begin{equation}
    \mathcal{U}^\dag(\Lambda) a(p)\mathcal{U}(\Lambda)=\int dn\, \big(\mathcal{U}^\dag_\phi(\Lambda)a(p,n)\mathcal{U}_\phi(\Lambda)\big) \otimes |\Lambda n\rangle \langle \Lambda n|
\end{equation}
which, comparing with \eqref{eq:atransf} yields
\begin{equation}\label{eq:apntrans}
\mathcal{U}^\dag_\phi(\Lambda)a(p,n)\mathcal{U}_\phi(\Lambda)=a(\Lambda p,\Lambda n)\,,
\end{equation}
a relation which can also be obtained from \eqref{eq:aexp} by fixing $n$. By employing \eqref{eq:aexp} one can also show that the different annihilation operators $a(p,n)$ and $a(p,n')$ are 
related by Bogoliubov transformations such that in general $[a(p,n), a(p',n')]\neq 0$ for $n\neq n'$. In particular, \eqref{eq:apntrans} does not preserve the particle number, only the complete transformation does. 

The previous also leads to 
the proper treatment of vacuum fluctuations.  
Notice first that a similar controlled expansion is assigned to 
$\mathcal{H}(\hat{n})=\int dn \, \mathcal{H}(n)\otimes |n\rangle \langle n|$
with $\mathcal{H}(n)$ acting on the matter sector. 
Then, each integrated $\mathcal{H}(n)$ can be diagonalized as a usual quadratic Hamiltonian, in analogy with the results in section \ref{sec:expart} but with a classical $n$: 
\begin{equation}
    \int d^Dx\, \mathcal{H}(n)=\int\frac{d^Dp}{(2\pi)^D} E_p(n) a^\dag(p,n)a(p,n) +\lambda(||n||)
\end{equation}
with $\lambda(||n||)$ a $n$-dependent constant that arises from having normal ordered the ladder operators $a(p,n)$. Surprisingly, when this ``constant'' is taken into account in the complete operator one has 
\begin{equation}\label{eq:cosm}
     \int d^Dx\, \mathcal{H}(\hat{n})=
     \int d^Dx\, :\mathcal{H}(\hat{n}):+\lambda(||\hat n||)\,,
  \end{equation} 
with $||\hat n||=\int dn\, ||n|| |n\rangle \langle n|$, i.e. rather than a constant shift, the (integrated) Hamiltonian picks a \emph{vacuum energy operator} acting on the foliation Hilbert space.

One can show that the associated vacuum energy density operator is just 
$\hat{\rho}_{\text{vac}}:=\rho_{\text{vac}} ||\hat{n}||$, with 
$\rho_{\text{vac}}$ the conventional vacuum energy density. In this simple scenario, the only quantum foliation effect on the vacuum density could be a statistical average over energy scales induced by $|\psi(n)|^2=|\langle n|\psi\rangle|^2\neq 0$ for $||n||\neq 1$.

\subsection{Matter-Foliation entangled states and explicit covariance of expectation values}
Another  interesting consequence of the operator expansion (\ref{eq:atimesn}) is that its eigenstates are in general entangled in the matter-foliation partition. The same is true for the quantum action $\mathcal{S}$ (see Eq.\ \eqref{eq:scontrol}). 
Consider in fact the concept of vacuum. For each fixed $n$, the operators $a(p,n)$ have a vacuum $|\Omega_n\rangle$ such that 
\begin{equation}
    a(p,n)|\Omega_n\rangle=0
\end{equation}
for all, in general off-shell, values of $p$. These vacua are all states in the matter sector and may be explicitly expanded  as $|\Omega_n\rangle=\int \mathcal{D}\phi(x) \Psi_n[\phi(x)]|\phi(x)\rangle$ in the basis \eqref{eq:fieldbas} with $\Psi_n[\phi(x)]$ their wavefunction.

On the other hand,  we may introduce a   history-like vacuum 
as
\begin{equation}\label{eq:globvac}
|\Omega\rangle=\int dn\,|\Omega_n\rangle\otimes |n\rangle \,,
\end{equation}
satisfying $a(p)|\Omega\rangle=0$, with $a(p)\equiv a(p,\hat{n})$. The state $|\Omega\rangle$ contains the information of the vacua of all possible time directions simultaneously. It achieves so through its entanglement with the quantum foliation. In fact, we can recover the vacuum of a given observer as $|\Omega_n\rangle=\langle n|\Omega\rangle$, i.e. by \emph{conditioning} on the foliation.  Remarkably, this feature and the structure of \eqref{eq:globvac} resemble the PaW formalism where evolution emerges from stationary history states by conditioning on internal ``clocks readings'' \cite{PaW.83, QT.15}.

Lorentz symmetry makes the entangled state
$|\Omega\rangle$ preferable over other eigenstates of $a(p)$: this state satisfies
 $\mathcal{U}(\Lambda)|\Omega\rangle=|\Omega\rangle$ or equivalently 
\begin{equation}\label{eq:jwdw}
\mathcal{J}_{\mu\nu}|\Omega\rangle=0\,,
\end{equation} which may be compared with the PaW universe equation \cite{QT.15}, a Wheeler-DeWitt like equation. This property is a direct direct consequence of \eqref{eq:apntrans} which implies $\mathcal{U}(\Lambda)|\Omega_n\rangle \otimes |n\rangle=|\Omega_{\Lambda^{-1}n}\rangle \otimes |\Lambda^{-1}n\rangle$. The integral in \eqref{eq:globvac} undoes this transformation via a trivial change of variables ($|\det(\Lambda)|=1$). Clearly, the invariance is not satisfied for more general superposition, i.e. by states which add weights to the sum \eqref{eq:globvac}. In particular, product states $|\Omega_n\rangle \otimes |n\rangle$ are annihilated by $a(p)$ but break Lorentz symmetry explicitly.

The previous structure holds for general states. This can be seen by considering a basis of Fock states which as usual can be obtained by acting  with creation operators $a^\dag(p)$ on the vacuum $|\Omega\rangle$ . For example, a two particle Fock state  can be written as 
\begin{equation}
    a^\dag(p_1) a^\dag(p_2) |\Omega\rangle= \int dn\, a^\dag(p_1,n) a^\dag(p_2,n) |\Omega_n\rangle \otimes |n\rangle\,.
\end{equation}
 In general, we have
$
    |\Psi\rangle=\int dn\, |\Psi_n\rangle \otimes |n\rangle
$
so that $\langle n|\Psi\rangle= |\Psi_n\rangle$ for $|\Psi_n\rangle$ the state for that particular choice of time, thus recovered by conditioning on the foliation. All of these states satisfy the constraint equation $\mathcal{J}_{\mu\nu}|\Psi\rangle=0$.
Interestingly, one can consider conditioning with respect to more general states $|\psi\rangle=\int dn\, \psi(n)|n\rangle$ which correspond to a quantum superposition of foliations. In this case, $\langle \psi|\Psi\rangle=\int dn\, \psi(n) |\Psi_n\rangle$ which induces a particular superposition of matter states.

On the other hand, we have seen in section \ref{sec:stcorr} that the correspondence between the extended approach and conventional QM is nontrivial, in the sense that it requires a sum (trace) over extended states (section \ref{sec:stcorr}). This trace can be purified (section \ref{sec:stgstates}) and rewritten as a generalized mean value in a duplicated Hilbert space.
We can then employ the states 
\begin{equation}\label{eq:purhist}
\begin{split}
      |\Omega_\tau\rangle \rangle&=\int dn\, |\Omega_{\tau n}\rangle \rangle \otimes   |n\rangle\\
      |\overline{\Omega}_\tau\rangle \rangle&=\int dn\, |\overline{\Omega}_{\tau n}\rangle \rangle \otimes   |n\rangle
\end{split}
\end{equation}
for $|\Omega_{\tau n}\rangle\rangle$, $|\overline{\Omega}_{\tau n}\rangle \rangle$ defined in \eqref{eq: states2} with the $n$ dependence codified in $p^0-E_\textbf{p}\to p^\mu n_\mu - E_p(n)$ and the vacuum $|\Omega_n\rangle \rangle=|\Omega_n\rangle |\tilde{\Omega}_n\rangle$  (the states in section \ref{sec:stgstates} should be written, in the notation of this section, with a  subindex $n\equiv \eta^{\mu 0}$). Notice that we use a single foliation Hilbert space for both the system and the environment. Lorentz transformations are defined as before with the complete ``angular momentum'' operator being
\begin{equation}\label{eq:totangm}
    \mathcal{J}_{\mu\nu}={\cal L}_{\mu\nu}+\tilde{\cal L}_{\mu\nu}+l_{\mu\nu}\,.
\end{equation}
  The history states \eqref{eq:purhist} satisfy the Lorentz invariance condition
    \begin{equation}\label{eq:purinv}
\mathcal{J}_{\mu\nu}|\Omega_\tau\rangle\rangle=\mathcal{J}_{\mu\nu}|\overline{\Omega}_\tau\rangle\rangle=0\,.
\end{equation}

With these definitions and notation, the representation \eqref{eq:weakv} of spacetime correlators  should be written as $\langle \mathcal{O}\rangle_n= \langle \langle \overline{\Omega}_{\tau n}| \mathcal{O}\otimes \mathbbm{1}_E|\Omega_{\tau n}\rangle \rangle/\langle \langle \overline{\Omega}_{\tau n}|\Omega_{\tau n}\rangle \rangle$. We add the subindex $n$ to indicate that the mean value corresponds to the fixed foliation $n^\mu$. In order to recover $\langle \mathcal{O}\rangle_n$ from the history states \eqref{eq:purhist} we resort to conditioning which can be written in compact form as 
\begin{equation}\label{eq:oncond}
    \langle \mathcal{O}\rangle_n= \frac{\langle \langle \overline{\Omega}_\tau|\mathcal{O}\otimes \mathbbm{1}_E\Pi_n |\Omega_\tau\rangle\rangle}{\langle \langle \overline{\Omega}_\tau|\Pi_n|\Omega_\tau\rangle\rangle}\,,
\end{equation}
with $\Pi_n:=|n\rangle \langle n|$ and $\mathcal{O}$ non necessarily a separable operator in the matter-foliation partition but commuting with $n^\mu$ (e.g. $\mathcal{O}=a^\dag(p)a(k)$). This may also be rewritten as $ \langle \mathcal{O}\rangle_n={\rm Tr} R_{\tau n} \mathcal{O}_n$ with
\begin{equation}
    R_{\tau n}:=
    \frac{\langle n|\Omega_\tau\rangle \rangle\langle \langle \overline{\Omega}_\tau| n\rangle}{
    \langle \langle \overline{\Omega}_\tau|\Pi_n|\Omega_\tau\rangle\rangle}\,,
\end{equation}
the generalized state conditioned to the foliation value $n^\mu$.
This is how one recovers conventional QM associated with fixed foliations in the complete formalism with quantum foliations. In other words, we have recovered the correspondence of the previous section \ref{sec:stcorr} between the spacetime approach and conventional QM by introducing the idea of conditioning with respect to eigenstates of $n^\mu$. 

In addition, as evidenced by Eq.\ \eqref{eq:totangm}, the foliation participates now in spacetime transformations. This is reflected in the transformation properties of mean values as well. What we find is that for relativistic theories the statement of Lorentz invariance becomes explicit:
\begin{equation}\label{eq:invobs}
\begin{split}
    \langle \mathcal{O}(\Lambda \textbf{x})\rangle_{\Lambda n} &=
 \frac{\langle \langle \overline{\Omega}_\tau|U^\dag(\Lambda)(\mathcal{O}\otimes \mathbbm{1}_E\Pi_n)  U(\Lambda) |\Omega_\tau\rangle\rangle}{\langle \langle \overline{\Omega}_\tau|U^\dag(\Lambda)\Pi_n U(\Lambda)|\Omega_\tau\rangle\rangle}\\&=
 \langle \mathcal{O}(\textbf{x})\rangle_n \,,
\end{split}
\end{equation}
  where we have considered an operator which depends explicitly on a certain number of spacetime points for concreteness (e.g. $ \mathcal{O}(\textbf{x})\equiv \mathcal{O}(x_1,x_2,\dots)=\pi(x_1)\phi(x_2)\dots\,$). Notice that the first equality is not a dynamical statement: it is just a consequence of the geometrical transformation rules \eqref{eq:qtransf}. As such it holds independently of the theory. Instead, the second equality holds only for relativistic actions satisfying \eqref{eq:actioninv} and implying \eqref{eq:purinv}.
The important result \eqref{eq:invobs} tells us that for relativistic theories the quantum expectation values are 
functions 
of both the spacetime coordinates $\textbf{x}$ and the foliation $n^\mu$ vector, combined in invariant ways. This includes functions such as momentum integrals containing terms $p^2-m^2$, as in the Feynman propagator, but in addition terms such as $p^\mu n_\mu$, $E_p(n)$ are allowed within these integrals, and in fact appear, e.g., in (regularized) momenta correlators $\langle \pi(x) \pi(y)\rangle_n$ which are now invariant quantities as well. The same holds true for any other mean value, non-necessarily localized in certain spacetime points. Thus in our approach, all \emph{physical predictions are explicitly covariant}.

The previous conclude the exposition on how the extended approach allows one to recover conventional physical predictions, while at the same time making their spacetime symmetries explicit. Let us now briefly comment on the possibility of going beyond conventional physics, by looking for genuine quantum foliation effect.
Notice that if we now replace the projector $\Pi_n$ with a statistical mixture of foliations $\Pi_n\to \rho_n=\int dn\, p(n) \Pi_n$ we obtain $\langle \langle \overline{\Omega}_\tau|\mathcal{O} \rho_n |\Omega_\tau\rangle\rangle=\langle \psi|{\rm Tr}e^{i\mathcal{S}_\tau} \mathcal{O} |\psi\rangle$ for $|\psi\rangle=\int dn\, e^{i\phi_n}\sqrt{p(n)}|n\rangle, \phi_n\in \mathbbm{R}$ which is just a statistical (``classical'') mixture of mean values. By employing $e^{i\mathcal{S}_\tau}$ in the current form (commuting with $n^\mu$) no genuine quantum effect arises from the foliation. This is in principle expected from a non-interacting matter-foliation theory.
We can however postulate that the proper generalization of the previous expression to a full quantum (pure) foliation is achieved by using other quantum projectors $\Pi_\psi= |\psi\rangle \langle \psi|$:
\begin{equation}\label{eq:qfweakv}
\begin{split}
     \langle \mathcal{O}\rangle_\psi&= \frac{ \langle \langle \overline{\Omega}_\tau|\mathcal{O}\otimes \mathbbm{1}_E \Pi_\psi|\Omega_\tau\rangle\rangle}{\langle \langle \overline{\Omega}_\tau|\Pi_\psi |\Omega_\tau\rangle\rangle}
\end{split}
\end{equation}
which corresponds to the conditioning $\langle \psi| \Omega_\tau\rangle\rangle$ with $|\psi\rangle$ an arbitrary state of the foliation (at least for $\mathcal{O}$ acting trivially in the foliation e.g. $\mathcal{O}=\phi(x)\phi(y)$; ladder operators may be assigned to the states themselves). These new mean values can be explicitly evaluated by using e.g. that $a(p,n)$ and $a(p,n')$ are related by a Bogoliubov transformation.

Notice that the individual terms $\langle \langle \overline{\Omega}_{\tau n'}| \mathcal{O}\otimes \mathbbm{1}_E|\Omega_{\tau n}\rangle \rangle$ arising from \eqref{eq:qfweakv}  cannot be written in terms of $e^{i\mathcal{S}_\tau}$ unless $n'=n$. Thus the partial trace over the environment now generalizes the action non-trivially.  In other words, we can access this quantum effect of the foliation only through the system \& environment representation of spacetime correlators (for matter non-interacting with the foliation). 
The validity of this generalization may depend on whether we attribute real physical existence to the environment or not. One can begin to address such a question by considering observables $\mathcal{O}$ 
 which do not ignore the environment as the one considered in section \ref{sec:stgstates} (i.e. use $\mathcal{O}\neq \mathcal{O}\otimes \mathbbm{1}_E$).

\section{Conclusions and outlook}\label{sec:concl}

We have shown that QM admits a spacetime symmetric Hilbert space formulation that treats all spacetime coordinates of matter fields as site indices and describes the possible foliations of spacetime through quantum states. 
We have obtained the formalism 
by quantizing an augmented 
classical phase space which keeps the time choice of the  Legendre transform as  dynamical and that 
yields an explicitly covariant version of Hamilton equations. The quantization process leads to off-shell actions and particle operators that are 
 nonseparable in the matter-foliation partition, 
 highlighting the necessity of a quantum foliation to preserve Lorentz symmetry.

The challenge of recovering conventional unitary evolution in a framework with field operators commuting for different spacetime points (even those causally connected in the conventional sense) has been raised and overcome. The crucial finding is the existence of a correspondence between the extended geometrical correlators, associated with the quantum action, and conventional propagators associated with the ground state of a given Hamiltonian and unitary evolution. Thermal propagators can also be obtained by compactifying time. Correlators at equal times (for a given foliation) correspond to conventional correlators such as the ones defining space-like entanglement, but for operators inserted at different times, the unitary evolution in the Heisenberg picture emerges. From these considerations, Feynman's rules and the classical limit (in the extended version) are also recovered. Some remarks about reinterpreting this map as a holographic-like correspondence, with the $d+1$ dimensional theory arising from  a $d+2$ dimensional theory have also been presented. In particular, the time scale $\tau$, which appears naturally when defining the map, might be identified with a holographic coordinate. These aspects, eventual relations with well-known holographic dualities, and whether the presence of a time scale $\tau$ yields some insight into the renormalization process, are left for future
investigations.

We have also shown that the previous emergence of time evolution can be understood in terms of correlations with an environment by using techniques recently introduced in the AdS/CFT (dS/CFT) context \cite{nak.21,tak.23}. From this point of view, the system $\&$ environment are globally described by a generalized pure state containing the causal information of the theory. 
This perspective also provides a direct operational meaning to time-like propagators in terms of weak values. 
This raises the natural question of whether one can regard the environment as a real unaccessible physical system whose correlations with the system induce its evolution in time. The situation resembles the PaW mechanism, according to which time evolution emerges from the entanglement \cite{b.16} between a system and ``the rest''. 
 Another similar proposal is the ``thermal time hypothesis'' which uses the thermalization of a statistical state to define ``internal'' time \cite{rov.93} (see also \cite{favalli2022peaceful}). 
While our formalism is in principle significantly different from these proposals, 
these previous ideas on the emergence of time, of current interest in the literature (see introduction), encourage one to investigate the issue of the environment further.  

The foliation-independent quantization of matter fields allows for a very simple and explicit definition of spacetime transformations. These preserve the geometrical character of Einstein's relativity as they are defined independently from the dynamics. In this sense, our proposal ``disentangles'' transformations that mix space and time from dynamical information, the latter being encoded in generalized states as described above.
In a very precise sense, 
the spacetime transformations appear again intertwined with the dynamics: the quantum action and particle operators are foliation-controlled operators. 
Moreover,  
we have seen that the Lorentz invariant eigenstates of invariant actions, such as the vacuum of the off-shell particles of the given theory, are entangled in the matter-foliation partition. 
They also have the same structure as in the PaW formalism \cite{PaW.83}, a similarity that has been used to introduce the concept of conditioning on the foliation. The conditioning specifies the observer relative to whom the dynamical description of the system is given (as opposed to the emergence of evolution in the PaW approach). 
By conditioning with respect to fixed ``classical'' states of the foliation one recovers conventional QM, in the sense of the previous correspondence. 
We have then discussed under what conditions quantum effects of the foliation could arise.

In this manuscript, we have focused on a constant foliation and Minkowski spacetime. Even if spacetime is flat one obvious generalization is to consider a non-constant foliation $n^\mu=n^\mu(x)$, e.g.  associated with Rindler coordinates. In other words, the formalism admits an obvious generalization to the case of a \emph{foliation field}. It is e.g. straightforward to see that by replacing $n^\mu \to n^\mu(x)$ in $\mathcal{S}$ for the scalar field, a version of Hamilton equations for a general curved foliation is obtained, equivalent to the Klein-Gordon equation. According to our proposal, one would also impose
$\{n^\mu(x),\kappa_\nu(y)\}=\delta^\mu_{\;\nu} \delta^{(D)}(x-y)$ classically and $[n^\mu(x),\kappa_\nu(y)]=i\delta^\mu_{\;\nu} \delta^{(D)}(x-y)$ in the quantum case so that the foliation Hilbert space would now be spanned by states $|n(x)\rangle$ representing field configurations in spacetime ($\hat{n}^\mu(x)|n(x)\rangle=n^\mu(x)|n(x)\rangle$). Moreover, one can introduce an angular momentum for this field $l_{\mu\nu}(x):=n_\nu(x) \kappa_\mu(x)-n_\mu (x) \kappa_\nu(x)$ so that one can \emph{unitarily} transform any foliation field eigenstate to another. In particular, in Minkowski spacetime any curved foliation eigenstate is unitarily related to $|\eta^{\mu 0}\rangle$, i.e., $|n(x)\rangle=\exp[i\int d^Dx\, l_{\mu\nu}(x)\Lambda^{\mu\nu}(x)]|\eta^{\mu 0}\rangle$ which is a \emph{quantum version} of the concept of \emph{momentarily comoving reference frame}. This unitary transformation is separable in spacetime reflecting our classical intuition, yet the quantum treatment of the foliation allows for many more exotic possibilities, as states of the foliation entangled across different spacetime points.

 As in the case of constant foliations, changing observers does not affect the algebra of fields in agreement with $[l_{\mu\nu}(x),\phi(x)]=[l_{\mu\nu}(x),\pi(x)]=0$. This should be contrasted with the usual treatment in QFT that requires to quantize on a given hypersurface, e.g. when considering a Rindler observer and deriving the Unruh effect \cite{fulling1973nonuniqueness,unruh1976notes}.
 A change of observer does however affect the action $\mathcal{S}$. In particular,  it is clear that a free action conditioned to a curved foliation is still a quadratic form but different from the one corresponding to an inertial observer. The normal modes then differ in general by a Bogoliubov transformation induced by the curvature of $n^\mu (x)$ which changes the vacuum state (we recall that for two constant foliations, no change in the vacuum arises). This would be the derivation of the Unruh effect from within the spacetime approach.

There is yet another interesting feature in flat spacetime to be considered, which is preliminary to extending our treatment to genuinely curved manifolds. While we have employed Minkowski coordinates to define our basic algebra of fields and quantization, it is feasible to describe e.g. the action in different curvilinear coordinates. In general, this leads to replacing conventional derivatives with covariant derivatives $\partial_\mu\to \nabla_\mu$ such that a contraction of the form $n^\mu \nabla_\mu$ corresponds to invariance under \emph{general} coordinate transformations. 
One can easily show that the Hamilton equations derived for general $n^\mu$ in other coordinate charts have precisely this form. It is in principle feasible to obtain these results by directly imposing algebras (of matter fields and foliations alike) with respect to other coordinate systems, which seems to indicate that the relation between charts might have a quantum  representation, another interesting possibility opened by working with spacetime algebras. 
Interestingly, a quantum treatment allowing for general parameterizations is also the main objective of the so-called ``parameterized field theories'' \cite{ish.85,hoehn2023matter}, an approach in which matter fields are functions of arbitrary curvilinear coordinates. These coordinates are associated with possible foliations of spacetime and are quantized as well \footnote{While this procedure is analogous to the PaW approach, as shown recently in \cite{hoehn2023matter}, therein matter fields are not quantized via an extended algebra as in our scheme. }. However, even for Minkowski spacetime, this approach suffers from problems in spacetime dimensions other than $1+1$. These difficulties might be  circumvented by developing further  our proposal since, as we mentioned above, the dynamical and geometrical information are decoupled.

Some remarks on the possibility of applying the formalism to gravity may also be appropriate (although currently speculative): beyond the mathematical and symmetry-based justifications for a dynamical foliation, we have also suggested that a dynamical spacetime could naturally 
lead to this concept. It would be interesting (and self-consistent) if this could be derived by applying the extended approach to gravity, at least working on a semi-classical level. We notice that the formalism employs actions defined in phase space variables, meaning that a Hamiltonian always needs to be introduced. This seems to lead directly to the conventional ADM approach \cite{adm.59} and its quantization. 
However, this is not the case: while the usual canonical quantization is based on unitarily evolving metrics on hypersurfaces, our formalism would treat the metric of each hypersurface as independent. This feature, together with a quantum foliation seems to lead instead to a description where the physical degree of freedom is the complete spacetime metric (with some eventual constraints). 
In such a construction, yet to be fully developed, the natural type of queries would not be of a dynamical nature but instead, intrinsically geometrical and associated with correlations, in analogy with the case of fields we developed in this manuscript.

Regarding matter fields, the main example we have employed 
is that of a scalar field. While most results and ideas hold for general field theories, some new considerations need to be made in the quantum treatment of spinorial fields related to how the coupling between the foliation and the spin affects the definition of momenta. This discussion, and the case of gauge theories, typically associated with fields with spin, is postponed. Nonetheless, several remarks are presented in  Appendix \ref{sec:apdirac}, where the classical treatment of a Dirac field is also fully developed.

Remarkably, most of the concepts we have developed for fields apply to any quantum mechanical system, including nonrelativistic ones. 
In fact, the idea of extending a conventional algebra to ``spacetime'' can always be applied. This allows one to construct the quantum action operator associated with a certain quantum Hamiltonian, a procedure that does not require a classical theory. Notably, unitary evolution is always recovered via its associated time-like correlators, as shown in \cite{diazp.21} and Appendix \ref{sec:Apmap}.  As a concrete example, we develop the spacetime extension of a qubit representation of the $\mathfrak{su}(2)$ algebra in Appendix \ref{sec:Apmap}.
A generalized purification, as the one we have employed for the Klein-Gordon free action, can also be introduced in general. While this is nontrivial (and we do not present the general case) we provide some ideas and examples in the 
same Appendix. By these means, one can replace the conventional notion of state and unitary evolution of QM with generalized states codifying not only the initial state but also the evolution and causal structure of a given theory.
These considerations pave the way to developing and applying new quantum informational and computational schemes, as we have discussed in section \ref{sec:stgstates}. At the same time, they provide a formulation capable of addressing the quantum foundational questions posed in the introduction.

	\section*{Acknowledgments}
 The authors would like to thank Marco Cerezo, Diego García-Martín and Federico Tomás Perez for fruitful discussions. N.L.D was supported by the Laboratory Directed Research and Development (LDRD) program of Los Alamos National Laboratory (LANL) under project numbers 20230049DR and 20230527ECR.
		We also acknowledge support from CONICET (N.L.D., J.M.M.) and  CIC (R.R.) of Argentina.  This work was supported  by CONICET PIP Grant No. 11220200101877CO.

\appendix

\section{Classical and quantum expressions for a general timelike $n^\mu$}\label{sec:Apa}
In this appendix we show explicitly how the classical and quantum relations involving the foliation are modified by a non normalized $n^\mu$. 

We define $||n||=\sqrt{n^\rho n_\rho}$ and assume $\sum_i||n_i||^2=-||n||^2$ (fixed speed of light) and $n_i^\mu n_\mu =0$. Then equation \eqref{eq:tensor} is replaced by 
\begin{equation}
    n^\mu n^\nu -\sum_{i=1}^d n_i^\mu n_i^\nu=||n||^2\eta^{\mu \nu}\,.
\end{equation}
Equations such as (\ref{eq:deriv}) need to be rescaled also, e.g.  $\partial_\rho \phi= ||n||^{-2}(n_\rho n^\mu\partial_\mu \phi- n_{i \rho}n^\mu_{i}\partial_\mu \phi)$.  For the Klein Gordon free case one obtains the classical equations
\begin{equation}
    n^\mu \partial_\mu \phi=||n||^2\pi
\end{equation}
and 
\begin{equation}
    \mathcal{H}=||n^2||\frac{1}{2}\pi^2+\frac{1}{2}\Big(\frac{n^\mu n^\nu}{||n||^2}-\eta^{\mu\nu}\Big)\partial_\mu\phi \partial_\nu \phi +\frac{1}{2}m^2\phi^2\,.
\end{equation}

The Hamilton equations are 
\begin{subequations}
    \begin{align}
        n^\mu \partial_\mu \pi &=\Big(\frac{n^\mu n^\nu}{||n||^2}-\eta^{\mu\nu}\Big)\partial_\mu \partial_\nu \phi -m^2 \phi\\
        n^\mu \partial_\mu \phi&= ||n||^2\pi
    \end{align}
\end{subequations}
which yield one again the Klein Gordon equation
\begin{equation}
(\eta^{\mu\nu}\partial_\mu\partial_\nu+m^2)\phi=0\,.
\end{equation}
An analogous expression is found with the addition of a potential.
 These can be recovered as before from $\mathcal{S}=\int d^Dx\, (\pi n^\mu \partial_\mu \phi-\mathcal{H})$ and setting $\{\phi,\mathcal{S}\}=\{\pi, \mathcal{S}\}\equiv 0$.

After quantization, the diagonalization of $\int d^Dx\, \mathcal{H}$ 
now leads to 
\begin{equation}\label{eq:epngen}
\begin{split}
      E_p(n)&=||n||\sqrt{\left(\frac{n^\mu n^\nu}{||n||^2}-\eta^{\mu\nu}\right)p_\mu p_\nu +m^2}\\
    &=||n||\sqrt{\sum_i\left(\frac{n^\mu_ip_\mu}{||n||}\right)^2+m^2}\,.
\end{split}
\end{equation} 
All equations of section \ref{sec:expart} hold with $E_p(n)$ given by (\ref{eq:epngen}) and $\pi n^\mu \partial_\mu \phi\to ||n^2||^{-1}\pi n^\mu \partial_\mu$, in agreement with the new definition of $\pi$, equivalent to a rescaling of $E_p$. In the normal expansion of the action, one should also replace $p^\mu n_\mu \to ||n^2||^{-1}p^\mu n_\mu $ accordingly. 
One can then show that physical quantities such as the Feynman propagator remain invariant, namely independent of $||n^2||$. 

 Notice that for $m=0$ and $D=1+1$ one has a conformal field theory (CFT) and any choice of $n^\mu$ can be made to diagonalize $\mathcal{S}$ through normal modes.

\section{The Dirac field case}\label{sec:apdirac}
In this Appendix we consider the application of the formalism of the main body to the case of a free Dirac Lagrangian density $\mathcal{L}_D=\bar{\psi}(i\gamma^\mu \partial_\mu -m)\psi$. While interactions may be introduced along the line developed in the free scalar case, we focus on this simple example as the aim is to show 
how the spin is treated in a framework with a dynamical foliation. Our main focus is the classical case, with some remarks on the quantization at the end of the section.

The generalized momentum for a general time-like $n^\mu$ is defined as before, yielding
\begin{equation}\label{eq:diracmomentum}
    \pi=\frac{\partial \mathcal{L}_D}{\partial (n^\mu \partial_\mu \psi)}=i \bar{\psi}\gamma^\mu n_\mu\,.
\end{equation}
For $n^\mu=\eta^{\mu0}$ we recover the usual relation $\pi=i\psi^\dag$, since $\bar{\psi}=\psi^\dag \gamma^0$ and $(\gamma^0)^2=1$ as it follows from the Clifford algebra of the gamma matrices 
$
\{\gamma^\mu,\gamma^\nu\}=2\eta^{\mu\nu}\,,
$
with the brackets indicating (only here) anticommutators.

On the other hand, by noting that $\gamma^\mu n_\mu \gamma^\nu n_\nu=\gamma'^0 \gamma'^0=1$, where we have defined $\gamma'^{0}:=\gamma^\mu n_\mu$ as the first  matrix from a new possible set of gamma matrices (satisfying the Clifford algebra as well), one can invert the momentum relation and write 
$\bar{\psi}=-i \pi \gamma^\mu n_\mu$. Then, taking the covariant Legendre transform ${\mathcal{H}_D=\pi n^\mu \partial_\mu \psi-\mathcal{L}_D}$ yields 
\begin{equation}
        \mathcal{H}_D=\pi [(n^\mu-\gamma^\nu \gamma^\mu n_\nu)\partial_\mu -im \gamma^\nu n_\nu]\psi\,.
\end{equation}
We can show that in this form $\mathcal{H}_D$ only depends on spatial derivatives: 
\begin{equation*}
    (n^\mu-\gamma^\nu \gamma^\mu n_\nu)\partial_\mu=\gamma^\rho \gamma^\mu n_\rho (n_\mu n_\nu -\eta_{\mu\nu})\partial^\nu\,,
\end{equation*}
 where we recall that $n_\mu n_\nu -\eta_{\mu\nu}$ projects onto the spatial hypersurfaces orthogonal to $n^\mu$.
Interestingly, Lorentz invariance is explicit in this form (see also below), while the conventional Dirac Hamiltonian density does not exhibit the symmetry explicitly.
The latter is recovered from $n^\mu=\eta^{\mu0}$ which implies $\gamma^\nu n_\nu=\gamma^0$ so that the first term becomes $(n^\mu-\gamma^0\gamma^\mu)i\partial_\mu\equiv -i \bm{\alpha}\cdot \bf{\nabla}$ (with $\alpha^i=\beta \gamma^i,\beta=\gamma^0$).

The Hamilton equation for $\psi$ yields
\begin{equation}
 n^\mu \partial_\mu \psi-   \frac{\partial \mathcal{H}}{\partial \pi}=-i\gamma^\nu n_\nu (\gamma^\mu i\partial_\mu-m)\psi=0
\end{equation}
which automatically implies the Dirac equation in its covariant form. Moreover, if we now introduce the classical spacetime algebra \eqref{eq:extpb}, i.e.,
\begin{equation}\label{eq:algd}
    \{\psi(x), \pi(y)\}=\delta^{(d+1)}(x-y)\,,
\end{equation}
we recover the previous from
\begin{equation}
\{\psi,\mathcal{S}_D\}=-i\gamma^\nu n_\nu (\gamma^\mu i\partial_\mu-m)\psi\,,
\end{equation}
after setting $\{\psi,\mathcal{S}_D\}\equiv 0$.
Here the Dirac action in spacetime phase space variables is
\begin{equation}
    \mathcal{S}_D=-\int d^{d+1}x\, i\pi \gamma^\nu n_\nu (\gamma^\mu i\partial_\mu -m)\psi\,,
\end{equation}
as immediately obtained by replacing $\bar{\psi}$ with the momentum using the inverse relation of \eqref{eq:diracmomentum} as before.

Let us now discuss more about the transformation properties of the fields. 
We assume the usual transformation rule $\psi(x)\to S_\Lambda \psi(\Lambda^{\shortminus 1} x)$ where we have introduced the matrix 
\begin{equation}
   S_\Lambda:=\exp(-i\omega_{\mu\nu}\sigma^{\mu \nu}/4)\,, 
\end{equation}
for $\sigma^{\mu\nu}=[\gamma^\mu,\gamma^
\nu]$. This implies as usual $S^{\shortminus 1}_\Lambda \gamma^\mu S_\Lambda=\Lambda^\mu_{\;\;\nu}\gamma^\nu$. 
We also impose, in agreement with the transformation of $x^\mu$,  $n^\mu\to (\Lambda^{\shortminus 1})^{\mu}_{\;\nu} n^\nu$.
If we now combine these rules with the definition of momentum in Eq.\ \eqref{eq:diracmomentum} we obtain
\begin{equation*}
    \begin{split}
          \pi(x)&\to i\bar{\psi}(\Lambda^{\shortminus 1}x)S^{\shortminus 1}_\Lambda\gamma^\mu \Lambda_\mu^{\;\nu}n_\nu=i\bar{\psi}(\Lambda^{\shortminus 1}x)\gamma^\nu n_\nu S^{\shortminus 1}_{\Lambda} \\
      &=\pi (\Lambda^{\shortminus 1}x)S^{\shortminus 1}_\Lambda\,.
    \end{split}
\end{equation*}
This allows us to summarize the transformation properties as 
\begin{subequations}\label{eq:transfdirac}
    \begin{align}
        \psi(x)&\to S_\Lambda\psi(\Lambda^{-1} x)\\
        n^\mu &\to (\Lambda^{-1})^{\mu}_{\;\;\nu}n^\nu\\
        \pi(x)&\to \pi(\Lambda^{-1} x)S_\Lambda^{-1}\,.
    \end{align}
\end{subequations}
Notice that as a consequence, the algebra \eqref{eq:algd} is explicitly preserved by a Lorentz transformation, both in the spacetime components and in the spinorial ones. Similarly, quantities such as $\mathcal{S}_D$ and $\int d^{d+1}x\, \mathcal{H}$ are in fact explicitly invariant: the transformations of $\psi,\pi$ imply $\gamma^\mu\to S^{-1}_\Lambda\gamma^\mu S_\Lambda$ for all gamma matrices, and they always appear contracted to $n_\mu$ or $\partial_\mu$. 

Equations \eqref{eq:transfdirac} are the spinorial generalization of the main body Eqs.\ \eqref{eq:transf} (the slight difference in convention regarding the coordinates is common for spinorial fields). In order to recover these transformations from the extended phase space algebra, we consider the total angular momentum 
\begin{equation}
\mathcal{J}^{\mu\nu}=\mathcal{L}^{\mu\nu}+\mathcal{S}^{\mu\nu}+l^{\mu\nu}
\end{equation}
with
\begin{align}
    \mathcal{L}^{\mu\nu}&:=-\int d^{d+1}x\,\pi(x^\mu\partial^\nu-x^\nu \partial^\mu)\psi\\
     \mathcal{S}^{\mu\nu}&:=\frac{i}{2}\int d^{d+1}x\,\pi\sigma^{\mu\nu}\psi\\
     l^{\mu\nu}&:=n^\mu \kappa^\nu-n^\nu \kappa^\mu\,,
\end{align}
The only novelty with respect to the scalar field case is the spinorial part $\mathcal{S}^{\mu\nu}$, as expected.
Then Eqs.\  \eqref{eq:transfdirac} can be obtained from the action of the exponential of the total angular momentum on the fields. Let us show this explicitly up to first order ($\Lambda=1+\omega+\mathcal{O}(\omega^2)$) for the spinorial part:
\begin{equation*}
\begin{split}
    \psi+\frac{\omega_{\mu\nu}}{2}\{\mathcal{S}^{\mu\nu},\psi\}&=\big(1-\tfrac{i}{4}\sigma^{\mu\nu}\omega_{\mu\nu}\big)\psi= S_\Lambda \psi\,+\mathcal{O}(\omega^2)\\
    \pi+\frac{\omega_{\mu\nu}}{2}\{\mathcal{S}^{\mu\nu},\pi\}&=\pi\big(1+\tfrac{i}{4}\sigma^{\mu\nu}\omega_{\mu\nu}\big)= \pi S^{-1}_\Lambda\!+\mathcal{O}(\omega^2)\,.
\end{split}
\end{equation*}
In addition, one can easily prove that $\{\mathcal{L}^{\mu\nu},\mathcal{S}^{\alpha \beta}\}=0$ since $\sigma^{\alpha\beta}$ does not depend on field coordinates while $(x^\mu\partial^\nu-x^\nu \partial^\mu)$ is independent of spinorial components. We also have $ \psi(x)+\frac{\omega_{\mu\nu}}{2}\{\mathcal{L}^{\mu\nu},\psi(x)\}=\psi(\Lambda^{-1}x)+\mathcal{O}(\omega^2)$ so that complete series of nested brackets yield

\begin{equation}
\begin{split}   
\psi(x)+\tfrac{\omega_{\mu\nu}}{2}\{\mathcal{J}^{\mu\nu},\psi(x)\}+\dots
    &=S_\Lambda \psi(\Lambda^{-1}x)\\
       n^\rho+\tfrac{\omega_{\mu\nu}}{2}\{\mathcal{J}^{\mu\nu},n^\alpha\}+\dots &=(\Lambda^{-1})_{\;\alpha}^{\rho} n^\alpha\\
       \pi(x)+\tfrac{\omega_{\mu\nu}}{2}\{\mathcal{J}^{\mu\nu},\pi(x)\}+\dots&= \pi(\Lambda^{-1}x)S_\Lambda^{-1}\,,
\end{split}
\end{equation}
where the dots indicate higher-order nested brackets, e.g., the next order being $\tfrac{1}{2!}\tfrac{\omega_{\mu\nu}\omega_{\alpha\beta}}{4}\{\mathcal{J}^{\mu\nu},\{\mathcal{J}^{\alpha\beta},...\}\}$.
These are precisely the transformations in Eqs.\ \eqref{eq:transfdirac}.

Let us now make a few comments regarding the quantization of Dirac's field according to the extended scheme. The first natural difference with the scalar field case is that anticommutation rules should be imposed, namely, now one promotes the spacetime algebra \eqref{eq:algd} to a spacetime anticommutator. This guarantees the positivity of the energy (as in the usual case). 
This also means that the map we have established in section \ref{sec:stcorr}, and that we discussed for more general quantum systems in the Appendix \ref{sec:Apmap},  needs to be modified: since fields at different times do not commute, no underlying product structure is present. Instead, one can construct a correspondence to conventional QM via Wick's theorem in analogy with the approach in section \ref{sec:stcorr} by replacing \eqref{eq:contr} with its fermionic version, with some additional proper modifications related to fermionic parity. 

Additional subtleties arise in the quantum case related to the fact that $-i\pi \neq \psi^\dag$ for general $n^\mu$, namely the anticommutation relation is not between the field and its conjugate. One can show that this is in perfect agreement with explicit Lorentz covariance and again a reason for introducing $n^\mu$: e.g., notice that the algebra $\{\psi(x),i\psi^\dag(y)\}=\delta^{(d+1)}(x-y)$ is not invariant since $S_\Lambda$ is not a unitary matrix, but \eqref{eq:algd} is. In fact, Lorentz transformations are unitary with respect to a proper definition of the inner product, induced by the previous. This is also related to the covariant inner product introduced in \cite{di.19} for Dirac particles. Finally, a generalized purification scheme may also be introduced for fermions. While none of these features pose a particular challenge, 
 their detailed expositions warrant a separate discussion, to be addressed in future work.

\section{Generalized purification for general free bosons}\label{sec:Appurif}
In this appendix, we discuss the generalized purification introduced in \eqref{eq:ptrace}.
We consider the case of ``generalized density'' operators of the form
\begin{equation}
    \rho=\frac{e^{-H}}{{\rm Tr}\,e^{-H}},\;\;\;\;\;H=\sum_k H_k=\sum_k \lambda_k a^\dag_k a_k
\end{equation}
where we let $\lambda_k$ be a general complex number with $\text{Re}(\lambda_k)>0$ (for $\lambda_k\in \mathbbm{R}$,  $\rho$ is a thermal state with quadratic diagonal Hamiltonian).

These operators can of course be written as
\begin{equation}
    \rho=\underset{k}{\otimes} \,Z_k^{-1}e^{-\lambda_k a_k^\dag a_k}=\underset{k}{\otimes} \,Z_k^{-1}\sum_{n_k}e^{-\lambda_k n_k}|n_k\rangle_k\langle n_k|
\end{equation}
so that we only need to purify each $\rho_k=e^{-H_k}$ (with $\rho=\otimes_k \rho_k$)
and take the tensor product at the end. Here we have also defined the ``partition functions'' $Z_k:={\rm Tr}e^{-H_k}$ such that $Z:={\rm Tr}\, e^{-H}=\prod_k Z_k$. Notice that the free Klein-Gordon action $e^{i\mathcal{S}_\tau}$ has precisely this form with the index $k$ corresponding to the $D$ dimensional momentum.

We now introduce for each $k$ the two distinct states (for ease of notation we omit the $k$ indices in the states)
\begin{align}
    |0_{\lambda_k}\rangle\rangle&= \sum_n e^{-\frac{\lambda_k n}{2} }|n\rangle|n\rangle=\exp\Big(e^{-\frac{\lambda_k}{2}} \,a_k^\dag \tilde{a}_k^\dag\Big)|0\rangle \rangle\nonumber\\
|\overline{0}_{\lambda_k}\rangle\rangle&= \sum_n e^{-\frac{\lambda^\ast_k}{2} n }|n\rangle|n\rangle=\exp\Big(e^{-\frac{\lambda^\ast_k}{2}}\, a_k^\dag \tilde{a}_k^\dag\Big)|0\rangle \rangle\,,
\end{align}
with $|0\rangle\rangle:=|0\rangle\otimes |\tilde{0}\rangle$.
We may refer to the states $|\tilde{n}\rangle$ as environment states with $|0_{\lambda_k}\rangle \rangle$, $|\overline{0}_{\lambda_k}\rangle\rangle$ vectors of a doubled Hilbert space spanned by $|n\rangle |\tilde{n}\rangle$, just as in a standard bosonic thermal purification.
Notice that $|\overline{0}_{\lambda_k}\rangle\rangle$ corresponds to the replacement $\lambda_k\to \lambda_k^\ast$ in $|0_{\lambda_k}\rangle \rangle$ so that 
\begin{equation}
    |0_{\lambda_k}\rangle\rangle_k \langle\langle \overline{0}_{\lambda_k}|=\sum_{n,n'}e^{-\lambda_k \frac{n+n'}{2}}|n\rangle \langle n'|\otimes |\tilde{n}\rangle \langle \tilde{n}'|\,.
\end{equation}
Then, the partial trace over the environment yields
\begin{equation}
    e^{-H_k}={\rm Tr}_E |0_{\lambda_k}\rangle\rangle \langle\langle \overline{0}_{\lambda_k}|
\end{equation}
where we used $\langle \tilde{n}|\tilde{n}'\rangle=\delta_{nn'}$. This also implies
\begin{equation}
    Z_k={}_k\langle\langle \overline{0}_{\lambda_k}|0_{\lambda_k}\rangle\rangle\,.
\end{equation}

  To obtain the complete $\rho$ we take the product of the previous states and define
\begin{align}\label{eq:appurif}
    |0_\lambda\rangle\rangle&= \underset{k}{\otimes} \,|0_{\lambda_k}\rangle\rangle_k=\exp\Big(\sum_k e^{-\frac{\lambda_k}{2}}\, a_k^\dag \tilde{a}_k^\dag\Big)|0\rangle \rangle\nonumber\\ |\overline{0}_{\lambda}\rangle\rangle&= \underset{k}{\otimes} \,|\overline{0}_{\lambda_k}\rangle\rangle= \exp\Big(\sum_ke^{-\frac{\lambda^\ast_k}{2}}\, a_k^\dag \tilde{a}_k^\dag\Big)|0\rangle \rangle\,,
\end{align}
for $|0\rangle \rangle= |0\rangle \otimes |\tilde{0}\rangle$ now the complete vacua ($|0\rangle=\otimes_k |0\rangle_k$, $|\tilde{0}\rangle=\otimes_k |\tilde{0}\rangle_k$). It is now clear that
\begin{equation}
    e^{-H}={\rm Tr}_E |0_{\lambda}\rangle\rangle \langle\langle \overline{0}_{\lambda}|
\end{equation}
and 
\begin{equation}
    Z=\langle\langle \overline{0}_\lambda|0_\lambda\rangle\rangle\,.
\end{equation}
It is worth noting that for $\lambda\in \mathbb{R}$, $|\overline{\Psi}\rangle\rangle=|\Psi\rangle\rangle$ and all the previous expressions reduce to the ones in conventional purification. 

Notice also that  $|0_\lambda\rangle\rangle$, $|\overline{0}_\lambda\rangle\rangle$ are Bogoliubov vacua of the annihilation operators
\begin{equation}
\begin{split}
a'_k&:=u(\lambda_k)a_k+v(\lambda_k)\tilde{a}^\dag_k\\
\tilde{a}'_k&:=u(\lambda_k)\tilde{a}_k+v(\lambda_k) a^\dag_k\\
\overline{a}'_k&:=u(\lambda^\ast_k)a_k+v(\lambda^\ast_k)\tilde{a}^\dag_k\\
\tilde{\overline{a}}'_k&:=u(\lambda^\ast_k)\tilde{a}_k+v(\lambda^\ast_k)a^\dag_k
\end{split} 
\end{equation}
respectively, for
\begin{equation}
    u(\lambda_k)=\frac{1}{\sqrt{1-e^{-{\rm Re}(\lambda_k)}}}\,\;\;\;v(\lambda_k)=-\frac{e^{-\lambda_k/2}}{\sqrt{1-e^{-{\rm Re}(\lambda_k)}}}
\end{equation}
satisfying $|u(\lambda_k)|^2-|v(\lambda_k)|^2=1$ (and hence $[a'_k,a'^\dag_{l}]=[\tilde{a}'_k,\tilde{a}'^\dag_{l}]=[\bar{a}'_k,\bar{a}'^\dag_{l}]=[\tilde{\bar{a}}'_k,\tilde{\bar{a}}'^\dag_{l}]=\delta_{kl}$ with the other commutators vanishing). This can be easily proven by explicitly showing that 
\begin{equation}
a'_k|0_\lambda\rangle\rangle=\tilde{a}'_k|0_\lambda\rangle\rangle=\bar{a}'_k|\overline{0}_\lambda\rangle\rangle=\tilde{\bar{a}}'_k|\overline{0}_\lambda\rangle\rangle=0\,.
\end{equation}
One may then apply the formalism developed in \cite{bal.69} to express the generalized mean values (such as Eq.\ (\ref{eq:weakv})) as a vacuum expectation value in biorthogonal bases.

In order to obtain equations \eqref{eq: states2} one can take the continuum limit of \eqref{eq:appurif} directly within the sums. This step can be further justified by considering first a finite spacetime volume which renders the momentum indices  $p$ discrete, with the algebra \eqref{eq:algpart} recovered as the large volume limit (see also the conventional thermofield dynamics approach \cite{kh.09}). Notice that these results apply directly to a finite $T$ and $\epsilon$ which allows the recovery of thermal correlators from the same purification scheme (see \ref{sec:stcorr}).

\section{Correspondence with conventional QM for discrete spacetime and general quantum systems}\label{sec:Apmap}
In this Appendix we discuss how the main text correspondence to conventional QM works for discrete time and for general systems and theories. The notion of spacetime generalized state, arising from the purification of the map, can be applied as well, as we show in a simple qubit system.

{\it Discrete formalism}. Let us first write a discrete spacetime version of the extended algebra \eqref{eq:extalg}:
\begin{equation}\label{eq:discextalg}
[\phi_{im},\pi_{jn}]=i\delta_{ij}\delta_{mn}\,,
\end{equation}
where $i,j$ represent time sites, and $m,n$ spatial sites. The corresponding conventional canonical algebra is  $[\phi_{m},\pi_{n}]=i\delta_{mn}$ at equal times. Notice that in the  standard approach field operators at different points in space commute meaning that the total Hilbert space has the product structure $\mathcal{H}=\otimes_m h_{m}$ with $h$ the Hilbert space of a single bosonic mode. When we  extended the algebra as in \eqref{eq:discextalg} this is generalized to time, with the new Hilbert space structure being
\begin{equation}
    \mathcal{H}=\otimes_i \mathcal{H}_i=\otimes_{i,m}h_{im}\,.
\end{equation}
 We see that a tensor product structure is applied to both space and time. In fact, there is nothing which  distinguishes time and space in \eqref{eq:discextalg}, we only fixed a convention in order to introduce dynamics below.  

\begin{figure}[t!]
    \centering
    \includegraphics[width=\columnwidth]{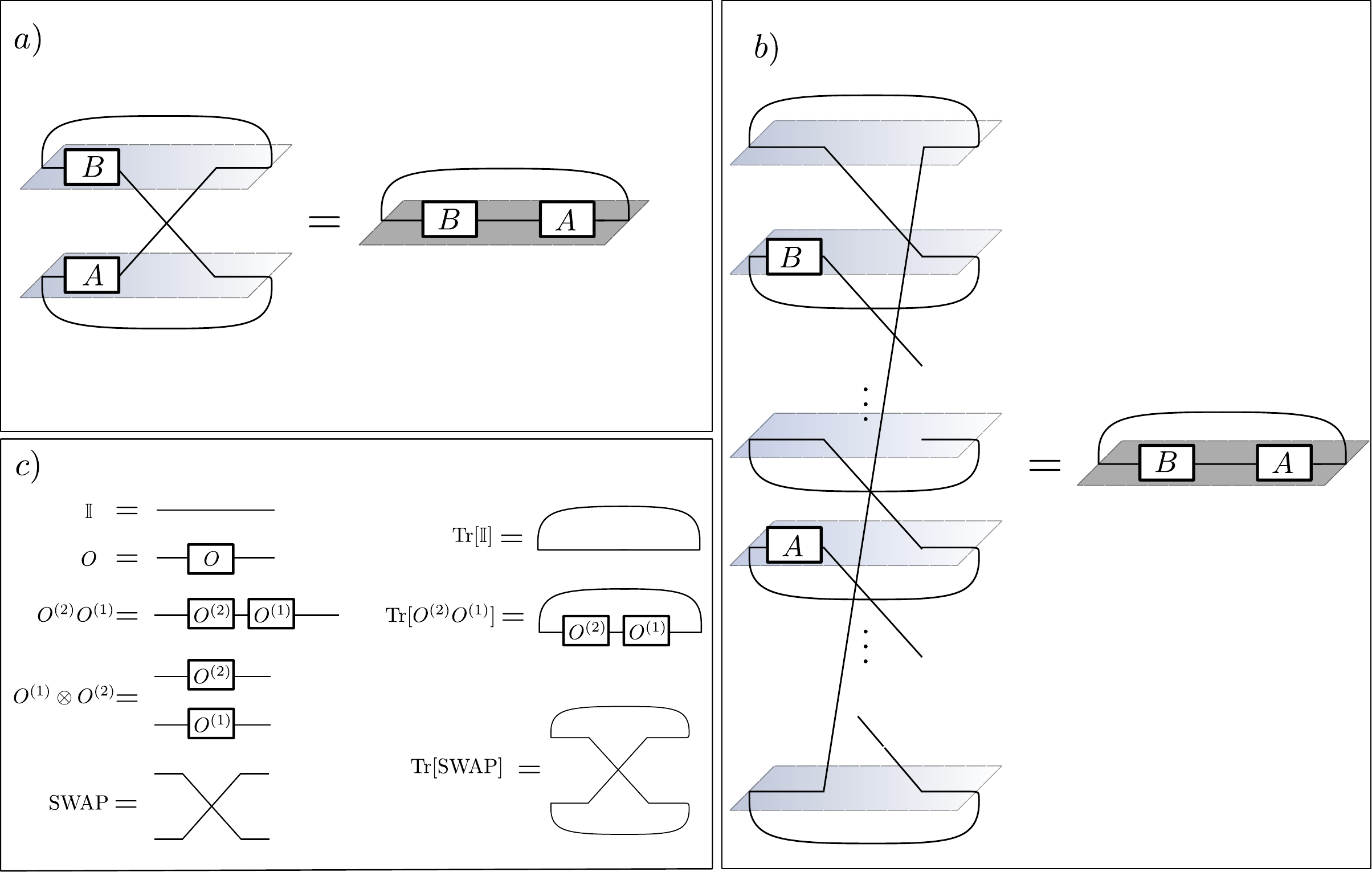}
    \caption{\textbf{Tensor network representation of the correspondence. } The operator $e^{i\epsilon\mathcal{P}_0}$ allows one to translate traces in $\mathcal{H}=\otimes_i h_i$ to traces in $h_i$, as easily seen in tensor network notation. The notation is introduced in c) while the planes in a) and b) have been added to emphasize that a Hilbert space is assigned to each time slice. For QFTs  each plane represents the Hilbert space of fields quantized in a given hypersurface. In a) we represent Eq.\ \eqref{eq:trab}. In b) we show the representation of the same trace with a larger number of time-slices. }
    \label{fig:tn}
\end{figure}

 Such a product structure in time can be defined for any quantum system (the fermionic case is more subtle as discussed in Appendix \ref{sec:apdirac}): one considers a Hilbert space $h$ and then constructs an extended Hilbert space $\mathcal{H}=\otimes_i h_i$ for a given number of times. If the Hilbert space $h$ has a basis of states $|n\rangle$, then 
 \begin{equation}
     \mathcal{H}=\text{span}\{|n_1n_2\dots n_N\rangle\}\,,
 \end{equation}
i.e., it has a ``quantum trajectory''-like basis, with $N$ the number of time-slices. Let us now define an extended operator $e^{i\epsilon\mathcal{P}_0}$ such that
\begin{equation}\label{eq:eipket}
e^{i\epsilon\mathcal{P}_0}|n_1n_2\dots n_N\rangle:=|n_N n_1 n_2\dots\rangle\,.
\end{equation}
It can be easily shown \cite{diazp.21} that
\begin{equation}\label{eq:disccorres}
    {\rm Tr}[e^{i\epsilon\mathcal{P}_0}\otimes_i O^{(i)}_i]={\rm Tr}[\hat{T}\,\Pi_i O^{(i)}]\,,
\end{equation}
where the first trace is taken in the extended Hilbert space while the second in the conventional one. The time ordering operator $\hat{T}$ indicates that the product of operators on the right should follow the time ordering (from larger to smaller) on the left. This is the essence behind the correspondence: the operator $e^{i\epsilon\mathcal{P}_0}$ is translating traces of tensor products of operators to traces of conventional composition of those same operators. For concreteness let us show this for $N=2$:
\begin{equation}
\begin{split}\label{eq:trab}
    {\rm Tr}[e^{i\epsilon\mathcal{P}_0}A\otimes B]&=\sum_{n_1,n_2}\langle n_2 n_1|A\otimes B |n_1 n_2\rangle\\
    &=\sum_{n_1,n_2}\langle n_1|B|n_2\rangle \langle n_2|A|n_1\rangle\\
    &=\sum_n \langle n|BA |n\rangle= {\rm Tr}[BA]\,,
\end{split} 
\end{equation}
which is Eq.\ \eqref{eq:disccorres} for $A=O^{(1)}$, $B=O^{(2)}$. Notice that for $N=2$, $e^{i\mathcal{P}_0}$ is just the SWAP operator, and the previous is essentially a SWAP test \cite{buhrman2001quantum}. Notice also that this same correlator can be represented in an extended Hilbert space with an arbitrary number of times, as depicted in Figure \ref{fig:tn} in tensor network's notation.

The next step is to relate the previous kinematic construction to actual correlators and to the action operator. In this scenario the proper definition of the action for a time step $\epsilon$  is
\begin{equation}
e^{i\mathcal{S}}:=e^{i\epsilon\mathcal{P}_0}\otimes _ie^{-i\epsilon H_i}=e^{i\epsilon(\mathcal{P}_0-\sum_i H_i)}\,.
\end{equation}
It was in fact proven in \cite{diazp.21} that
\begin{equation}\label{eq:evmap}
    {\rm Tr}\big[|\psi\rangle_0 \langle \psi| \,e^{i\mathcal{S}}\otimes_i O^{(i)}_i\big]=\langle \psi,T|\hat{T}\,\Pi_i O^{(i)}_H(t_i)|\psi\rangle
\end{equation}
with $t_i=\epsilon i$, $T=\epsilon N$ and  $|\psi,t\rangle:=e^{iHt}|\psi\rangle$.
In other words, replacing $\mathcal{P}_0$ with the action $\mathcal{S}$ in \eqref{eq:disccorres} corresponds to adding evolution. The amount of evolution of each operator is determined by the Hilbert space in which they act. In addition, to specify the initial state we insert it in the initial time slice ($|\psi\rangle_0 \langle \psi|\equiv |\psi\rangle_0 \langle \psi|\otimes \mathbbm{1}$).

Let us also notice that it is straightforward to extend \eqref{eq:evmap} to time-dependent Hamiltonians by defining the action as $e^{i\mathcal{S}}=e^{i\epsilon\mathcal{P}_0}\otimes_i U_i[(i+1)\epsilon,i\epsilon]$ with $U_i(t',t)$ the time evolution operator, evolving from $t$ to $t'$, acting on the slice $i$.
An explicit time dependence might be also added to the operators $O^{(i)}$. The time ordering is then preserved as long as the position in time and the external time dependence are consistent (e.g. if one is considering an operator $O=j(t)\phi_m$, with $j(t)$ a function, one should insert $O_i=j(t)\phi_{im}$ for $i\epsilon=t$).

If one is only interested in the ground state, one can omit the initial state by adding an imaginary part to time (all the previous holds for a non-hermitian $H$ \cite{diaz.21}) and considering the large time limit, just as it is usually done in the PI formulation. Our main body example might be reinterpreted in this way. One can also consider  thermal correlation functions by making the replacement (Wick rotation) $H\to -iH$ in the definition of the operator  $\mathcal{S}$. A simple relabeling of the previous equations gives
$ {\rm Tr}\big[e^{i\mathcal{S}}\otimes_i O^{(i)}_i\big]={\rm Tr} \big[e^{-\beta H} \hat{T}_\theta\,\Pi_i O^{(i)}_H(\theta_i)\big]$ where $O_H(\theta)\equiv e^{H \theta} O e^{-H\theta}$ indicating ``thermal evolution'', namely the operator ``evolved'' up to inverse temperature $\theta$ and $\theta_i=i\epsilon$.
This also implies  Eq.\ \eqref{eq:ptrace} in the main body and ${\rm Tr} \,[e^{i\mathcal{S}}]={\rm Tr}\,[e^{-\beta H}]$.

{\it Spacetime treatment of qubit systems.}  Let us now show in a very simple example how the generalized purification applies beside the bosonic field case.
We first introduce a conventional situation of two qubits separated in space at a given time for comparison. In this scenario we describe the associated state of the system through a density matrix which can be written as \begin{equation}\label{eq:rho}
\rho=\sum_{i,j=0}^3 \langle P_i\otimes P_j\rangle_\rho \, P_i\otimes P_j
\end{equation} 
with 
$
    \langle P_i\otimes P_j\rangle_\rho ={\rm Tr} [\rho P_i\otimes P_j]
$
and $P_i$ Pauli matrices for $P_0=\mathbbm{1}$. This means that the state completely defines the correlators at a given time and viceversa. If the state is not pure we can always consider a purification and rewrite the previous as a pure state mean value. For example, for $\rho=p |00\rangle\langle 00| +(1-p)|11\rangle\langle 11|$ we can write 
\begin{equation}
    \langle P_i\otimes P_j\rangle_\rho =\langle \langle \Psi| P_i\otimes P_j\otimes \mathbbm{1}_E|\Psi\rangle\rangle
\end{equation}
where the global state involving the ``environment'' can be choosen as $|\Psi\rangle\rangle:=\sqrt{p}|000\rangle+\sqrt{1-p}|111\rangle$ such that
\begin{equation}\label{eq:purif}
    \rho={\rm Tr}_E |\Psi\rangle\rangle\langle \langle \Psi||\,.
\end{equation}

Now let us consider the case of a single qubit and two times in the new approach. The new formalism describes the situation through a Hilbert space which is isomorphic to the one of the previous example involving two qubits. 
In fact, we conventionally describe the Hilbert space of a qubit as the smallest (irreducible) representation of the algebra $[P_i,P_j]=2i\epsilon_{ijk}P_k$, where $i,j,k=1,2,3$ and $\epsilon_{ijk}$ is the Levi-Civita symbol. Our formalism imposes then,
\begin{equation}\label{eq:algpauli}
[P_{ti},P_{t'j}]=\delta_{tt'}2i\epsilon_{ijk}P_{tk}\,,
\end{equation}
with the prescription of employing the conventional Hilbert space representation for each fixed time slice.

On the other hand,
according to the previous discussion,
the operator of interest, namely the one yielding the correlators in spacetime (see Eq.\ \eqref{eq:evmap}) is not $\rho$ but 
\begin{equation}
    \bar{\rho}:=|\psi\rangle\langle \psi|\otimes \mathbbm{1}e^{i\mathcal{S}}= \sum_{i=0}^1|\psi\rangle  \langle i,\epsilon| \otimes  | i\rangle\langle  \psi,\epsilon|\,,
\end{equation}
 where we assumed an initial pure state $|\psi\rangle$ for simplicity and used that for two times $e^{i\mathcal{S}}=e^{i \mathcal{P}_0}e^{-i\epsilon H}\otimes e^{-i\epsilon H}$. We also recall that $|\psi,\epsilon\rangle\equiv e^{i\epsilon H }|\psi\rangle$. 
As any other operator, $\bar{\rho}$ can be written in terms of the correlators as 
 \begin{equation}\label{eq:rhobar}
     \bar{\rho}=\sum_{i,j=0}^3 \langle P_i\otimes P_j\rangle_{\bar{\rho}}\, P_i\otimes P_j\,,
 \end{equation} 
 where we know by construction that the mean values satisfy $\langle P_i\otimes P_j\rangle={\rm Tr} [\bar{\rho} P_i\otimes P_j]=\langle \psi,T|P_j(\epsilon)P_i |\psi\rangle$, with $T=2\epsilon$. We can verify this explicitly: 
\begin{align*}
    \langle P_i\otimes P_j\rangle_{\bar{\rho}}&=\!\sum_i \langle \psi,\epsilon|P_j |i\rangle \langle i,\epsilon |P_j|\psi \rangle\\&=\sum_i \langle \psi,2\epsilon|P_j(\epsilon)|i,\epsilon\rangle \langle i,\epsilon |P_j|\psi \rangle\\ &=\langle \psi,T|P_j(\epsilon)P_i |\psi\rangle\,,
\end{align*}
 where we rearranged the terms in the first equality and used the completeness relation in the last. Notice how the operators appear according to the temporal order on the left hand side, in agreement with (\ref{eq:evmap}).
The previous implies that $\bar{\rho}$ is the unique operator in this Hilbert space whose correlators in time are the conventional propagators. 
Notice that correlators in time are now treated exactly as in our previous spatial example. The difference between the two situations is codified in the different characteristics of $\rho$, $\bar{\rho}$, with the first being a state but not the second.

Remarkably, we can interpret $\bar{\rho}$ as arising from a generalized state: without loss of generality let us write $|\psi\rangle=|0\rangle$. 
We now introduce a couple of states correlated with a qubit environment:
\begin{align}
|\Psi\rangle\rangle&:=\frac{|000\rangle+|011\rangle}{\sqrt{2}}\nonumber\\
|\Phi\rangle\rangle&:=(e^{i\epsilon H}\otimes e^{i\epsilon H}\otimes \mathbbm{1})\frac{|000\rangle+|101\rangle}{\sqrt{2}}\,.
\end{align}
One can easily verify that
    \begin{equation}
    \frac{ \langle P_i\otimes P_j\rangle_{\bar{\rho}}}{\rm Tr \,\bar{\rho}}
    =\frac{\langle \langle \Phi| P_i\otimes P_j\otimes \mathbbm{1}_E|\Psi\rangle\rangle}{\langle \langle \Phi|\Psi\rangle\rangle}=\frac{\langle 0,T|P_j(\epsilon)P_i |0\rangle}{\langle 0,T |0\rangle}\,,
\end{equation}
as it follows from 
\begin{equation}\label{eq:purifgen}
\frac{\bar{\rho}}{\rm Tr \,\bar{\rho}}={\rm Tr}_E \,R\,,\;\;\;\;R:=\frac{|\Psi\rangle\rangle\langle \langle \Phi|}{\langle \langle \Phi|\Psi\rangle\rangle}\,,
\end{equation}
 with $R$ a generalized state, i.e., a non-orthogonal projector. We also have $2\,\langle \langle \Phi|\Psi\rangle\rangle={\rm Tr} \,\bar{\rho}=\langle 0,T|0\rangle$, where the factor $2$ may be absorbed in the states without changing the condition $R^2=R$.
We have thus obtained a generalized purification of $\bar{\rho}$, as the ones we discussed in the main body's section \ref{sec:stgstates} and Appendix \ref{sec:Appurif} for a bosonic field. In fact, one can show that obtaining $R$ is possible for any system and evolution (see also the recent discussion in the context of time-dependent holographic spacetimes \cite{tak.23, har.23, nar.22, chu2023time} ).

The previous example shows once again how the new formalism treats space and time equally, with the Hilbert space of one qubit and two times having dimension $2^2$ and being the same as the one of two qubits separated in space at a single time. Equations \eqref{eq:rho} and \eqref{eq:rhobar} are formally the same, with the differences codified in the correlators. We also see that the generalized purification \eqref{eq:purifgen} is analogous to the traditional one shown in \eqref{eq:purif}. Of course, considering e.g. two qubits and two times leads to a space of dimension $2^4$ with all variables on equal footing. 
The differences between space and time are not apparent at the Hilbert space level, instead, they manifest in the properties of ``states'' with $|\Psi\rangle\rangle\langle \langle \Psi|$ being an orthogonal projector but not $R$. The latter codifies not only the initial state but also the evolution and causal structure of the theory.

{\it Bosonic case and continuum limit.} It is interesting to see the consequences of this map in the case of the bosonic field in discrete spacetime. As an example, the previous let us write
\begin{equation}
  \!\!\!{\rm Tr}\big[|{\phi}\rangle_0 \langle {\phi}|\, e^{i\mathcal{S}}\phi_{im}\phi_{jn}\big]=\langle {\phi},T|\hat{T} \phi_{H m}(t_i)\phi_{H n}(t_j)|{\phi}\rangle
\end{equation}
which is the ``Feynman propagator'' for a finite time window $T$ and for an initial and finite configuration of the field in space $|{\phi}\rangle:=\otimes_m |\phi_m\rangle$ ($\hat{\phi}_m |{\phi}\rangle=\phi_m |{\phi}\rangle$ in a given time slice, where we have introduced the hat for clarity).

Notice that the right-hand side can be naturally written as a path integral between the configurations $\vec{\phi}$ and inserting two field operators. The left-hand side looks suspiciously similar to such a construction, except for the fact that involves operators and a trace in (the extended) Hilbert space. To understand the relation between the two, one must expand the trace in some basis. While infinite choices are possible, the field spacetime eigenbasis $|\bm{\phi}\rangle:=\otimes_{i,m}|\phi_{im}\rangle$ ($\hat{\phi}_{im}|\bm{\phi}\rangle=\phi_{im}|\bm{\phi}\rangle$ \footnote{In the continuum version of main text, $|\phi\rangle$ corresponds to $|\phi({\bf x})\rangle$ whereas $|\bm\phi\rangle$ to $|\phi(x)\rangle$}) leads directly to Feynman PI \cite{diazp.21}. In this sense, the PI formulation emerges from the formalism as well. For example,  it is easily seen that
\begin{equation}\label{eq:legendremv}
  \langle \bm{\phi}|e^{i \epsilon\mathcal{P}_0}|\bm{\pi}\rangle=\exp\Big[i\epsilon\sum_{i,m} \pi_{im}\frac{(\phi_{i+1,m}-\phi_{im})}{\epsilon}\Big]\langle \bm{\phi}|\bm{\pi}\rangle\,,
\end{equation}
revealing that $\mathcal{P}_0$ is related to the Legendre transform, where $|\bm{\pi}\rangle:=\otimes_{i,m} |\pi_{im}\rangle$ is the field momentum eigenbasis.

One can further justify the appearance of the Legendre transform by employing a normal mode representation of $\mathcal{P}_0$. Given annihilation (creation) operators $a_{im}$, $a_{im}^\dag$ satisfying $[a_{im},a^\dag_{jn}]=\delta_{ij}\delta_{mn}$ (linearly related with $\phi_{im}$, $\pi_{im}$) one can define Fourier modes in time via $a_{km}:=\frac{1}{\sqrt{N}}\sum_j e^{i\omega_k j\epsilon}a_{jm}$ where $\omega_k=2\pi k /T$ and $k$ takes $N=T/\epsilon$ different values. Then
\begin{equation}\label{eq:discreteFTp}
    \mathcal{P}_0=\sum_{k,m} \omega_k a^\dag_{km} a_{km}
\end{equation}
yields $e^{i\epsilon \mathcal{P}_0}a_{jm}e^{-\epsilon \mathcal{P}_0}=a_{j+1,m}$ in agreement with \eqref{eq:eipket}. If we now rewrite $\mathcal{P}_0$ in the time basis we have $\mathcal{P}_0=\sum_{m,j,j'}a^\dag_{j'm}iD_{jj'}a_{jm}$ with $D:=-\frac{1}{N}\sum_k i\omega_k e^{i\omega_k (j-j')\epsilon}$ which is a discrete version of a derivative in time of the Kronecker delta $\delta_{jj'}$. The form is once again that of the Legendre transform, now in the ``variables'' $a_{jm},a_{jm}^\dag$. Notice also that Eq.\ \eqref{eq:discreteFTp} is a discrete version of the main body expression \eqref{eq:p0diagonal} for canonical foliation.

 Moreover, consider the continuum spacetime limit in Hilbert space. The procedure is the same as it is usually done in space, so let us first review the conventional spacelike scenario. 
Given a constant spacing $a$, one considers operators $\phi(\textbf{x}):=\frac{\phi_\textbf{m}}{a^{d/2}}$, $\pi(\textbf{x}):=\frac{\pi_\textbf{m}}{a^{d/2}}$ for $\textbf{x}=a \textbf{m}$ (assuming $d$ dimensions) and $\bf m$ a vector of integer entries.  Then, the canonical algebra $[\phi_{\textbf{m}},\pi_{\textbf{m}'}]=i\delta_{\bf m,\bf m'}$ implies $$[\phi(\textbf{x}),\pi(\textbf{x}')]= i a^{-d}\delta_{\bf m,\bf m'}\to i\delta^{(d)}(\textbf{x}-\textbf{x}')$$ in the limit $a\to 0$. The same treatment can be applied to obtain  \eqref{eq:extalg} from \eqref{eq:discextalg}  by defining $\phi(x):=\frac{\phi_{j\textbf{m}}}{\sqrt{\epsilon}a^{d/2}}$, $\pi(x):=\frac{\pi_{j\textbf{m}}}{\sqrt{\epsilon}a^{d/2}}$, with $x=(\epsilon j, a\textbf{m})$, so that 
$$[\phi(x),\pi(x')]=i\epsilon^{-1}a^{-d}\delta_{jj'}\delta_{\bf m,\bf m'}\to i\delta^{(d+1)}(x-x')\,.$$
While there is no longer a countable number of time slices (nor of spatial slices) one can still define a time translation operator and relate it to the previous version. In fact, one can rigorously show \cite{diaz.21,diazp.21} that the expansion \eqref{eq:discreteFTp} (holding with the index $k$ taking arbitrary integer values) leads to $\mathcal{P}_0 \to \int d^Dx\, \pi(x) \dot{\phi}(x)$ for $\epsilon,a \to 0$, as suggested by \eqref{eq:legendremv} but holding at the operator level and in agreement with the main body results. The limit of small spacing also assumes $\epsilon N =T$ constant, as well as the usual spatial condition $a M=L$ (for $M$ the number of spatial slices and $L$ the total length of the ``box''). One can then take the limits $T,L\to \infty$ to recover the formalism of the main body. In this case, the Fourier creation (annihilation) operators satisfy a continuum algebra as well, according to 
Eq.\ \eqref{eq:algpart}: e.g.,   keeping $L$ finite one defines $a_m(p^0):=\sqrt{T}a_{km}$ with $p^0=2\pi k/T$ so that $[a_m(p^0),a^\dag_{m'}(p'^0)]=T\delta_{kk'}\delta_{mm'}\to 2\pi\delta(p^0-p'^0)\delta_{mm'}$.

Notice that the continuum time limit is well defined for the operator $\mathcal{P}_0$ itself, so that $e^{i\tau \mathcal{P}_0}$ implements geometric time translations such as $\phi(x)\to \phi(x^0+\tau,\textbf{x})$ for $\tau\in \mathbbm{R}$. 
On the other hand, if one would like to consider the limit $N\to \infty$ first (keeping $\epsilon$ finite, such that $j=-\frac{N}{2},\ldots,\frac{N}{2}$ and $T=\epsilon N\rightarrow\infty$) the FT now leads to continuous $p^0\in (-\Lambda, \Lambda)$ with $\Lambda:= (2\epsilon)^{-1}$ functioning as a natural cutoff, so that 
the proper definition of the generator of translations in time becomes $\mathcal{P}_0:=\sum_m\int_{-\Lambda}^{\Lambda} dp^0\,  p^0 a^\dag_m(p^0)a_m(p^0)$, which takes the place of \eqref{eq:discreteFTp}. Then  $e^{i\epsilon \mathcal{P}_0}a_{jm}e^{-i\epsilon \mathcal{P}_0}=a_{j+1,m}$ for any integer $j$. 
In conclusion, in all of these limits the generator of translations in time is properly defined. These definitions constitute a natural extension of the case of a compactified discrete time and agree with the possible limits (namely small spacing and/or large $T$) of that basic scenario.

Notice that a similar treatment might be employed for other spacetime algebras, such as the one in Eq.\ \eqref{eq:algpauli}, leading to the replacement $\delta_{tt'}\to \delta(t-t')$.


\begin{thebibliography}{73}%
\makeatletter
\providecommand \@ifxundefined [1]{%
 \@ifx{#1\undefined}
}%
\providecommand \@ifnum [1]{%
 \ifnum #1\expandafter \@firstoftwo
 \else \expandafter \@secondoftwo
 \fi
}%
\providecommand \@ifx [1]{%
 \ifx #1\expandafter \@firstoftwo
 \else \expandafter \@secondoftwo
 \fi
}%
\providecommand \natexlab [1]{#1}%
\providecommand \enquote  [1]{``#1''}%
\providecommand \bibnamefont  [1]{#1}%
\providecommand \bibfnamefont [1]{#1}%
\providecommand \citenamefont [1]{#1}%
\providecommand \href@noop [0]{\@secondoftwo}%
\providecommand \href [0]{\begingroup \@sanitize@url \@href}%
\providecommand \@href[1]{\@@startlink{#1}\@@href}%
\providecommand \@@href[1]{\endgroup#1\@@endlink}%
\providecommand \@sanitize@url [0]{\catcode `\\12\catcode `\$12\catcode
  `\&12\catcode `\#12\catcode `\^12\catcode `\_12\catcode `\%12\relax}%
\providecommand \@@startlink[1]{}%
\providecommand \@@endlink[0]{}%
\providecommand \url  [0]{\begingroup\@sanitize@url \@url }%
\providecommand \@url [1]{\endgroup\@href {#1}{\urlprefix }}%
\providecommand \urlprefix  [0]{URL }%
\providecommand \Eprint [0]{\href }%
\providecommand \doibase [0]{https://doi.org/}%
\providecommand \selectlanguage [0]{\@gobble}%
\providecommand \bibinfo  [0]{\@secondoftwo}%
\providecommand \bibfield  [0]{\@secondoftwo}%
\providecommand \translation [1]{[#1]}%
\providecommand \BibitemOpen [0]{}%
\providecommand \bibitemStop [0]{}%
\providecommand \bibitemNoStop [0]{.\EOS\space}%
\providecommand \EOS [0]{\spacefactor3000\relax}%
\providecommand \BibitemShut  [1]{\csname bibitem#1\endcsname}%
\let\auto@bib@innerbib\@empty
\bibitem [{\citenamefont {Dyson}(1949)}]{dy.94}%
  \BibitemOpen
  \bibfield  {author} {\bibinfo {author} {\bibfnamefont {F.~J.}\ \bibnamefont
  {Dyson}},\ }\bibfield  {title} {\bibinfo {title} {The radiation theories of
  tomonaga, schwinger, and feynman},\ }\href
  {https://journals.aps.org/pr/abstract/10.1103/PhysRev.75.486} {\bibfield
  {journal} {\bibinfo  {journal} {Phys. Rev.}\ }\textbf {\bibinfo {volume}
  {75}},\ \bibinfo {pages} {486} (\bibinfo {year} {1949})}\BibitemShut
  {NoStop}%
\bibitem [{\citenamefont {Feynman}(1948)}]{Feynm.1948}%
  \BibitemOpen
  \bibfield  {author} {\bibinfo {author} {\bibfnamefont {R.~P.}\ \bibnamefont
  {Feynman}},\ }\bibfield  {title} {\bibinfo {title} {Space-time approach to
  non-relativistic quantum mechanics},\ }\href
  {https://journals.aps.org/rmp/abstract/10.1103/RevModPhys.20.367} {\bibfield
  {journal} {\bibinfo  {journal} {Rev. Mod. Phys.}\ }\textbf {\bibinfo {volume}
  {20}},\ \bibinfo {pages} {367} (\bibinfo {year} {1948})}\BibitemShut
  {NoStop}%
\bibitem [{\citenamefont {Feynman}(2005)}]{feyn.05}%
  \BibitemOpen
  \bibfield  {author} {\bibinfo {author} {\bibfnamefont {R.~P.}\ \bibnamefont
  {Feynman}},\ }\bibfield  {title} {\bibinfo {title} {The principle of least
  action in quantum mechanics, {PhD} dissertation (1942)},\ }in\ \href@noop {}
  {\emph {\bibinfo {booktitle} {Feynman's Thesis—A New Approach To Quantum
  Theory}}}\ (\bibinfo  {publisher} {World Scientific},\ \bibinfo {year}
  {2005})\BibitemShut {NoStop}%
\bibitem [{\citenamefont {Fitzsimons}\ \emph {et~al.}(2015)\citenamefont
  {Fitzsimons}, \citenamefont {Jones},\ and\ \citenamefont {Vedral}}]{fit.15}%
  \BibitemOpen
  \bibfield  {author} {\bibinfo {author} {\bibfnamefont {J.~F.}\ \bibnamefont
  {Fitzsimons}}, \bibinfo {author} {\bibfnamefont {J.~A.}\ \bibnamefont
  {Jones}},\ and\ \bibinfo {author} {\bibfnamefont {V.}~\bibnamefont
  {Vedral}},\ }\bibfield  {title} {\bibinfo {title} {Quantum correlations which
  imply causation},\ }\href {https://www.nature.com/articles/srep18281}
  {\bibfield  {journal} {\bibinfo  {journal} {Sci. Rep.}\ }\textbf {\bibinfo
  {volume} {5}},\ \bibinfo {pages} {18281} (\bibinfo {year}
  {2015})}\BibitemShut {NoStop}%
\bibitem [{\citenamefont {Horsman}\ \emph {et~al.}(2017)\citenamefont
  {Horsman}, \citenamefont {Heunen}, \citenamefont {Pusey}, \citenamefont
  {Barrett},\ and\ \citenamefont {Spekkens}}]{ho.17}%
  \BibitemOpen
  \bibfield  {author} {\bibinfo {author} {\bibfnamefont {D.}~\bibnamefont
  {Horsman}}, \bibinfo {author} {\bibfnamefont {C.}~\bibnamefont {Heunen}},
  \bibinfo {author} {\bibfnamefont {M.~F.}\ \bibnamefont {Pusey}}, \bibinfo
  {author} {\bibfnamefont {J.}~\bibnamefont {Barrett}},\ and\ \bibinfo {author}
  {\bibfnamefont {R.~W.}\ \bibnamefont {Spekkens}},\ }\bibfield  {title}
  {\bibinfo {title} {Can a quantum state over time resemble a quantum state at
  a single time?},\ }\href
  {https://royalsocietypublishing.org/doi/10.1098/rspa.2017.0395} {\bibfield
  {journal} {\bibinfo  {journal} {Proc. R. Soc. A}\ }\textbf {\bibinfo {volume}
  {473}},\ \bibinfo {pages} {20170395} (\bibinfo {year} {2017})}\BibitemShut
  {NoStop}%
\bibitem [{\citenamefont {Cotler}\ \emph {et~al.}(2018)\citenamefont {Cotler},
  \citenamefont {Jian}, \citenamefont {Qi},\ and\ \citenamefont
  {Wilczek}}]{cot.18}%
  \BibitemOpen
  \bibfield  {author} {\bibinfo {author} {\bibfnamefont {J.}~\bibnamefont
  {Cotler}}, \bibinfo {author} {\bibfnamefont {C.~M.}\ \bibnamefont {Jian}},
  \bibinfo {author} {\bibfnamefont {X.}~\bibnamefont {Qi}},\ and\ \bibinfo
  {author} {\bibfnamefont {F.}~\bibnamefont {Wilczek}},\ }\bibfield  {title}
  {\bibinfo {title} {Superdensity operators for spacetime quantum mechanics},\
  }\href {https://link.springer.com/article/10.1007/JHEP09(2018)093} {\bibfield
   {journal} {\bibinfo  {journal} {J. High Energy Phys.}\ }\textbf {\bibinfo
  {volume} {2018}},\ \bibinfo {pages} {93}}\BibitemShut {NoStop}%
\bibitem [{\citenamefont {Giacomini}\ \emph {et~al.}(2019)\citenamefont
  {Giacomini}, \citenamefont {Castro-Ruiz},\ and\ \citenamefont
  {Brukner}}]{giac.19}%
  \BibitemOpen
  \bibfield  {author} {\bibinfo {author} {\bibfnamefont {F.}~\bibnamefont
  {Giacomini}}, \bibinfo {author} {\bibfnamefont {E.}~\bibnamefont
  {Castro-Ruiz}},\ and\ \bibinfo {author} {\bibfnamefont
  {{\v{C}}.}~\bibnamefont {Brukner}},\ }\bibfield  {title} {\bibinfo {title}
  {Quantum mechanics and the covariance of physical laws in quantum reference
  frames},\ }\href {https://www.nature.com/articles/s41467-018-08155-0}
  {\bibfield  {journal} {\bibinfo  {journal} {Nat.\ Commun.}\ }\textbf
  {\bibinfo {volume} {10}},\ \bibinfo {pages} {494} (\bibinfo {year}
  {2019})}\BibitemShut {NoStop}%
\bibitem [{\citenamefont {Diaz}\ \emph {et~al.}(2019)\citenamefont {Diaz},
  \citenamefont {Matera},\ and\ \citenamefont {Rossignoli}}]{dia.19}%
  \BibitemOpen
  \bibfield  {author} {\bibinfo {author} {\bibfnamefont {N.~L.}\ \bibnamefont
  {Diaz}}, \bibinfo {author} {\bibfnamefont {J.~M.}\ \bibnamefont {Matera}},\
  and\ \bibinfo {author} {\bibfnamefont {R.}~\bibnamefont {Rossignoli}},\
  }\bibfield  {title} {\bibinfo {title} {History state formalism for scalar
  particles},\ }\href
  {https://journals.aps.org/prd/abstract/10.1103/PhysRevD.100.125020}
  {\bibfield  {journal} {\bibinfo  {journal} {Phys. Rev. D}\ }\textbf {\bibinfo
  {volume} {100}},\ \bibinfo {pages} {125020} (\bibinfo {year}
  {2019})}\BibitemShut {NoStop}%
\bibitem [{\citenamefont {Diaz}\ \emph
  {et~al.}(2021{\natexlab{a}})\citenamefont {Diaz}, \citenamefont {Matera},\
  and\ \citenamefont {Rossignoli}}]{diaz.21}%
  \BibitemOpen
  \bibfield  {author} {\bibinfo {author} {\bibfnamefont {N.~L.}\ \bibnamefont
  {Diaz}}, \bibinfo {author} {\bibfnamefont {J.~M.}\ \bibnamefont {Matera}},\
  and\ \bibinfo {author} {\bibfnamefont {R.}~\bibnamefont {Rossignoli}},\
  }\bibfield  {title} {\bibinfo {title} {Spacetime quantum actions},\ }\href
  {https://journals.aps.org/prd/abstract/10.1103/PhysRevD.103.065011}
  {\bibfield  {journal} {\bibinfo  {journal} {Phys. Rev. D}\ }\textbf {\bibinfo
  {volume} {103}},\ \bibinfo {pages} {065011} (\bibinfo {year}
  {2021}{\natexlab{a}})}\BibitemShut {NoStop}%
\bibitem [{\citenamefont {Diaz}\ \emph
  {et~al.}(2021{\natexlab{b}})\citenamefont {Diaz}, \citenamefont {Matera},\
  and\ \citenamefont {Rossignoli}}]{diazp.21}%
  \BibitemOpen
  \bibfield  {author} {\bibinfo {author} {\bibfnamefont {N.~L.}\ \bibnamefont
  {Diaz}}, \bibinfo {author} {\bibfnamefont {J.}~\bibnamefont {Matera}},\ and\
  \bibinfo {author} {\bibfnamefont {R.}~\bibnamefont {Rossignoli}},\ }\bibfield
   {title} {\bibinfo {title} {Path integrals from spacetime quantum actions},\
  }\href {https://arxiv.org/abs/2111.05383} {\bibfield  {journal} {\bibinfo
  {journal} {arXiv:2111.05383}\ } (\bibinfo {year}
  {2021}{\natexlab{b}})}\BibitemShut {NoStop}%
\bibitem [{\citenamefont {H\"ohn}\ \emph {et~al.}(2021)\citenamefont {H\"ohn},
  \citenamefont {Smith},\ and\ \citenamefont {Lock}}]{hoh.21}%
  \BibitemOpen
  \bibfield  {author} {\bibinfo {author} {\bibfnamefont {P.~A.}\ \bibnamefont
  {H\"ohn}}, \bibinfo {author} {\bibfnamefont {A.~R.~H.}\ \bibnamefont
  {Smith}},\ and\ \bibinfo {author} {\bibfnamefont {M.~P.~E.}\ \bibnamefont
  {Lock}},\ }\bibfield  {title} {\bibinfo {title} {Trinity of relational
  quantum dynamics},\ }\href {https://doi.org/10.1103/PhysRevD.104.066001}
  {\bibfield  {journal} {\bibinfo  {journal} {Phys. Rev. D}\ }\textbf {\bibinfo
  {volume} {104}},\ \bibinfo {pages} {066001} (\bibinfo {year}
  {2021})}\BibitemShut {NoStop}%
\bibitem [{\citenamefont {Foti}\ \emph {et~al.}(2021)\citenamefont {Foti},
  \citenamefont {Coppoli}, \citenamefont {Barni}, \citenamefont {Cuccoli},\
  and\ \citenamefont {Verrucchi}}]{fot.21}%
  \BibitemOpen
  \bibfield  {author} {\bibinfo {author} {\bibfnamefont {C.}~\bibnamefont
  {Foti}}, \bibinfo {author} {\bibfnamefont {A.}~\bibnamefont {Coppoli}},
  \bibinfo {author} {\bibfnamefont {G.}~\bibnamefont {Barni}}, \bibinfo
  {author} {\bibfnamefont {A.}~\bibnamefont {Cuccoli}},\ and\ \bibinfo {author}
  {\bibfnamefont {P.}~\bibnamefont {Verrucchi}},\ }\bibfield  {title} {\bibinfo
  {title} {Time and classical equations of motion from quantum entanglement via
  the {P}age and {W}ootters mechanism with generalized coherent states},\
  }\href {https://doi.org/10.1038/s41467-021-21782-4} {\bibfield  {journal}
  {\bibinfo  {journal} {Nat. Commun.}\ }\textbf {\bibinfo {volume} {12}},\
  \bibinfo {pages} {1787} (\bibinfo {year} {2021})}\BibitemShut {NoStop}%
\bibitem [{\citenamefont {Favalli}\ and\ \citenamefont
  {Smerzi}(2022{\natexlab{a}})}]{fav.22}%
  \BibitemOpen
  \bibfield  {author} {\bibinfo {author} {\bibfnamefont {T.}~\bibnamefont
  {Favalli}}\ and\ \bibinfo {author} {\bibfnamefont {A.}~\bibnamefont
  {Smerzi}},\ }\bibfield  {title} {\bibinfo {title} {A model of quantum
  spacetime},\ }\href@noop {} {\bibfield  {journal} {\bibinfo  {journal} {AVS
  Quantum Science}\ }\textbf {\bibinfo {volume} {4}},\ \bibinfo {pages}
  {044403} (\bibinfo {year} {2022}{\natexlab{a}})}\BibitemShut {NoStop}%
\bibitem [{\citenamefont {Paiva}\ \emph {et~al.}(2022)\citenamefont {Paiva},
  \citenamefont {Te’eni}, \citenamefont {Peled}, \citenamefont {Cohen},\ and\
  \citenamefont {Aharonov}}]{pai.22}%
  \BibitemOpen
  \bibfield  {author} {\bibinfo {author} {\bibfnamefont {I.~L.}\ \bibnamefont
  {Paiva}}, \bibinfo {author} {\bibfnamefont {A.}~\bibnamefont {Te’eni}},
  \bibinfo {author} {\bibfnamefont {B.~Y.}\ \bibnamefont {Peled}}, \bibinfo
  {author} {\bibfnamefont {E.}~\bibnamefont {Cohen}},\ and\ \bibinfo {author}
  {\bibfnamefont {Y.}~\bibnamefont {Aharonov}},\ }\bibfield  {title} {\bibinfo
  {title} {Non-inertial quantum clock frames lead to non-hermitian dynamics},\
  }\href {https://doi.org/10.1038/s42005-022-01081-0} {\bibfield  {journal}
  {\bibinfo  {journal} {Commun.\ Phys.}\ }\textbf {\bibinfo {volume} {5}},\
  \bibinfo {pages} {298} (\bibinfo {year} {2022})}\BibitemShut {NoStop}%
\bibitem [{\citenamefont {Van~Raamsdonk}(2010)}]{van.10}%
  \BibitemOpen
  \bibfield  {author} {\bibinfo {author} {\bibfnamefont {M.}~\bibnamefont
  {Van~Raamsdonk}},\ }\bibfield  {title} {\bibinfo {title} {Building up
  spacetime with quantum entanglement},\ }\href
  {https://link.springer.com/article/10.1007/s10714-010-1034-0} {\bibfield
  {journal} {\bibinfo  {journal} {Gen. Relativ. Gravit.}\ }\textbf {\bibinfo
  {volume} {42}},\ \bibinfo {pages} {2323} (\bibinfo {year}
  {2010})}\BibitemShut {NoStop}%
\bibitem [{\citenamefont {Cao}\ \emph {et~al.}(2017)\citenamefont {Cao},
  \citenamefont {Carroll},\ and\ \citenamefont {Michalakis}}]{Ca.17}%
  \BibitemOpen
  \bibfield  {author} {\bibinfo {author} {\bibfnamefont {C.}~\bibnamefont
  {Cao}}, \bibinfo {author} {\bibfnamefont {S.~M.}\ \bibnamefont {Carroll}},\
  and\ \bibinfo {author} {\bibfnamefont {S.}~\bibnamefont {Michalakis}},\
  }\bibfield  {title} {\bibinfo {title} {Space from {H}ilbert space: Recovering
  geometry from bulk entanglement},\ }\href
  {https://journals.aps.org/prd/abstract/10.1103/PhysRevD.95.024031} {\bibfield
   {journal} {\bibinfo  {journal} {Phys. Rev. D}\ }\textbf {\bibinfo {volume}
  {95}},\ \bibinfo {pages} {024031} (\bibinfo {year} {2017})}\BibitemShut
  {NoStop}%
\bibitem [{\citenamefont {Giovannetti}\ \emph {et~al.}(2015)\citenamefont
  {Giovannetti}, \citenamefont {Lloyd},\ and\ \citenamefont {Maccone}}]{QT.15}%
  \BibitemOpen
  \bibfield  {author} {\bibinfo {author} {\bibfnamefont {V.}~\bibnamefont
  {Giovannetti}}, \bibinfo {author} {\bibfnamefont {S.}~\bibnamefont {Lloyd}},\
  and\ \bibinfo {author} {\bibfnamefont {L.}~\bibnamefont {Maccone}},\
  }\bibfield  {title} {\bibinfo {title} {Quantum time},\ }\href
  {https://journals.aps.org/prd/abstract/10.1103/PhysRevD.92.045033} {\bibfield
   {journal} {\bibinfo  {journal} {Phys. Rev. D}\ }\textbf {\bibinfo {volume}
  {92}},\ \bibinfo {pages} {045033} (\bibinfo {year} {2015})}\BibitemShut
  {NoStop}%
\bibitem [{\citenamefont {Boette}\ \emph {et~al.}(2016)\citenamefont {Boette},
  \citenamefont {Rossignoli}, \citenamefont {Gigena},\ and\ \citenamefont
  {Cerezo}}]{b.16}%
  \BibitemOpen
  \bibfield  {author} {\bibinfo {author} {\bibfnamefont {A.}~\bibnamefont
  {Boette}}, \bibinfo {author} {\bibfnamefont {R.}~\bibnamefont {Rossignoli}},
  \bibinfo {author} {\bibfnamefont {N.}~\bibnamefont {Gigena}},\ and\ \bibinfo
  {author} {\bibfnamefont {M.}~\bibnamefont {Cerezo}},\ }\bibfield  {title}
  {\bibinfo {title} {System-time entanglement in a discrete-time model},\
  }\href {https://journals.aps.org/pra/abstract/10.1103/PhysRevA.93.062127}
  {\bibfield  {journal} {\bibinfo  {journal} {Phys. Rev. A}\ }\textbf {\bibinfo
  {volume} {93}},\ \bibinfo {pages} {062127} (\bibinfo {year}
  {2016})}\BibitemShut {NoStop}%
\bibitem [{\citenamefont {Boette}\ and\ \citenamefont
  {Rossignoli}(2018)}]{b.18}%
  \BibitemOpen
  \bibfield  {author} {\bibinfo {author} {\bibfnamefont {A.}~\bibnamefont
  {Boette}}\ and\ \bibinfo {author} {\bibfnamefont {R.}~\bibnamefont
  {Rossignoli}},\ }\bibfield  {title} {\bibinfo {title} {History states of
  systems and operators},\ }\href
  {https://journals.aps.org/pra/abstract/10.1103/PhysRevA.98.032108} {\bibfield
   {journal} {\bibinfo  {journal} {Phys. Rev. A}\ }\textbf {\bibinfo {volume}
  {98}},\ \bibinfo {pages} {032108} (\bibinfo {year} {2018})}\BibitemShut
  {NoStop}%
\bibitem [{\citenamefont {Barison}\ \emph {et~al.}(2022)\citenamefont
  {Barison}, \citenamefont {Vicentini}, \citenamefont {Cirac},\ and\
  \citenamefont {Carleo}}]{barison2022variational}%
  \BibitemOpen
  \bibfield  {author} {\bibinfo {author} {\bibfnamefont {S.}~\bibnamefont
  {Barison}}, \bibinfo {author} {\bibfnamefont {F.}~\bibnamefont {Vicentini}},
  \bibinfo {author} {\bibfnamefont {I.}~\bibnamefont {Cirac}},\ and\ \bibinfo
  {author} {\bibfnamefont {G.}~\bibnamefont {Carleo}},\ }\bibfield  {title}
  {\bibinfo {title} {Variational dynamics as a ground-state problem on a
  quantum computer},\ }\href {https://doi.org/10.1103/PhysRevResearch.4.043161}
  {\bibfield  {journal} {\bibinfo  {journal} {Phys.\ Rev.\ Res.}\ }\textbf
  {\bibinfo {volume} {4}},\ \bibinfo {pages} {043161} (\bibinfo {year}
  {2022})}\BibitemShut {NoStop}%
\bibitem [{\citenamefont {Diaz}\ \emph {et~al.}(2023)\citenamefont {Diaz},
  \citenamefont {Braccia}, \citenamefont {Larocca}, \citenamefont {Matera},
  \citenamefont {Rossignoli},\ and\ \citenamefont {Cerezo}}]{diaz2023parallel}%
  \BibitemOpen
  \bibfield  {author} {\bibinfo {author} {\bibfnamefont {N.~L.}\ \bibnamefont
  {Diaz}}, \bibinfo {author} {\bibfnamefont {P.}~\bibnamefont {Braccia}},
  \bibinfo {author} {\bibfnamefont {M.}~\bibnamefont {Larocca}}, \bibinfo
  {author} {\bibfnamefont {J.}~\bibnamefont {Matera}}, \bibinfo {author}
  {\bibfnamefont {R.}~\bibnamefont {Rossignoli}},\ and\ \bibinfo {author}
  {\bibfnamefont {M.}~\bibnamefont {Cerezo}},\ }\bibfield  {title} {\bibinfo
  {title} {Parallel-in-time quantum simulation via {P}age and {W}ootters
  quantum time},\ }\href {https://arxiv.org/abs/2308.12944} {\bibfield
  {journal} {\bibinfo  {journal} {arXiv:2308.12944}\ } (\bibinfo {year}
  {2023})}\BibitemShut {NoStop}%
\bibitem [{\citenamefont {Page}\ and\ \citenamefont {Wootters}(1983)}]{PaW.83}%
  \BibitemOpen
  \bibfield  {author} {\bibinfo {author} {\bibfnamefont {D.~N.}\ \bibnamefont
  {Page}}\ and\ \bibinfo {author} {\bibfnamefont {W.~K.}\ \bibnamefont
  {Wootters}},\ }\bibfield  {title} {\bibinfo {title} {Evolution without
  evolution: Dynamics described by stationary observables},\ }\href
  {https://journals.aps.org/prd/abstract/10.1103/PhysRevD.27.2885} {\bibfield
  {journal} {\bibinfo  {journal} {Phys. Rev. D}\ }\textbf {\bibinfo {volume}
  {27}},\ \bibinfo {pages} {2885} (\bibinfo {year} {1983})}\BibitemShut
  {NoStop}%
\bibitem [{\citenamefont {Isham}\ and\ \citenamefont {Kuchar}(1985)}]{ish.85}%
  \BibitemOpen
  \bibfield  {author} {\bibinfo {author} {\bibfnamefont {C.}~\bibnamefont
  {Isham}}\ and\ \bibinfo {author} {\bibfnamefont {K.}~\bibnamefont {Kuchar}},\
  }\bibfield  {title} {\bibinfo {title} {Representations of spacetime
  diffeomorphisms. {I}. {C}anonical parametrized field theories},\ }\href
  {https://doi.org/https://doi.org/10.1016/0003-4916(85)90018-1} {\bibfield
  {journal} {\bibinfo  {journal} {Ann. Phys. (N.Y.)}\ }\textbf {\bibinfo
  {volume} {164}},\ \bibinfo {pages} {288} (\bibinfo {year}
  {1985})}\BibitemShut {NoStop}%
\bibitem [{\citenamefont {Isham}(1994)}]{ish.93}%
  \BibitemOpen
  \bibfield  {author} {\bibinfo {author} {\bibfnamefont {C.~J.}\ \bibnamefont
  {Isham}},\ }\bibfield  {title} {\bibinfo {title} {Quantum logic and the
  histories approach to quantum theory},\ }\href
  {https://doi.org/10.1063/1.530544} {\bibfield  {journal} {\bibinfo  {journal}
  {J.\ Math.\ Phys.}\ }\textbf {\bibinfo {volume} {35}},\ \bibinfo {pages}
  {2157} (\bibinfo {year} {1994})}\BibitemShut {NoStop}%
\bibitem [{\citenamefont {Isham}\ \emph {et~al.}(1998)\citenamefont {Isham},
  \citenamefont {Linden}, \citenamefont {Savvidou},\ and\ \citenamefont
  {Schreckenberg}}]{Ish.98}%
  \BibitemOpen
  \bibfield  {author} {\bibinfo {author} {\bibfnamefont {C.~J.}\ \bibnamefont
  {Isham}}, \bibinfo {author} {\bibfnamefont {N.}~\bibnamefont {Linden}},
  \bibinfo {author} {\bibfnamefont {K.}~\bibnamefont {Savvidou}},\ and\
  \bibinfo {author} {\bibfnamefont {S.}~\bibnamefont {Schreckenberg}},\
  }\bibfield  {title} {\bibinfo {title} {Continuous time and consistent
  histories},\ }\href {https://doi.org/10.1063/1.532265} {\bibfield  {journal}
  {\bibinfo  {journal} {J.\ Math.\ Phys.}\ }\textbf {\bibinfo {volume} {39}},\
  \bibinfo {pages} {1818} (\bibinfo {year} {1998})}\BibitemShut {NoStop}%
\bibitem [{\citenamefont {Hartle}\ and\ \citenamefont {Marolf}(1997)}]{Ma.97}%
  \BibitemOpen
  \bibfield  {author} {\bibinfo {author} {\bibfnamefont {J.~B.}\ \bibnamefont
  {Hartle}}\ and\ \bibinfo {author} {\bibfnamefont {D.}~\bibnamefont
  {Marolf}},\ }\bibfield  {title} {\bibinfo {title} {Comparing formulations of
  generalized quantum mechanics for reparametrization-invariant systems},\
  }\href {https://link.aps.org/doi/10.1103/PhysRevD.56.6247} {\bibfield
  {journal} {\bibinfo  {journal} {Phys. Rev. D}\ }\textbf {\bibinfo {volume}
  {56}},\ \bibinfo {pages} {6247} (\bibinfo {year} {1997})}\BibitemShut
  {NoStop}%
\bibitem [{\citenamefont {Gambini}\ \emph {et~al.}(2009)\citenamefont
  {Gambini}, \citenamefont {Porto}, \citenamefont {Pullin},\ and\ \citenamefont
  {Torterolo}}]{g.09}%
  \BibitemOpen
  \bibfield  {author} {\bibinfo {author} {\bibfnamefont {R.}~\bibnamefont
  {Gambini}}, \bibinfo {author} {\bibfnamefont {R.~A.}\ \bibnamefont {Porto}},
  \bibinfo {author} {\bibfnamefont {J.}~\bibnamefont {Pullin}},\ and\ \bibinfo
  {author} {\bibfnamefont {S.}~\bibnamefont {Torterolo}},\ }\bibfield  {title}
  {\bibinfo {title} {Conditional probabilities with {D}irac observables and the
  problem of time in quantum gravity},\ }\href
  {https://doi.org/10.1103/PhysRevD.79.041501} {\bibfield  {journal} {\bibinfo
  {journal} {Phys. Rev. D}\ }\textbf {\bibinfo {volume} {79}},\ \bibinfo
  {pages} {041501(R)} (\bibinfo {year} {2009})}\BibitemShut {NoStop}%
\bibitem [{\citenamefont {Kucha{\v{r}}}(2011)}]{Ku.11}%
  \BibitemOpen
  \bibfield  {author} {\bibinfo {author} {\bibfnamefont {K.~V.}\ \bibnamefont
  {Kucha{\v{r}}}},\ }\bibfield  {title} {\bibinfo {title} {Time and
  interpretations of quantum gravity},\ }\href
  {https://www.worldscientific.com/doi/abs/10.1142/S0218271811019347}
  {\bibfield  {journal} {\bibinfo  {journal} {Int. J. Mod. Phys. D}\ }\textbf
  {\bibinfo {volume} {20}},\ \bibinfo {pages} {3} (\bibinfo {year}
  {2011})}\BibitemShut {NoStop}%
\bibitem [{\citenamefont {Hoehn}\ \emph {et~al.}(2023)\citenamefont {Hoehn},
  \citenamefont {Russo},\ and\ \citenamefont {Smith}}]{hoehn2023matter}%
  \BibitemOpen
  \bibfield  {author} {\bibinfo {author} {\bibfnamefont {P.~A.}\ \bibnamefont
  {Hoehn}}, \bibinfo {author} {\bibfnamefont {A.}~\bibnamefont {Russo}},\ and\
  \bibinfo {author} {\bibfnamefont {A.~R.}\ \bibnamefont {Smith}},\ }\bibfield
  {title} {\bibinfo {title} {Matter relative to quantum hypersurfaces},\ }\href
  {https://arxiv.org/abs/2308.12912} {\bibfield  {journal} {\bibinfo  {journal}
  {arXiv:2308.12912}\ } (\bibinfo {year} {2023})}\BibitemShut {NoStop}%
\bibitem [{\citenamefont {Savvidou}(1999)}]{Sav.99}%
  \BibitemOpen
  \bibfield  {author} {\bibinfo {author} {\bibfnamefont {K.}~\bibnamefont
  {Savvidou}},\ }\bibfield  {title} {\bibinfo {title} {The action operator for
  continuous-time histories},\ }\href
  {https://pubs.aip.org/aip/jmp/article/40/11/5657/398849/The-action-operator-for-continuous-time-histories}
  {\bibfield  {journal} {\bibinfo  {journal} {J.\ Math.\ Phys.}\ }\textbf
  {\bibinfo {volume} {40}},\ \bibinfo {pages} {5657} (\bibinfo {year}
  {1999})}\BibitemShut {NoStop}%
\bibitem [{Note1()}]{Note1}%
  \BibitemOpen
  \bibinfo {note} {Notice that the variation in phase space variables of the
  action ${S\protect \!=\protect \!\DOTSI \intop \ilimits@ dt dx dy(\pi \phi
  _{y} -{\protect \cal H})}$, $\protect \!\protect \!{\protect \text {with}
  \protect \tmspace +\thickmuskip {.2777em}\protect \mathcal {H}}$ in \protect
  \eqref {1}, is well defined (here ${\phi _{x^\mu }\protect \!\equiv \protect
  \!\partial _\mu \phi }$). In fact, one obtains ${\delta S=\DOTSI \intop
  \ilimits@ dtdxdy\protect \,[ (\phi _y+\pi )\delta \pi - (\pi _y+\phi
  _{tt}-\phi _{xx})\delta \phi ]}+\left .\DOTSI \intop \ilimits@ dxdt\protect
  \, \pi \delta \phi \right |_{y_i}^{y_f}+\left .\DOTSI \intop \ilimits@
  dxdy\protect \, \phi _t \delta \phi \right |_{t_i}^{t_f}-\left .\DOTSI \intop
  \ilimits@ dydt\protect \, \phi _x \delta \phi \right |_{x_i}^{x_f}$. All the
  boundary terms vanish under standard assumptions, namely $\delta \phi
  (t_i)=\delta \phi (t_f)=0$ and $\protect \frac {\partial \protect \mathcal
  {L}}{\partial \phi _i}\to 0$ for large $|x_i|$ in noncompact manifolds, or
  $\delta \phi =0$ in all the boundaries of a compact spacetime \cite
  {banados2016short}. Thus no ``differentiability'' problem \cite
  {regge1974role} arises. Notice, however, that $\protect \cal H$ in \protect
  \eqref {1} is not positive definite.}\BibitemShut {Stop}%
\bibitem [{\citenamefont {Doi}\ \emph {et~al.}(2023{\natexlab{a}})\citenamefont
  {Doi}, \citenamefont {Harper}, \citenamefont {Mollabashi}, \citenamefont
  {Takayanagi},\ and\ \citenamefont {Taki}}]{har.23}%
  \BibitemOpen
  \bibfield  {author} {\bibinfo {author} {\bibfnamefont {K.}~\bibnamefont
  {Doi}}, \bibinfo {author} {\bibfnamefont {J.}~\bibnamefont {Harper}},
  \bibinfo {author} {\bibfnamefont {A.}~\bibnamefont {Mollabashi}}, \bibinfo
  {author} {\bibfnamefont {T.}~\bibnamefont {Takayanagi}},\ and\ \bibinfo
  {author} {\bibfnamefont {Y.}~\bibnamefont {Taki}},\ }\bibfield  {title}
  {\bibinfo {title} {Timelike entanglement entropy},\ }\href
  {https://link.springer.com/article/10.1007/JHEP05(2023)052} {\bibfield
  {journal} {\bibinfo  {journal} {J.\ High Energy Phys.}\ }\textbf {\bibinfo
  {volume} {2023}},\ \bibinfo {pages} {52} (\bibinfo {year}
  {2023}{\natexlab{a}})}\BibitemShut {NoStop}%
\bibitem [{\citenamefont {de~Donder}(1930)}]{de.30}%
  \BibitemOpen
  \bibfield  {author} {\bibinfo {author} {\bibfnamefont {T.}~\bibnamefont
  {de~Donder}},\ }\href {https://books.google.com.ar/books?id=3kI7AQAAIAAJ}
  {\emph {\bibinfo {title} {Th{\'e}orie invariantive du calcul des
  variations}}}\ (\bibinfo  {publisher} {Gauthier-Villars, Paris},\ \bibinfo
  {year} {1930})\BibitemShut {NoStop}%
\bibitem [{\citenamefont {Weyl}(1935)}]{wey.35}%
  \BibitemOpen
  \bibfield  {author} {\bibinfo {author} {\bibfnamefont {H.}~\bibnamefont
  {Weyl}},\ }\bibfield  {title} {\bibinfo {title} {Geodesic fields in the
  calculus of variation for multiple integrals},\ }\href
  {http://www.jstor.org/stable/1968645} {\bibfield  {journal} {\bibinfo
  {journal} {Ann.\ Math.}\ }\textbf {\bibinfo {volume} {36}},\ \bibinfo {pages}
  {607} (\bibinfo {year} {1935})}\BibitemShut {NoStop}%
\bibitem [{\citenamefont {Gotay}(1991)}]{got.91}%
  \BibitemOpen
  \bibfield  {author} {\bibinfo {author} {\bibfnamefont {M.~J.}\ \bibnamefont
  {Gotay}},\ }\bibfield  {title} {\bibinfo {title} {A multisymplectic framework
  for classical field theory and the calculus of variations {II}: Space+ time
  decomposition},\ }\href
  {https://www.sciencedirect.com/science/article/pii/092622459190014Z}
  {\bibfield  {journal} {\bibinfo  {journal} {Differ.\ Geom.\ Appl.}\ }\textbf
  {\bibinfo {volume} {1}},\ \bibinfo {pages} {375} (\bibinfo {year}
  {1991})}\BibitemShut {NoStop}%
\bibitem [{\citenamefont {Kanatchikov}(1998)}]{kan.98}%
  \BibitemOpen
  \bibfield  {author} {\bibinfo {author} {\bibfnamefont {I.~V.}\ \bibnamefont
  {Kanatchikov}},\ }\bibfield  {title} {\bibinfo {title} {Canonical structure
  of classical field theory in the polymomentum phase space},\ }\href
  {https://doi.org/10.1016/S0034-4877%2898%2980182-1} {\bibfield  {journal}
  {\bibinfo  {journal} {Rep.\ Math.\ Phys.}\ }\textbf {\bibinfo {volume}
  {41}},\ \bibinfo {pages} {49} (\bibinfo {year} {1998})}\BibitemShut {NoStop}%
\bibitem [{\citenamefont {Isham}\ and\ \citenamefont
  {Savvidou}(2002)}]{ish.02}%
  \BibitemOpen
  \bibfield  {author} {\bibinfo {author} {\bibfnamefont {C.}~\bibnamefont
  {Isham}}\ and\ \bibinfo {author} {\bibfnamefont {K.}~\bibnamefont
  {Savvidou}},\ }\bibfield  {title} {\bibinfo {title} {The foliation operator
  in history quantum field theory},\ }\href
  {https://pubs.aip.org/aip/jmp/article/43/11/5493/378072/The-foliation-operator-in-history-quantum-field}
  {\bibfield  {journal} {\bibinfo  {journal} {J.\ Math.\ Phys.}\ }\textbf
  {\bibinfo {volume} {43}},\ \bibinfo {pages} {5493} (\bibinfo {year}
  {2002})}\BibitemShut {NoStop}%
\bibitem [{\citenamefont {Chester}\ \emph {et~al.}(2024)\citenamefont
  {Chester}, \citenamefont {Arsiwalla}, \citenamefont {Kauffman}, \citenamefont
  {Planat},\ and\ \citenamefont {Irwin}}]{chester2024quantization}%
  \BibitemOpen
  \bibfield  {author} {\bibinfo {author} {\bibfnamefont {D.}~\bibnamefont
  {Chester}}, \bibinfo {author} {\bibfnamefont {X.~D.}\ \bibnamefont
  {Arsiwalla}}, \bibinfo {author} {\bibfnamefont {L.~H.}\ \bibnamefont
  {Kauffman}}, \bibinfo {author} {\bibfnamefont {M.}~\bibnamefont {Planat}},\
  and\ \bibinfo {author} {\bibfnamefont {K.}~\bibnamefont {Irwin}},\ }\bibfield
   {title} {\bibinfo {title} {Quantization of a new canonical, covariant, and
  symplectic {H}amiltonian density},\ }\href
  {https://doi.org/10.3390/sym16030316} {\bibfield  {journal} {\bibinfo
  {journal} {Symmetry}\ }\textbf {\bibinfo {volume} {16}},\ \bibinfo {pages}
  {316} (\bibinfo {year} {2024})}\BibitemShut {NoStop}%
\bibitem [{Note2()}]{Note2}%
  \BibitemOpen
  \bibinfo {note} {Assuming standard boundary conditions, namely asymptotically
  vanishing fields for large $|\protect \textbf {x}|$, in agreement with $n^\mu
  $ timelike}\BibitemShut {NoStop}%
\bibitem [{Note3()}]{Note3}%
  \BibitemOpen
  \bibinfo {note} {From the mathematical perspective we can identify the
  ensuing phase space $\Omega $ with the limit $N\to \infty $ of the direct
  product $\Omega \equiv \omega _t^{\times N}$ for $\omega _t$ the traditional
  phase space defined at fixed $t$. This is precisely the mathematical
  structure conventionally applied to fields in space such that $\omega
  _t\equiv \omega _{tx}^{\times M}$ for $M$ spatial slices and $\omega _{tx}$
  the phase space of a single oscillator. In summary, we may write $\Omega
  \equiv \omega _{tx}^{\times N\cdot M}$.}\BibitemShut {Stop}%
\bibitem [{\citenamefont {Giovannetti}\ \emph {et~al.}(2023)\citenamefont
  {Giovannetti}, \citenamefont {Lloyd},\ and\ \citenamefont
  {Maccone}}]{gio.23}%
  \BibitemOpen
  \bibfield  {author} {\bibinfo {author} {\bibfnamefont {V.}~\bibnamefont
  {Giovannetti}}, \bibinfo {author} {\bibfnamefont {S.}~\bibnamefont {Lloyd}},\
  and\ \bibinfo {author} {\bibfnamefont {L.}~\bibnamefont {Maccone}},\
  }\bibfield  {title} {\bibinfo {title} {Geometric event-based quantum
  mechanics},\ }\href {https://doi.org/10.1088/1367-2630/acb793} {\bibfield
  {journal} {\bibinfo  {journal} {New J.\ Phys.}\ }\textbf {\bibinfo {volume}
  {25}},\ \bibinfo {pages} {023027} (\bibinfo {year} {2023})}\BibitemShut
  {NoStop}%
\bibitem [{\citenamefont {Schubert}(2001)}]{Schb.01}%
  \BibitemOpen
  \bibfield  {author} {\bibinfo {author} {\bibfnamefont {C.}~\bibnamefont
  {Schubert}},\ }\bibfield  {title} {\bibinfo {title} {Perturbative quantum
  field theory in the string-inspired formalism},\ }\href
  {https://doi.org/10.1016/S0370-1573%2801%2900013-8} {\bibfield  {journal}
  {\bibinfo  {journal} {Phys. Rep.}\ }\textbf {\bibinfo {volume} {355}},\
  \bibinfo {pages} {73} (\bibinfo {year} {2001})}\BibitemShut {NoStop}%
\bibitem [{\citenamefont {Diaz}\ and\ \citenamefont
  {Rossignoli}(2019)}]{di.19}%
  \BibitemOpen
  \bibfield  {author} {\bibinfo {author} {\bibfnamefont {N.~L.}\ \bibnamefont
  {Diaz}}\ and\ \bibinfo {author} {\bibfnamefont {R.}~\bibnamefont
  {Rossignoli}},\ }\bibfield  {title} {\bibinfo {title} {History state
  formalism for {D}irac’s theory},\ }\href
  {https://journals.aps.org/prd/abstract/10.1103/PhysRevD.99.045008} {\bibfield
   {journal} {\bibinfo  {journal} {Phys. Rev. D}\ }\textbf {\bibinfo {volume}
  {99}},\ \bibinfo {pages} {045008} (\bibinfo {year} {2019})}\BibitemShut
  {NoStop}%
\bibitem [{\citenamefont {Srednicki}(2007)}]{srednicki2007}%
  \BibitemOpen
  \bibfield  {author} {\bibinfo {author} {\bibfnamefont {M.}~\bibnamefont
  {Srednicki}},\ }\href@noop {} {\emph {\bibinfo {title} {Quantum field
  theory}}}\ (\bibinfo  {publisher} {Cambridge Univ.\ Press},\ \bibinfo {year}
  {2007})\BibitemShut {NoStop}%
\bibitem [{\citenamefont {Khanna}(2009)}]{kh.09}%
  \BibitemOpen
  \bibfield  {author} {\bibinfo {author} {\bibfnamefont {F.}~\bibnamefont
  {Khanna}},\ }\href {https://books.google.com.ar/books?id=iIRpDQAAQBAJ} {\emph
  {\bibinfo {title} {Thermal Quantum Field Theory: Algebraic Aspects and
  Applications}}}\ (\bibinfo  {publisher} {World Scientific},\ \bibinfo {year}
  {2009})\BibitemShut {NoStop}%
\bibitem [{Note4()}]{Note4}%
  \BibitemOpen
  \bibinfo {note} {In \cite {diazp.21} it is shown that for large $\tau $ one
  can rewrite the map as an asymptotic mean value of more complicated ($\tau $
  evolved) operators. Here we are more concerned with a notion of state for
  arbitrary (even small) $\tau $.}\BibitemShut {Stop}%
\bibitem [{\citenamefont {Nakata}\ \emph {et~al.}(2021)\citenamefont {Nakata},
  \citenamefont {Takayanagi}, \citenamefont {Taki}, \citenamefont {Tamaoka},\
  and\ \citenamefont {Wei}}]{nak.21}%
  \BibitemOpen
  \bibfield  {author} {\bibinfo {author} {\bibfnamefont {Y.}~\bibnamefont
  {Nakata}}, \bibinfo {author} {\bibfnamefont {T.}~\bibnamefont {Takayanagi}},
  \bibinfo {author} {\bibfnamefont {Y.}~\bibnamefont {Taki}}, \bibinfo {author}
  {\bibfnamefont {K.}~\bibnamefont {Tamaoka}},\ and\ \bibinfo {author}
  {\bibfnamefont {Z.}~\bibnamefont {Wei}},\ }\bibfield  {title} {\bibinfo
  {title} {New holographic generalization of entanglement entropy},\ }\href
  {https://doi.org/10.1103/PhysRevD.103.026005} {\bibfield  {journal} {\bibinfo
   {journal} {Phys. Rev. D}\ }\textbf {\bibinfo {volume} {103}},\ \bibinfo
  {pages} {026005} (\bibinfo {year} {2021})}\BibitemShut {NoStop}%
\bibitem [{\citenamefont {Doi}\ \emph {et~al.}(2023{\natexlab{b}})\citenamefont
  {Doi}, \citenamefont {Harper}, \citenamefont {Mollabashi}, \citenamefont
  {Takayanagi},\ and\ \citenamefont {Taki}}]{tak.23}%
  \BibitemOpen
  \bibfield  {author} {\bibinfo {author} {\bibfnamefont {K.}~\bibnamefont
  {Doi}}, \bibinfo {author} {\bibfnamefont {J.}~\bibnamefont {Harper}},
  \bibinfo {author} {\bibfnamefont {A.}~\bibnamefont {Mollabashi}}, \bibinfo
  {author} {\bibfnamefont {T.}~\bibnamefont {Takayanagi}},\ and\ \bibinfo
  {author} {\bibfnamefont {Y.}~\bibnamefont {Taki}},\ }\bibfield  {title}
  {\bibinfo {title} {Pseudoentropy in d{S/CFT} and timelike entanglement
  entropy},\ }\href {https://doi.org/10.1103/PhysRevLett.130.031601} {\bibfield
   {journal} {\bibinfo  {journal} {Phys.\ Rev.\ Lett.}\ }\textbf {\bibinfo
  {volume} {130}},\ \bibinfo {pages} {031601} (\bibinfo {year}
  {2023}{\natexlab{b}})}\BibitemShut {NoStop}%
\bibitem [{\citenamefont {Narayan}(2023)}]{nar.22}%
  \BibitemOpen
  \bibfield  {author} {\bibinfo {author} {\bibfnamefont {K.}~\bibnamefont
  {Narayan}},\ }\bibfield  {title} {\bibinfo {title} {de {S}itter space,
  extremal surfaces, and time entanglement},\ }\href
  {https://doi.org/10.1103/PhysRevD.107.126004} {\bibfield  {journal} {\bibinfo
   {journal} {Phys.\ Rev.\ D}\ }\textbf {\bibinfo {volume} {107}},\ \bibinfo
  {pages} {126004} (\bibinfo {year} {2023})}\BibitemShut {NoStop}%
\bibitem [{\citenamefont {Chu}\ and\ \citenamefont
  {Parihar}(2023)}]{chu2023time}%
  \BibitemOpen
  \bibfield  {author} {\bibinfo {author} {\bibfnamefont {C.-S.}\ \bibnamefont
  {Chu}}\ and\ \bibinfo {author} {\bibfnamefont {H.}~\bibnamefont {Parihar}},\
  }\bibfield  {title} {\bibinfo {title} {Time-like entanglement entropy in
  ads/bcft},\ }\href {https://doi.org/10.1007/JHEP06%282023%29173} {\bibfield
  {journal} {\bibinfo  {journal} {arXiv:2304.10907}\ } (\bibinfo {year}
  {2023})}\BibitemShut {NoStop}%
\bibitem [{\citenamefont {Maldacena}(1999)}]{mal.99}%
  \BibitemOpen
  \bibfield  {author} {\bibinfo {author} {\bibfnamefont {J.}~\bibnamefont
  {Maldacena}},\ }\bibfield  {title} {\bibinfo {title} {The large-n limit of
  superconformal field theories and supergravity},\ }\href
  {https://link.springer.com/article/10.1023/A:1026654312961} {\bibfield
  {journal} {\bibinfo  {journal} {Int.\ J.\ Theor.\ Phys.}\ }\textbf {\bibinfo
  {volume} {38}},\ \bibinfo {pages} {1113} (\bibinfo {year}
  {1999})}\BibitemShut {NoStop}%
\bibitem [{Note5()}]{Note5}%
  \BibitemOpen
  \bibinfo {note} {There are other useful extensions of the notion of state
  associated with ``quantum time'' (see e.g \cite {QT.15,b.16,di.19,dia.19}).
  However, none of these allow a partial trace over time regions \cite
  {diazp.21}.}\BibitemShut {Stop}%
\bibitem [{\citenamefont {Ryu}\ and\ \citenamefont
  {Takayanagi}(2006)}]{ryu.06}%
  \BibitemOpen
  \bibfield  {author} {\bibinfo {author} {\bibfnamefont {S.}~\bibnamefont
  {Ryu}}\ and\ \bibinfo {author} {\bibfnamefont {T.}~\bibnamefont
  {Takayanagi}},\ }\bibfield  {title} {\bibinfo {title} {Holographic derivation
  of entanglement entropy from the anti--de {S}itter space/conformal field
  theory correspondence},\ }\href
  {https://doi.org/10.1103/PhysRevLett.96.181602} {\bibfield  {journal}
  {\bibinfo  {journal} {Phys.\ Rev.\ Lett.}\ }\textbf {\bibinfo {volume}
  {96}},\ \bibinfo {pages} {181602} (\bibinfo {year} {2006})}\BibitemShut
  {NoStop}%
\bibitem [{\citenamefont {Casini}\ and\ \citenamefont {Huerta}(2009)}]{cas.09}%
  \BibitemOpen
  \bibfield  {author} {\bibinfo {author} {\bibfnamefont {H.}~\bibnamefont
  {Casini}}\ and\ \bibinfo {author} {\bibfnamefont {M.}~\bibnamefont
  {Huerta}},\ }\bibfield  {title} {\bibinfo {title} {Entanglement entropy in
  free quantum field theory},\ }\href
  {https://doi.org/10.1088/1751-8113/42/50/504007} {\bibfield  {journal}
  {\bibinfo  {journal} {J.\ Phys.\ A}\ }\textbf {\bibinfo {volume} {42}},\
  \bibinfo {pages} {504007} (\bibinfo {year} {2009})}\BibitemShut {NoStop}%
\bibitem [{\citenamefont {Aharonov}\ \emph {et~al.}(1988)\citenamefont
  {Aharonov}, \citenamefont {Albert},\ and\ \citenamefont {Vaidman}}]{yak.88}%
  \BibitemOpen
  \bibfield  {author} {\bibinfo {author} {\bibfnamefont {Y.}~\bibnamefont
  {Aharonov}}, \bibinfo {author} {\bibfnamefont {D.~Z.}\ \bibnamefont
  {Albert}},\ and\ \bibinfo {author} {\bibfnamefont {L.}~\bibnamefont
  {Vaidman}},\ }\bibfield  {title} {\bibinfo {title} {How the result of a
  measurement of a component of the spin of a spin-1/2 particle can turn out to
  be 100},\ }\href {https://doi.org/10.1103/PhysRevLett.60.1351} {\bibfield
  {journal} {\bibinfo  {journal} {Phys. Rev. Lett.}\ }\textbf {\bibinfo
  {volume} {60}},\ \bibinfo {pages} {1351} (\bibinfo {year}
  {1988})}\BibitemShut {NoStop}%
\bibitem [{\citenamefont {Dressel}\ \emph {et~al.}(2014)\citenamefont
  {Dressel}, \citenamefont {Malik}, \citenamefont {Miatto}, \citenamefont
  {Jordan},\ and\ \citenamefont {Boyd}}]{dres.14}%
  \BibitemOpen
  \bibfield  {author} {\bibinfo {author} {\bibfnamefont {J.}~\bibnamefont
  {Dressel}}, \bibinfo {author} {\bibfnamefont {M.}~\bibnamefont {Malik}},
  \bibinfo {author} {\bibfnamefont {F.~M.}\ \bibnamefont {Miatto}}, \bibinfo
  {author} {\bibfnamefont {A.~N.}\ \bibnamefont {Jordan}},\ and\ \bibinfo
  {author} {\bibfnamefont {R.~W.}\ \bibnamefont {Boyd}},\ }\bibfield  {title}
  {\bibinfo {title} {Colloquium: Understanding quantum weak values: Basics and
  applications},\ }\href {https://doi.org/10.1103/RevModPhys.86.307} {\bibfield
   {journal} {\bibinfo  {journal} {Rev.\ Mod.\ Phys.}\ }\textbf {\bibinfo
  {volume} {86}},\ \bibinfo {pages} {307} (\bibinfo {year} {2014})}\BibitemShut
  {NoStop}%
\bibitem [{\citenamefont {Wagner}\ \emph {et~al.}(2023)\citenamefont {Wagner},
  \citenamefont {Schwartzman-Nowik}, \citenamefont {Paiva}, \citenamefont
  {Te'eni}, \citenamefont {Ruiz-Molero}, \citenamefont {Barbosa}, \citenamefont
  {Cohen},\ and\ \citenamefont {Galv{\~a}o}}]{wag.23}%
  \BibitemOpen
  \bibfield  {author} {\bibinfo {author} {\bibfnamefont {R.}~\bibnamefont
  {Wagner}}, \bibinfo {author} {\bibfnamefont {Z.}~\bibnamefont
  {Schwartzman-Nowik}}, \bibinfo {author} {\bibfnamefont {I.~L.}\ \bibnamefont
  {Paiva}}, \bibinfo {author} {\bibfnamefont {A.}~\bibnamefont {Te'eni}},
  \bibinfo {author} {\bibfnamefont {A.}~\bibnamefont {Ruiz-Molero}}, \bibinfo
  {author} {\bibfnamefont {R.~S.}\ \bibnamefont {Barbosa}}, \bibinfo {author}
  {\bibfnamefont {E.}~\bibnamefont {Cohen}},\ and\ \bibinfo {author}
  {\bibfnamefont {E.~F.}\ \bibnamefont {Galv{\~a}o}},\ }\bibfield  {title}
  {\bibinfo {title} {Quantum circuits measuring weak values and
  {K}irkwood-{D}irac quasiprobability distributions, with applications},\
  }\href {https://doi.org/10.48550/arXiv.2302.00705} {\bibfield  {journal}
  {\bibinfo  {journal} {arXiv:2302.00705}\ } (\bibinfo {year}
  {2023})}\BibitemShut {NoStop}%
\bibitem [{\citenamefont {Aharonov}\ and\ \citenamefont
  {Kaufherr}(1984)}]{aha.84}%
  \BibitemOpen
  \bibfield  {author} {\bibinfo {author} {\bibfnamefont {Y.}~\bibnamefont
  {Aharonov}}\ and\ \bibinfo {author} {\bibfnamefont {T.}~\bibnamefont
  {Kaufherr}},\ }\bibfield  {title} {\bibinfo {title} {Quantum frames of
  reference},\ }\href {https://doi.org/10.1103/PhysRevD.30.368} {\bibfield
  {journal} {\bibinfo  {journal} {Phys. Rev. D}\ }\textbf {\bibinfo {volume}
  {30}},\ \bibinfo {pages} {368} (\bibinfo {year} {1984})}\BibitemShut
  {NoStop}%
\bibitem [{\citenamefont {Rovelli}(1991)}]{rov.91}%
  \BibitemOpen
  \bibfield  {author} {\bibinfo {author} {\bibfnamefont {C.}~\bibnamefont
  {Rovelli}},\ }\bibfield  {title} {\bibinfo {title} {Time in quantum gravity:
  An hypothesis},\ }\href {https://doi.org/10.1103/PhysRevD.43.442} {\bibfield
  {journal} {\bibinfo  {journal} {Phys. Rev. D}\ }\textbf {\bibinfo {volume}
  {43}},\ \bibinfo {pages} {442} (\bibinfo {year} {1991})}\BibitemShut
  {NoStop}%
\bibitem [{\citenamefont {Castro-Ruiz}\ \emph {et~al.}(2018)\citenamefont
  {Castro-Ruiz}, \citenamefont {Giacomini},\ and\ \citenamefont
  {Brukner}}]{cas.18}%
  \BibitemOpen
  \bibfield  {author} {\bibinfo {author} {\bibfnamefont {E.}~\bibnamefont
  {Castro-Ruiz}}, \bibinfo {author} {\bibfnamefont {F.}~\bibnamefont
  {Giacomini}},\ and\ \bibinfo {author} {\bibfnamefont {{\v{C}}.}~\bibnamefont
  {Brukner}},\ }\bibfield  {title} {\bibinfo {title} {Dynamics of quantum
  causal structures},\ }\href {https://doi.org/10.1103/PhysRevX.8.011047}
  {\bibfield  {journal} {\bibinfo  {journal} {Phys. Rev. X}\ }\textbf {\bibinfo
  {volume} {8}},\ \bibinfo {pages} {011047} (\bibinfo {year}
  {2018})}\BibitemShut {NoStop}%
\bibitem [{\citenamefont {Favalli}\ and\ \citenamefont
  {Smerzi}(2020)}]{fav.20}%
  \BibitemOpen
  \bibfield  {author} {\bibinfo {author} {\bibfnamefont {T.}~\bibnamefont
  {Favalli}}\ and\ \bibinfo {author} {\bibfnamefont {A.}~\bibnamefont
  {Smerzi}},\ }\bibfield  {title} {\bibinfo {title} {Time observables in a
  timeless universe},\ }\href
  {https://quantum-journal.org/papers/q-2020-10-29-354/} {\bibfield  {journal}
  {\bibinfo  {journal} {Quantum}\ }\textbf {\bibinfo {volume} {4}},\ \bibinfo
  {pages} {354} (\bibinfo {year} {2020})}\BibitemShut {NoStop}%
\bibitem [{\citenamefont {Weinberg}(1972)}]{Wein.72}%
  \BibitemOpen
  \bibfield  {author} {\bibinfo {author} {\bibfnamefont {S.}~\bibnamefont
  {Weinberg}},\ }\href@noop {} {\emph {\bibinfo {title} {Gravitation and
  Cosmology: Principles and Applications of the General Theory of
  Relativity}}}\ (\bibinfo  {publisher} {New York: Wiley},\ \bibinfo {year}
  {1972})\BibitemShut {NoStop}%
\bibitem [{\citenamefont {Rovelli}(1993)}]{rov.93}%
  \BibitemOpen
  \bibfield  {author} {\bibinfo {author} {\bibfnamefont {C.}~\bibnamefont
  {Rovelli}},\ }\bibfield  {title} {\bibinfo {title} {Statistical mechanics of
  gravity and the thermodynamical origin of time},\ }\href
  {https://doi.org/10.1088/0264-9381/10/8/015} {\bibfield  {journal} {\bibinfo
  {journal} {Class.\ Quantum Gravity}\ }\textbf {\bibinfo {volume} {10}},\
  \bibinfo {pages} {1549} (\bibinfo {year} {1993})}\BibitemShut {NoStop}%
\bibitem [{\citenamefont {Favalli}\ and\ \citenamefont
  {Smerzi}(2022{\natexlab{b}})}]{favalli2022peaceful}%
  \BibitemOpen
  \bibfield  {author} {\bibinfo {author} {\bibfnamefont {T.}~\bibnamefont
  {Favalli}}\ and\ \bibinfo {author} {\bibfnamefont {A.}~\bibnamefont
  {Smerzi}},\ }\bibfield  {title} {\bibinfo {title} {Peaceful coexistence of
  thermal equilibrium and the emergence of time},\ }\href
  {https://journals.aps.org/prd/abstract/10.1103/PhysRevD.105.023525}
  {\bibfield  {journal} {\bibinfo  {journal} {Phys.\ Rev.\ D}\ }\textbf
  {\bibinfo {volume} {105}},\ \bibinfo {pages} {023525} (\bibinfo {year}
  {2022}{\natexlab{b}})}\BibitemShut {NoStop}%
\bibitem [{\citenamefont {Fulling}(1973)}]{fulling1973nonuniqueness}%
  \BibitemOpen
  \bibfield  {author} {\bibinfo {author} {\bibfnamefont {S.~A.}\ \bibnamefont
  {Fulling}},\ }\bibfield  {title} {\bibinfo {title} {Nonuniqueness of
  canonical field quantization in {R}iemannian space-time},\ }\href
  {https://doi.org/10.1103/PhysRevD.7.2850} {\bibfield  {journal} {\bibinfo
  {journal} {Phys.\ Rev.\ D}\ }\textbf {\bibinfo {volume} {7}},\ \bibinfo
  {pages} {2850} (\bibinfo {year} {1973})}\BibitemShut {NoStop}%
\bibitem [{\citenamefont {Unruh}(1976)}]{unruh1976notes}%
  \BibitemOpen
  \bibfield  {author} {\bibinfo {author} {\bibfnamefont {W.~G.}\ \bibnamefont
  {Unruh}},\ }\bibfield  {title} {\bibinfo {title} {Notes on black-hole
  evaporation},\ }\href {https://doi.org/10.1103/PhysRevD.14.870} {\bibfield
  {journal} {\bibinfo  {journal} {Phys.\ Rev.\ D}\ }\textbf {\bibinfo {volume}
  {14}},\ \bibinfo {pages} {870} (\bibinfo {year} {1976})}\BibitemShut
  {NoStop}%
\bibitem [{Note6()}]{Note6}%
  \BibitemOpen
  \bibinfo {note} {While this procedure is analogous to the PaW approach, as
  shown recently in \cite {hoehn2023matter}, therein matter fields are not
  quantized via an extended algebra as in our scheme.}\BibitemShut {Stop}%
\bibitem [{\citenamefont {Arnowitt}\ \emph {et~al.}(1959)\citenamefont
  {Arnowitt}, \citenamefont {Deser},\ and\ \citenamefont {Misner}}]{adm.59}%
  \BibitemOpen
  \bibfield  {author} {\bibinfo {author} {\bibfnamefont {R.}~\bibnamefont
  {Arnowitt}}, \bibinfo {author} {\bibfnamefont {S.}~\bibnamefont {Deser}},\
  and\ \bibinfo {author} {\bibfnamefont {C.~W.}\ \bibnamefont {Misner}},\
  }\bibfield  {title} {\bibinfo {title} {Dynamical structure and definition of
  energy in general relativity},\ }\href
  {https://doi.org/10.1103/PhysRev.116.1322} {\bibfield  {journal} {\bibinfo
  {journal} {Phys. Rev.}\ }\textbf {\bibinfo {volume} {116}},\ \bibinfo {pages}
  {1322} (\bibinfo {year} {1959})}\BibitemShut {NoStop}%
\bibitem [{\citenamefont {Balian}\ and\ \citenamefont {Brezin}(1969)}]{bal.69}%
  \BibitemOpen
  \bibfield  {author} {\bibinfo {author} {\bibfnamefont {R.}~\bibnamefont
  {Balian}}\ and\ \bibinfo {author} {\bibfnamefont {E.}~\bibnamefont
  {Brezin}},\ }\bibfield  {title} {\bibinfo {title} {Nonunitary {B}ogoliubov
  transformations and extension of {W}ick’s theorem},\ }\href
  {https://doi.org/10.1007/BF02710281} {\bibfield  {journal} {\bibinfo
  {journal} {Il Nuovo Cimento B (1965-1970)}\ }\textbf {\bibinfo {volume}
  {64}},\ \bibinfo {pages} {37} (\bibinfo {year} {1969})}\BibitemShut {NoStop}%
\bibitem [{\citenamefont {Buhrman}\ \emph {et~al.}(2001)\citenamefont
  {Buhrman}, \citenamefont {Cleve}, \citenamefont {Watrous},\ and\
  \citenamefont {De~Wolf}}]{buhrman2001quantum}%
  \BibitemOpen
  \bibfield  {author} {\bibinfo {author} {\bibfnamefont {H.}~\bibnamefont
  {Buhrman}}, \bibinfo {author} {\bibfnamefont {R.}~\bibnamefont {Cleve}},
  \bibinfo {author} {\bibfnamefont {J.}~\bibnamefont {Watrous}},\ and\ \bibinfo
  {author} {\bibfnamefont {R.}~\bibnamefont {De~Wolf}},\ }\bibfield  {title}
  {\bibinfo {title} {Quantum fingerprinting},\ }\href
  {https://doi.org/10.1103/PhysRevLett.87.167902} {\bibfield  {journal}
  {\bibinfo  {journal} {Phys.\ Rev.\ Lett.}\ }\textbf {\bibinfo {volume}
  {87}},\ \bibinfo {pages} {167902} (\bibinfo {year} {2001})}\BibitemShut
  {NoStop}%
\bibitem [{Note7()}]{Note7}%
  \BibitemOpen
  \bibinfo {note} {In the continuum version of main text, $|\phi \rangle $
  corresponds to $|\phi ({\protect \bf x})\rangle $ whereas $|\protect \bm
  {\phi }\rangle $ to $|\phi (x)\rangle $}\BibitemShut {NoStop}%
\bibitem [{\citenamefont {Banados}\ and\ \citenamefont
  {Reyes}(2016)}]{banados2016short}%
  \BibitemOpen
  \bibfield  {author} {\bibinfo {author} {\bibfnamefont {M.}~\bibnamefont
  {Banados}}\ and\ \bibinfo {author} {\bibfnamefont {I.}~\bibnamefont
  {Reyes}},\ }\bibfield  {title} {\bibinfo {title} {A short review on
  {N}oether’s theorems, gauge symmetries and boundary terms},\ }\href
  {https://doi.org/10.1142/S0218271816300214} {\bibfield  {journal} {\bibinfo
  {journal} {Int.\ J.\ Mod.\ Phys.\ D}\ }\textbf {\bibinfo {volume} {25}},\
  \bibinfo {pages} {1630021} (\bibinfo {year} {2016})}\BibitemShut {NoStop}%
\bibitem [{\citenamefont {Regge}\ and\ \citenamefont
  {Teitelboim}(1974)}]{regge1974role}%
  \BibitemOpen
  \bibfield  {author} {\bibinfo {author} {\bibfnamefont {T.}~\bibnamefont
  {Regge}}\ and\ \bibinfo {author} {\bibfnamefont {C.}~\bibnamefont
  {Teitelboim}},\ }\bibfield  {title} {\bibinfo {title} {Role of surface
  integrals in the {H}amiltonian formulation of general relativity},\ }\href
  {https://doi.org/10.1016/0003-4916(74)90404-7} {\bibfield  {journal}
  {\bibinfo  {journal} {Ann.\ Phys.\ (NY)}\ }\textbf {\bibinfo {volume} {88}},\
  \bibinfo {pages} {286} (\bibinfo {year} {1974})}\BibitemShut {NoStop}%
\end{thebibliography}
\end{document}